\pdfoutput=1
\documentclass[12pt]{article}
\usepackage[slantedGreek]{mathptmx}
\usepackage{fullpage}
\usepackage{tikz-cd}

\usepackage{graphicx}        
\usepackage{amssymb,amsmath,amsthm}
\usepackage{amsfonts}
\usepackage{stmaryrd}
\usepackage{comment}
\usepackage{subfigure}
\usepackage{enumitem}
\usepackage{pdflscape}
\usepackage{caption}
\renewcommand{\vec}[1]{\textbf{\textit{#1}}}
\renewcommand{\d}{\text{d}}

\DeclareSymbolFont{newfont}{OML}{cmm}{m}{it}
\DeclareMathSymbol{\epsilon}{3}{newfont}{15}

 \theoremstyle{theorem} 
 \newtheorem{thm}{Theorem}[subsection]
 \newtheorem{cor}{Corollary}[subsection]
 \newtheorem{prop}{Proposition}[subsection]
 
 \theoremstyle{definition}
\newtheorem*{pf}{Proof}
\newtheorem{rmk}{Remark}[subsection]

\numberwithin{equation}{section}
\usepackage[numbers,square,comma,sort&compress]{natbib}
\title{Generalized Finsler geometry and the anisotropic tearing of skin} 

\author{J.D. Clayton$^{1}$ \\ 
$^1$DEVCOM ARL, Aberdeen, MD 21005, USA 
}

\date
\nopagebreak

\begin{document}
\graphicspath{{figures/}}
\maketitle

\begin{abstract}
A continuum mechanical theory with foundations in generalized Finsler geometry 
describes the complex anisotropic behavior of skin.
A fiber bundle approach, encompassing
total spaces with assigned linear and nonlinear connections, geometrically characterizes evolving 
configurations of a deformable body with microstructure.
An internal state vector is introduced on each configuration, describing
subscale physics.
A generalized Finsler metric depends on position and the state vector,
where the latter dependence allows for both direction (i.e., as in Finsler
geometry) as well as magnitude.
Equilibrium equations are derived using a variational method,
extending concepts of finite-strain hyperelasticity coupled to
phase-field mechanics to generalized Finsler space.
For application to skin tearing, state vector components
represent microscopic damage processes (e.g., fiber rearrangements and
ruptures) in different directions with respect to intrinsic orientations
(e.g., parallel or perpendicular to Langer's lines).
Nonlinear potentials, motivated from soft-tissue mechanics and
 phase-field fracture theories, are assigned with
 orthotropic material symmetry pertinent to properties of skin.
Governing equations are derived for one- and two-dimensional
base manifolds.
Analytical solutions capture 
experimental force-stretch data, toughness, and observations
on evolving microstructure, in a more geometrically and
physically descriptive way than
prior phenomenological models.
\end{abstract}
\noindent \textbf{Mathematics Subject Classification (MSC) 2020}: 53Z05 (primary), 53B40, 74B20

\noindent

\tableofcontents

\section{Introduction}
Finsler geometry and its generalizations suggest the possibility of
enriched descriptions of numerous phenomena in mathematical physics,
albeit at the likely expense of greater complexity of analysis and
calculations compared to Riemannian geometry.
Fundamentals of Finsler geometry, aptly credited to Finsler \cite{finsler1918}, are discussed in
the classic monograph of Rund and the more recent text of Bao et al.~\cite{rund1959,bao2000}.
See also the overview article by Eringen \cite{eringen1971}.
Extensions to pseudo- and generalized Finsler geometries are described
in the monograph of Bejancu \cite{bejancu1990} and research cited therein \cite{miron1983,watanabe1983,matsumoto1986}, as
well as several more recent works \cite{bejancu2000,minguzzi2014}.

Generalized Finsler geometry is predominantly used
herein, since strict classical Finsler geometry is insufficient to describe all phenomena pertinent to the present class of materials physics.
Applications of (generalized) Finsler geometry in the broad physical sciences are vast and diverse; a thorough 
review is outside the scope of the present article.
Available books discuss applications in optics, thermodynamics, and biology \cite{antonelli1993}
and spinor structures and other topics in modern physics \cite{vacaru2005}.
Finsler geometry and its generalizations have also been used for describing anisotropic space-time, general relativity, quantum fields, gravitation, electromagnetism, and diffusion \cite{brandt1992,voicu2010,balan2011,ma2021,hoh2022,popov2022,abbo2022}.
The current work implements a continuum mechanical theory for the
behavior of solid materials.  

\subsection{Background}

Classical continuum mechanics, encompassing
nonlinear elasticity and plasticity theories as example constitutive frameworks, is couched in the
context of Riemannian-Cartan manifolds \cite{truesdell1960,kroner1960,marsden1983}. 
Non-vanishing torsion and/or curvature tensors may emerge depending on linear connections introduced
to describe various incompatibilities and possible sources of residual stresses
including dislocations and disclinations in crystals \cite{kroner1960,toupin1960,yavari2012,clayton2015}, 
inhomogeneous temperature distributions \cite{stojanovitch1969,ozakin2010},
or biological growth processes \cite{taka1991,rodriguez1994,yavari2010}.

In the classical Riemannian context, a continuous material body is viewed as a differentiable manifold $\mathcal{M}$ of dimension $n$, parameterized by 
coordinate chart(s) $\{X^A\}$ ($A=1,\ldots,n$).  A Riemannian metric is introduced on $\mathcal{M}$: in components, $G_{AB} = G_{AB}(X)$ populate a symmetric and positive-definite $n \times n$ tensor field, where argument $X$ implies functional dependence on the $X^A$ \cite{marsden1983}.  A covariant derivative $\nabla$ (i.e., the linear connection on $\mathcal{M}$) completes the geometric description.
 The corresponding linear connection coefficients most generally consist of $n^3$ independent field components $\Gamma^A_{BC} = \Gamma^A_{BC}(X)$.  For usual solid bodies, $n=3$, but other dimensions are permissible.
Coordinate descriptions may be framed in terms of holonomic or anholonomic bases, where the latter do not correspond to continuous curves parameterizing the body \cite{schouten1954,clayton2012}; anholonomic coordinates emerge for a multiplicative decomposition of the deformation gradient \cite{clayton2011,steinmann2015}.

In Finsler geometry and its generalizations, a base manifold $\mathcal{M}$, parameterized by one or more charts $\{X^A\}$ ($A=1,\ldots,n$), is again introduced.  A fiber bundle $(\mathcal{Z},\mathcal{M},\Pi,\mathcal{U})$ of total (generalized Finsler) space $\mathcal{Z}$ of dimension $n+m$ is constructed.  The total space in Finsler geometry is typically identified with the slit tangent bundle, i.e., $\mathcal{Z} \rightarrow T \! \mathcal{M} \backslash 0$ \cite{bao2000} with $m=n$, but this is not essential in more general formulations \cite{minguzzi2014,bejancu2000,bejancu1990}.
Auxiliary coordinates $\{D^K\}$ ($K=1,\ldots, m$) cover each fiber $\mathcal{U}$, such that $\mathcal{Z}$ is
parameterized by $\{X^A,D^K\}$.  Particular transformation laws are assigned for changes of coordinates associated with $X$ and $D$.  Nonlinear connection coefficients define non-holonomic bases that transform conventionally on $T \mathcal{M}$ under $X$-coordinate changes.  Furthermore, at least two, and up to four \cite{bejancu1990,hushmandi2012}, additional fields of connection coefficients  are needed to enable horizontal and vertical covariant derivatives with respect to $X$ and $D$ on $\mathcal{Z}$ \cite{rund1959,bejancu1990,minguzzi2014}.  

The generalized pseudo-Finsler metric tensor
is of the form $G_{AB} = G_{AB}(X,D)$, always symmetric. The metric is positive definite for Finsler geometry, but not always so for the pseudo-Finsler case \cite{minguzzi2014}. For strict Finsler geometry (but not necessarily its generalizations \cite{watanabe1982,miron1983,bejancu1990,popov2022}), $G_{AB}$ are second derivatives of a (squared) fundamental function $ \frac{1}{2} \mathcal{F}^2$ with respect to $D^A$ \cite{rund1959,bao2000}.  In Finsler geometry, $\mathcal{F}$ is positively homogeneous of degree one in $D$; the resulting metric is homogeneous of degree zero in $D$ \cite{rund1959,bao2000} and thus should not depend only on the magnitude of a vector comprised of components $\{ D^A \}$.  In (generalized) Finsler space, the $(X,D)$-coordinate dependencies of the metric and various linear and nonlinear connections enter other geometric objects and tensorial operations: torsion and curvature forms, volume and area forms, the divergence theorem \cite{rund1975}, etc.

Motivation for Finsler geometry is
description of detailed physics via the set of auxiliary coordinates $\{ D^A \}$ attached to each position $X$ in real physical space.
In solid mechanics, the idea can be interpreted as an extension of micropolar, micromorphic, or Cosserat-type theory \cite{toupin1964,mindlin1964,eringen1968,regueiro2010,eremeyev2013,eremeyev2020} from Riemannian (and more often, Euclidean) space to generalized Finsler space. 
However, in classical micromorphic theories, a Riemannian, rather than Finsler, metric is used, the material domain
with microstructure is fully parameterized by the $\{X^A\}$; basis vectors and
coordinate transformation laws are those of classical continuum mechanics, 
as are integral theorems.  The director triads of micromorphic theories
enter the balance laws and constitutive functions, but they do not affect geometric objects and covariant derivatives in the same way as $D$ of generalized Finsler space.

\subsection{Prior work}

The first application of Finsler geometry in the context of continuum mechanics of solids
appears to be the treatment of ferromagnetic elastic-plastic crystals of Amari \cite{amari1962}.
Conservation laws and field theories, with application to ferromagnetics, were further developed
by Ikeda \cite{ikeda1973,ikeda1981,ikeda1981b,ikeda1981c}.
Bejancu \cite{bejancu1990} gives a generalized Finsler treatment of kinematics of
deformable bodies.
More contemporary theories includes those of Saczuk and Stumpf \cite{saczuk1997,stumpf2000,stumpf2002},
with underpinnings in a monograph \cite{saczuk1996}. 
Different physical phenomena (e.g, different physical meanings of $\{ D^K \}$ \cite{stumpf2002}) are encompassed by their models that include kinematics, balance laws,
and thermodynamics, but their focus is most often on mechanics of elastic-plastic crystals and dislocations \cite{saczuk1996,saczuk1997,stumpf2000}.
See also recent theory \cite{yajima2020} that applies generalized Finsler geometry to topological defects and the comprehensive review \cite{claytonMMS22} of prior works on generalized Finsler geometry in continuum physics.

A new theory of Finsler-geometric continuum mechanics was developed for nonlinear elastic solids with evolving
microstructure, first published in the article \cite{clayton2017g} with a preliminary version in a technical report \cite{clayton2016}.
This variational theory was extended to allow for explicit inelastic deformation and applied to study phase transitions and shear localization
in crystalline solids \cite{clayton2017g,clayton2018p}. The theory has also been broadened for dynamics and shock waves \cite{clayton2017f,clayton2018c}, and most recently has been used to describe ferromagnetic solids \cite{claytonMMS22}, enriching the governing equations of Maugin and Eringen \cite{maugin1972a,maugin1972b} with 
pertinent aspects arising from Finsler geometry \cite{amari1962,ikeda1981c} .

Prior to this theory \cite{clayton2017g,claytonMMS22}, pragmatic
solutions of boundary value problems using 
continuum mechanical models incorporating generalized Finsler geometry appeared intractable due to complexity of governing equations and unwieldy parameterizations (e.g., uncertain constitutive functions and material constants).  
Most aforementioned work \cite{amari1962,ikeda1973,ikeda1981,ikeda1981b,ikeda1981c,bejancu1990,saczuk1997,stumpf2002,yajima2020} presented purely theoretical constructions without 
attempt to formulate or solve physical boundary value problems.
A material response was calculated by Saczuk and Stumpf \cite{saczuk1996,stumpf2000}, but motion and internal state coordinates were
prescribed a priori, without apparent solution of governing conservation laws for macroscopic and microscopic momentum and energy.
In contrast, the present theory \cite{clayton2017g,clayton2016} appears to be the first Finsler geometry-based continuum mechanics theory for which analytical and numerical solutions to the governing equations have been found, as evidenced by solutions to numerous problems for (non)linear elastic materials with evolving microstructure (e.g, fractures, twinning, phase transitions, dislocations), as evidenced in those and subsequent works \cite{clayton2017g,clayton2016,clayton2017z,clayton2017f,clayton2018p,clayton2018c,claytonMMS22}.

All prior applications considered stiff crystalline solids or generic materials.
The current research newly applies the theory to soft biological tissues, specifically the skin.
Furthermore, prior applications in fracture and cavitation \cite{clayton2017g,clayton2017z,clayton2018c,claytonMMS22}
were limited to either locally isotropic damage or to local material separation on a single cleavage plane.
The current treatment advances the description to anisotropic fractures or ruptures on multiple material surfaces at a single point $X$.
Most cited prior applications invoked only a single non-trivial state vector component in $D$ (an 
exception being a multi-component $D$ for twinning and fracture \cite{clayton2018c})
and most often conformal Weyl-type rescaling of $G_{AB}$ with canonically vanishing nonlinear connection
(with a few exceptions studied \cite{clayton2017z,clayton2018p}). 
The current research incorporates an anisotropic generalized Finsler metric for multi-dimensional problems and non-trivial nonlinear connections to show utility by example.

\subsection{Purpose and scope}

\noindent The scope of this paper covers two primary purposes:
\begin{itemize}[noitemsep,topsep=0pt] 
\item Demonstration of utility of the generalized Finsler geometric theory for describing anisotropic elasticity and anisotropic structural rearrangements
in soft biological tissue;
\item Consolidation and refinement of the theory for the
equilibrium (i.e., quasi-static) case.
\end{itemize} 
The first item furnishes the first known application of Finsler geometry-based continuum theory
 to analyze finite-strain mechanics of soft biological tissue. Prior work of others \cite{takano2017,mitsuhashi2018} used ideas from Finsler geometry to reproduce nonlinear stress-strain to failure responses of biologic solids, but that work used a discrete, rather than continuum, theory with material points represented as vertices linked by bonds; interaction potentials comprised bonding energies within a Hamiltonian. In that approach \cite{koibuchi2014,koibuchi2016,koibuchi2019}, a Finsler metric for bond stretch depends on orientation of local microstructure entities (e.g., molecular chains or collagen fibers) described by the Finsler director vector field $D$.
 Instead, the current continuum theory considers, in a novel way, effects of microstructure on anisotropy (elastic and damage-induced) in both a geometric and constitutive sense. The second item includes a renewed examination of Rund's divergence theorem \cite{rund1975} in the
context of an osculating Riemannian metric. It is shown that certain choices of
metric and connection coefficients, with possible addition of a source term to the energy conservation law,
can recover governing equations for biologic tissue growth \cite{yavari2010} in the quasi-static limit (Appendix B).
 
 \subsubsection{Soft tissue and skin mechanics}

Most soft tissues have inherent directionality
due to their collagen fiber-based and/or aligned cellular microstructures \cite{fung1993,cowin2007}, toward which tools of analysis from Finsler geometry might be anticipated to aptly apply.
The mechanics of skin deformation \cite{lanir1974,fung1993,ni2012}, degradation \cite{munoz2008,tubon2022}, and tearing \cite{yang2015,tubon2022} are investigated herein.
Like most biological materials, microstructure of skin is complex. The respective middle and outer layers of skin are the dermis and epidermis, with elastin and collagen fibers and cells embedded in a ground matrix. Underlying hypodermis (i.e., adipose) can be labeled an inner layer of the skin. The microstructure dictates nonlinear, anisotropic, viscoelastic, and tearing behaviors \cite{yang2015,jood2018,oft2018}. 
Mechanical behavior at small strains is primarily controlled by the elastin and ground substance, whereby the collagen fibers are coiled or slack \cite{jood2018}. Under increasing tensile stretch, the collagen fibers straighten and tighten, supporting most of the load, and compliance decreases. Under more severe stretching, fibers slide, delaminate, and rupture, leading to reduced stiffness, strain softening, and material failure \cite{munoz2008,gasser2011,yang2015,tubon2022}.

Experiments indicate that skin elasticity has orthotropic symmetry \cite{lanir1974,fung1993,ni2012,jood2018}.
Orthotropy arises from preferred arrangements of the collagen fibers, leading to greater stiffness along directions along which more fibers are aligned. In the plane of the dermis, fibers tend to be dispersed about a primary axis along which stiffness is greatest. In vivo, resting
skin tension is greatest along this axis, parallel to Langer's lines \cite{jood2018}.
In typical uniaxial and biaxial tests \cite{lanir1974,fung1993,ni2012,yang2015}, extracted skin is unstretched initially, but the greater stiffness along the primary direction persists, with differences in stiffness also emerging between orthogonal in-plane and out-of-plane directions \cite{lanir1974,jood2018}. As might be expected, damage processes are also anisotropic due to fiber degradation that differs with respect to direction of loading relative to the microstructure \cite{yang2015,tubon2022}.

Skin, as is most biological tissue, is simultaneously nonlinear elastic, viscoelastic, and poroelastic \cite{rubin2002,lanir1979,fung1993,oft2018}; pertinence of mechanisms  depends on the time scale of loading.
The present application considers only monotonic loading at a constant rate (e.g, no cycling or rate fluctuations).
Loading rates are assumed much slower or faster than viscous relaxation times.
Thus, the pseudo elastic approach is justified to study these experiments \cite{fung1993}, whereby hyperelastic models are deemed reasonable \cite{holzapfel2000,balzani2006,holzapfel2009,ni2012,nolan2014}, albeit noting that different elastic constants (e.g., static and dynamic moduli) are needed to fit data at vastly different limiting low and high loading rates \cite{lim2011,claytonMOSM20}. 
In future applications to problems with time dependence, internal state variables can be extended, leading to kinetic laws with explicit viscous dissipation \cite{rubin2002,holz1996a}.
The current study is limited to relatively small samples, tested in vitro, under uniaxial or biaxial extension \cite{yang2015,fung1993,lanir1974,ani2012}. The material is modeled as unstressed initially and homogeneous with regard to elastic properties. In the future, the current theory can be extended to study residual stress due to growth or heterogeneous material features, as well as heterogeneous elastic properties.  Residual stresses can be addressed, in the context of Riemannian manifolds, using a material metric having a non-vanishing Riemann-Christoffel curvature of its Levi-Civita connection \cite{yavari2010,ozakin2010} or an anholonomic multiplicative term in the deformation gradient \cite{rodriguez1994,skalak1996}. These ideas may be extended to generalized Finsler space (e.g., invoking the current fiber bundle approach) in future.

An early nonlinear elastic model described orthotropic symmetry using a phenomenological pseudo-strain energy potential \cite{tong1976}.
Another early model delineated contributions of elastin and collagen fibers \cite{lanir1979}.
More recently, a class of nonlinear elastic models accounting for anisotropy from fiber arrangements using structure tensors
has been successful for representing many soft tissues, including arterial walls \cite{holzapfel2000,balzani2012}, myocardium \cite{holzapfel2009,gultekin2016}, and skin \cite{ni2012}.
Polyconvex energy potentials can be incorporated for stability and to facilitate existence of (unique) solutions to nonlinear elastic problems \cite{balzani2006,balzani2012}. Fiber dispersion can be incorporated to modulate the degree of anisotropy \cite{gasser2006,ni2012}.
To date, most damage models accounting for softening and failure have been phenomenological,
whether implemented at the macroscopic scale (either isotropic or along preferred fiber directions) or at the scale of individual fibers and their distributions \cite{rodriguez2006,gasser2011,balzani2012,tubon2022}.
These damage models, with a basis in continuum damage mechanics \cite{simo1987}, are thermodynamically consistent in the sense that damage is dissipative, but their particular kinetic laws and (often numerous) parameters are calibrated to experimental data without much physical meaning. In contrast, the phase-field approach has been recently implemented for soft-tissue fracture or rupture, incorporating
relatively few parameters with physical origin (e.g., surface energy) and regularization facilitating unique solutions to problems involving material softening \cite{gultekin2019,chit2022}. The kinetic law or equilibrium equation for damage is derived
from fundamental principles \cite{gurtin1996} and drives material to a local minimum-energy state, in contrast to ad hoc equations simply selected to match data. 

\subsubsection{Overview of the current work}

Implementation of the present generalized Finsler theory consists of four key elements: definition of the internal state $D$, assignment of the metric tensor, assignment of the linear and nonlinear connections, and prescription of the local free energy potential.
For soft tissue mechanics, the state vector represents the fiber rearrangements.
Damage anisotropy is monitored via its direction, with different components of $D$ reflecting fiber reorganization and rupture with respect to orientations of microstructure features \cite{yang2015,tubon2022}; the magnitude of each component of $D$ measures local intensity of damage in a given material direction.
The metric tensor with components ${G}_{AB} (X,D)$ depends on position $X$ as well as direction and magnitude of $D$ in generalized Finsler space; novel $D$-dependence captures rescaling of the material manifold as damage entities open, close, or rearrange in different directions \cite{clayton2017z,claytonMMS22}.
The preferred linear connection is that of Chern and Rund \cite{bao2000}, ensuring compatibility with the divergence theorem
used to derive the Euler-Lagrange equations \cite{clayton2017g,claytonMMS22}.
The generalized Finslerian $D$-dependence of both the metric and linear connection explicitly affect the governing equations.
Roles of nonlinear connections are newly examined; a non-trivial prescription is shown to influence the fracture energy and stress-strain response.


The free energy density consists of nonlinear elastic contribution and an internal structure contribution. The nonlinear elastic potential enriches the orthotropic theory of Holzapfel, Ogden, Gasser, and others \cite{holzapfel2000,gasser2006,holzapfel2009,ni2012,nolan2014}
with implicit contributions from the generalized Finsler metric as well as anisotropic degradation from $D$.
The structural contribution is motivated from phase-field mechanics \cite{gultekin2019,clayton2014}.
A previous model for arterial dissection \cite{gultekin2019} accounted for fiber-scale damage anisotropy using a scalar order parameter. The current theory invokes a more physically descriptive, vector-valued order parameter (i.e., normalized $D$) of generalized Finsler type.
With regard to skin experiments, solutions obtained for the current model are shown
to admirably match extension and failure data, including stress-strain behavior and fracture toughness \cite{per1997,yang2015,tubon2022} with parameters having physical or geometric origins.
The general theory is thus potentially more physically realistic, and considered more descriptive from a geometric perspective, than past models based on phenomenological damage mechanics \cite{simo1987,balzani2012,li2016,claytonBM20}. 

This paper is organized as follows. Mathematical preliminaries (e.g., notation and definitions for objects in referential and spatial configurations) are provided in \S2. 
The Finsler-geometric theory of continuum mechanics is presented in \S3, including kinematics of finite deformation and equilibrium equations derived with a variational approach.
The next two sections specialize the theory to model soft tissue, specifically skin.
In \S4, a one-dimensional (1-D) model for the base manifold $\mathcal{M}$ is formulated. Analytical and semi-numerical solutions
are obtained for uniaxial extension and compared to experimental data.
In \S5, a two-dimensional (2-D) model for $\mathcal{M}$ is formulated, whereby the skin has orthotropic symmetry; solutions are obtained for
biaxial extension with anisotropic damage in orthogonal material directions.
Conclusions follow in \S6.

\section{Generalized Finsler space}

Content of \S2 consolidates a more thorough exposition given in a recent review \cite{claytonMMS22}, from which notation is adopted.
Other extensive texts include those of Rund, Bejancu, and Bao et al. \cite{rund1959,bejancu1990,bao2000}.   
A new contribution in the present \S2 is interpretation of the divergence theorem \cite{rund1975,claytonMMS22} using an osculating Riemannian metric, whereby for the further simplifying assumption of vanishing nonlinear connection, a representation akin to that of classical Riemannian geometry is obtained.

\subsection{Reference configuration}


The very general fiber bundle approach of Bejancu \cite{bejancu1990} encompasses geometric fundamentals of the theory.  
The reference configuration is linked to a particular instant in time at which
a deformable solid body is undeformed relative to some intrinsic state.  A differential manifold $\mathcal{M}$ of 
 dimension $n$ is physically identified with a body 
embedded in ambient Euclidean space of dimension $N \geq n$.  
\begin{rmk}
Such an embedding only applies to base manifold $\mathcal{M}$. Neither the total space of the fiber bundle $\mathcal{Z}$, to be introduced in what follows, nor its specialization to a Finsler space $F_n$ discussed in \S2.1.5, can generally be embedded in Euclidean space \cite{deicke1955,yano1954,rund1959}. 
\end{rmk}
Let $X \in \mathcal{M}$ denote a material
point or particle, and let $\{X^A\}(A=1,2,\ldots,n)$ denote a coordinate chart
that may partially or completely cover $\mathcal{M}$.  
Attached to each material point is a vector $\vec{D}$;
chart(s) of secondary coordinates $\{D^K\}(K=1,2,\ldots,m)$ are assigned over $\mathcal{M}$.  
Fields $\{D^K\}$ are smooth over $\mathcal{M}$: $\vec{D}$ is
as many times continuously differentiable with respect to $\{X^A\}$ as needed.

Define 
$\mathsf{Z} = (\mathcal{Z},\Pi,\mathcal{M},\mathcal{U})$ as a fiber bundle of total
space $\mathcal{Z}$ (dimension $n+m$), 
where $\Pi:\mathcal{Z} \rightarrow \mathcal{M}$
is the projection and $\mathcal{U} = \mathcal{Z}_X = \Pi^{-1} (X) $ is the fiber at $X$.  
A chart over a region of
$\mathcal{Z}$ is $\{X^A,D^K\}$. 
Each fiber is a vector space of dimension $m$; ($\mathcal{Z},\Pi,\mathcal{M})$ constitutes 
a vector bundle.
 Let $\mathcal{M}' \subset \mathcal{M}$ be an open neighborhood of any $X \in \mathcal{M}$, 
 $\Phi$ an isomorphism of vector spaces, and $P_1$ a projection operator onto the first factor.  Then the following diagram is commutative \cite{bejancu1990}:
\newline  \\
\begin{tikzcd}
\qquad   \Pi^{-1}(\mathcal{M}') \arrow[d, "\Pi"]  \arrow[r, "\Phi"] & \mathcal{M}' \times \mathbb{R}^m
  \arrow[ld,"P_1" ] \\
  \mathcal{M}'
\end{tikzcd}
\\
\subsubsection{Basis vectors and nonlinear connections}

Coordinate transformations
from $\{X,D\}$ to another chart $\{\tilde{X},\tilde{D}\}$ on $\mathcal{Z}$ are of the form 
\cite{bejancu1990,minguzzi2014}
\begin{equation}
\label{eq:trans1}
\tilde{X}^A = \tilde{X}^A(X),
\qquad \tilde{D}^J(X,D) = Q^J_K (X) D^K,
\end{equation}
where $Q^J_K$ is non-singular and differentiable, with 
inverse obeying $\tilde{Q}^I_K  Q^K_J = \delta^I_J$.
As usual, $\delta^I_J = 1 \, \forall \, I=J, \delta^I_J = 0 \, \forall \, I \neq J$.The holonomic basis for the tangent bundle $ T \mathcal{Z}$ is the field of frames
$\{\frac{\partial}{\partial X^A},\frac{\partial}{\partial D^K}\}$.  
The holonomic basis for the cotangent bundle $ T^* \! \mathcal{Z}$
is $\{d X^A, d D^K \}$.
Under a change of coordinates $(X,D) \rightarrow (\tilde{X},\tilde{D})$ 
on $\mathcal{Z}$ induced by
 $X \rightarrow \tilde{X}$ on $\mathcal{M}$, holonomic basis vectors on $T \mathcal{Z}$ 
transform from \eqref{eq:trans1} as \cite{bejancu1990,minguzzi2014}
\begin{equation}
\label{eq:trans2}
\frac{\partial}{\partial \tilde{X}^A}  = \frac{\partial X^B}{\partial \tilde{X}^A}\frac{\partial}{\partial X^B}
+ \frac{\partial D^K}{\partial \tilde{X}^A  }  \frac{\partial}{\partial D^K}
= \frac{\partial X^B}{\partial \tilde{X}^A}\frac{\partial}{\partial X^B}
+ \frac{\partial \tilde{Q}^K_J}{\partial \tilde{X}^A } \tilde{D}^J  \frac{\partial}{\partial D^K}, 
\end{equation}
\begin{equation}
\label{eq:trans2b}
\frac{\partial}{\partial \tilde{D}^J}  = \frac{\partial X^B}{\partial \tilde{D}^J}\frac{\partial}{\partial X^B} +
\frac{\partial D^K}{\partial \tilde{D}^J}\frac{\partial}{\partial D^K}
= \tilde{Q}^K_J \frac{\partial}{\partial D^K}.
\end{equation}
Similarly, for the holonomic basis on $T^* \! \mathcal{Z}$,
\begin{equation}
\label{eq:trans3}
d \tilde{X}^A =  \frac{\partial \tilde{X}^A}{\partial {X}^B} d X^B + \frac{\partial \tilde{X}^A}{\partial {D}^K} d D^K
=  \frac{\partial \tilde{X}^A}{\partial {X}^B} d X^B,
\end{equation}
\begin{equation}
\label{eq:trans3b}
d \tilde{D}^J  = \frac{\partial \tilde{D}^J}{\partial {X}^B} dX^B +
\frac{\partial \tilde{D}^J}{\partial {D}^K} dD^K
=  \frac{\partial {Q}^J_K}{\partial {X}^B } D^K d X^B + {Q}^J_K d D^K.
\end{equation}
Given \eqref{eq:trans1}, $\{ \frac{\partial}{\partial X^A} \}$ and $\{ d D^K \}$ do not transform as conventional basis
vectors on $\mathcal{Z}$.  Define \cite{bejancu1990,bejancu2000}
\begin{equation}
\label{eq:nonhol}
\frac{\delta}{\delta X^A} = \frac{\partial}{\partial X^A} - N^K_A \frac{\partial}{\partial D^K},
\qquad \delta D^K = dD^K + N^K_B dX^B.
\end{equation} 
Non-holonomic basis vectors $\{ \frac{\delta}{\delta X^A} \}$ and $\{ \delta D^K \}$ obey \cite{minguzzi2014}
\begin{equation}
\label{eq:dtrans}
\frac{\delta }{\delta \tilde{X}^A} = \frac{\partial X^B }{\partial \tilde{X}^A} \frac{\delta }{\delta {X}^B},
\quad
\delta \tilde{D}^J = Q^J_K \delta D^K; 
\qquad
\bigr{\langle} \frac{\delta}{\delta X^B}, dX^A \bigr{\rangle} = \delta^A_B, \quad
\bigr{\langle} \frac{\partial}{\partial D^K}, \delta D^J \bigr{\rangle}= \delta^J_K.
\end{equation}
The set $ \{ \frac{\delta}{\delta X^A}, \frac{\partial}{\partial D^K} \}$ are used as a convenient local basis on $T \mathcal{Z}$, and
the dual set $ \{ dX^A, \delta D^K \}$ on $T^* \mathcal{Z}$ \cite{bao2000,bejancu2000}.
The $N^K_B(X,D)$ are the nonlinear connection coefficients;
$N^K_B$ are presumed differentiable with respect to $(X,D)$.  
These do not obey coordinate transformation rules for linear connections
nor always correspond to a covariant derivative with the properties of a linear connection.
 For \eqref{eq:dtrans} to hold under coordinate transformations $X \rightarrow \tilde{X}$ \cite{bejancu1990,bao2000},
\begin{equation}
\label{eq:Ntrans}
\tilde{N}^J_A = \left(Q^J_K  N^K_B - \frac{\partial {Q}^J_K}{\partial {X}^B } D^K \right) \frac{\partial X^B} {\partial \tilde{X^A}}.
\end{equation}
\begin{rmk}
The geometry of tangent bundle $T \mathcal{Z}$ with nonlinear connection admits an orthogonal decomposition  $T \mathcal{Z} = V \mathcal{Z} \oplus H \mathcal{Z}$ into a vertical vector bundle $V \mathcal{Z}$ with local field of frames $ \{ \frac{\partial}{\partial D^A} \}$ and a horizontal distribution $H \mathcal{Z}$ with local field of frames $ \{ \frac{\delta}{\delta X^A} \}$ \cite{bejancu1990}.
\end{rmk}
\noindent Fibers of $V \mathcal{Z}$ and $H \mathcal{Z}$ are of respective dimensions $m$ and $n$. Henceforth, vertical and horizontal subspaces are of the same dimension: $m=n$.
Indices $J,K,\ldots$ can thus be replaced with $A,B,\ldots$ in the summation convention, which runs from 1 to $n$.
In \eqref{eq:trans1}, let
\begin{equation}
\label{eq:subseq}
Q^A_B = \frac{\partial \tilde{D}^A}{\partial D^B} =  \frac{\partial \tilde{X}^A}{\partial X^B} .
\end{equation}
A formal way of achieving \eqref{eq:subseq} via soldering forms is given by Minguzzi \cite{minguzzi2014}.
Coordinate differentiation operations are expressed as follows, with $f$ a differentiable function of arguments $(X,D)$:
\begin{equation}
\label{eq:diffnot}
\partial_A f(X,D) = \frac{\partial f (X,D)}{\partial X^A},  \quad
\bar{\partial}_A f(X,D) = \frac{\partial f(X,D)}{\partial D^A};
\quad \delta_A (\cdot) = \frac{\delta (\cdot)}{\delta X^A}
= \partial_A (\cdot) - N_A^B \bar{\partial}_B (\cdot).
\end{equation}
The special cases $f \rightarrow X$ and $f \rightarrow D$ are written \cite{clayton2017g,claytonMMS22}
\begin{equation}
\label{eq:partialD}
\partial_B X^A  = \frac{\partial X^A }{ \partial X^B} = \delta^A_B, \quad \bar{\partial}_B X^A  = 0; 
\qquad \partial_B D^A  = \frac{\partial D^A }{ \partial X^B}, \quad \bar{\partial}_B D^A  = \delta^A_B.
\end{equation}

\subsubsection{Length, area, and volume} 
The Sasaki metric tensor
\cite{yano1973,bao2000,hushmandi2012} enables a natural inner product of vectors over $\mathcal{Z}$:
\begin{equation}
\label{eq:Sasaki}
\pmb{\mathcal{G}}(X,D) = {\vec{G}}(X,D) + \check{\vec{G}}(X,D) = G_{AB}(X,D) dX^A \otimes dX^B + \check{G}_{AB}(X,D) \delta D^A \otimes \delta D^B;
\end{equation}
\begin{equation}
\label{eq:Sasaki2}
\mathcal{G}_{AB} = G_{AB}  =
\vec{G} \left( \frac{\delta}{\delta X^A} , \frac{\delta}{\delta X^B} \right) = \check{G}_{AB} = 
\check{\vec{G}} \left( \frac{\partial}{\partial D^A} , \frac{\partial}{\partial D^B} \right) = \check{G}_{BA} = G_{BA} = \mathcal{G}_{BA}.
\end{equation}
Components of $\vec{G}$ and $\check{\vec{G}}$ are equal and simply hereafter referred to as $G_{AB}$, but their bases span orthogonal subspaces.
Components $G_{AB}$ and inverse components $G^{AB}$ lower and raise indices
in the usual manner, and $G$ denotes the determinant of the $n \times n$ non-singular matrices of components of
$\vec{G}$ or $\check{\vec{G}}$:
\begin{equation}
\label{eq:detG}
G^{AB} G_{BC} = \delta^A_C; \qquad G(X,D) = \det [G_{AB}(X,D)] = \det [\check{G}_{AB}(X,D)]. 
\end{equation}
\begin{rmk}
Let $\vec{V} = V^A \frac{\delta}{\delta X^A} \in H \mathcal{Z}$ be a vector field over $\mathcal{Z}$;
its magnitude at point $(X,D)$ is $| \vec{V} | = \langle \vec{V},\pmb{\mathcal{G}} \vec{V} \rangle^{1/2} =
\langle \vec{V},\vec{G} \vec{V} \rangle^{1/2}
= |\vec{V} \cdot \vec{V}|^{1/2} = |V^A G_{AB} V^B |^{1/2} = |V^A V_A|^{1/2} \geq 0$, where $V^A$ and $G_{AB}$ are evaluated at $(X,D)$.
\end{rmk}
\noindent When interpreted as a block diagonal $2 n \times 2 n$ matrix, the determinant of $\pmb{\mathcal{G}}$ is \cite{saczuk1996,saczuk1997,stumpf2000}
\begin{equation}
\label{eq:detGS}
\mathcal{G}(X,D) = \det [G_{AB}(X,D)] \, \det [\check{G}_{AB}(X,D)] = | \det [G_{AB}(X,D)] |^2 = | G(X,D) |^2.
\end{equation}
Let $\d \vec{X}$ denote a differential line element of $\mathcal{M}$ referred to non-holonomic horizontal basis $ \{ \frac{\delta}{\delta X^A} \}$, and let $\d \vec{D}$ denote a line element of $\mathcal{U}$ referred to vertical basis $ \{ \frac{\partial}{\partial D^A} \}$.  Squared lengths are
\begin{equation}
\label{eq:lengths}
|\d \vec{X}|^2 = \langle \d \vec{X}, \pmb{\mathcal{G}} \d \vec{X} \rangle = G_{AB} \d X^A \d X^B,
\qquad
|\d \vec{D}|^2 = \langle \d \vec{D}, \pmb{\mathcal{G}} \d \vec{D} \rangle = G_{AB} \d D^A \d D^B.
\end{equation}
The respective volume element $\d V$ and volume form $d \Omega$ of the $n$-dimensional base manifold $\mathcal{M}$, and
the area form $\Omega$ for its boundary $\partial \! \mathcal{M}$, are defined by \cite{rund1975}
\begin{equation}
\label{eq:volforms}
\d V = \sqrt{G} \, \d X^1 \d X^2 \ldots \d X^n, \qquad d \Omega = \sqrt{G} \, d X^1 \wedge d X^2 \wedge \ldots \wedge d X^n,
\end{equation}
\begin{equation}
\label{eq:areaform}
\Omega = \sqrt{B} \, d U^1 \wedge \ldots \wedge d U^{n-1}.
\end{equation}
Local coordinates on the $(n-1)$-dimensional oriented hypersurface $\partial \! \mathcal{M}$ are given by parametric equations $X^A = X^A(U^\alpha) \, (\alpha = 1,\ldots,n-1)$, $B^A_\alpha =
\frac{\partial X^A}{\partial U^\alpha}$, and $B = \det (B^A_\alpha G_{AB} B^B_\beta)$. 

\subsubsection{Covariant derivatives}
Horizontal gradients of basis vectors are
determined by generic affine connection coefficients $H^A_{BC}$ and $K^A_{BC}$, where
$\nabla (\cdot)$ is the covariant derivative:
\begin{equation}
\label{eq:horizgrad}
\nabla_{\delta / \delta X^B} \frac{\delta}{\delta X^C} = H^A_{BC} \frac{\delta}{\delta X^A}, \qquad
\nabla_{\delta / \delta X^B} \frac{\partial}{\partial D^C} = K^A_{BC} \frac{\partial}{\partial D^A}.
\end{equation}
Analogously, vertical gradients employ generic connection coefficients $V^A_{BC}$ and $Y^A_{BC}$:
\begin{equation}
\label{eq:vertgrad}
\nabla_{\partial / \partial D^B} \frac{\partial}{\partial D^C} = V^A_{BC} \frac{\partial}{\partial D^A}, \qquad
\nabla_{\partial / \partial D^B} \frac{\delta}{\delta X^C} = Y^A_{BC} \frac{\delta}{\delta X^A}.
\end{equation}
 For example, let $\vec{V} = V^A \frac{\delta}{\delta X^A} \in H \mathcal{Z}$ be a vector field. Then the (total) covariant derivative of $\vec{V}$ is
\begin{equation}
\label{eq:excov}
\begin{split}
\nabla \vec{V} & = \nabla_{\delta / \delta X^B} \vec{V} \otimes d X^B + \nabla_{\partial / \partial D^B} \vec{V} \otimes \delta D^B \\ & = 
(\delta_B V^A + H^A_{BC} V^C) \frac{\delta}{\delta X^A} \otimes d X^B
+ (\bar{\partial}_B V^A + Y^A_{BC} V^C) \frac{\delta}{\delta X^A} \otimes \delta D^B
\\ & = V^A_{\, |B} \frac{\delta}{\delta X^A} \otimes d X^B + V^A|_B \frac{\delta}{\delta X^A} \otimes \delta D^B.
\end{split}
\end{equation}
Denoted by $(\cdot)_{|A}$ and $(\cdot)|_B$ are horizontal and vertical covariant derivatives with respect to 
 $\{X^A\}$ and $\{ D^B \}$. 
 \begin{rmk}
 The sequence of covariant indices on connections
follows some works \cite{schouten1954,marsden1983,eringen1971,clayton2014} and is the transpose of others \cite{rund1959,bejancu1990,steinmann2015}.
For symmetric connections, it is inconsequential.
\end{rmk}
\noindent Components of the horizontal covariant derivative of metric tensor $\vec{G} = G_{AB} \, d X^A \otimes d X^B $ 
(i.e., the horizontal part of $\pmb{\mathcal{G}}$) are
\begin{equation}
\label{eq:horizG1}
G_{AB|C} = \delta_C G_{AB} - H^D_{CA} G_{DB} -H^D_{CB} G_{AD} 
= \partial_C G_{AB} - N^D_C \bar{\partial}_D G_{AB} - H^D_{CA} G_{DB} - H^D_{CB} G_{DA}.
\end{equation}
The following identity is also noted for $G = \det (G_{AB})$, a scalar density \cite{rund1975}:
\begin{equation}
\label{eq:Gids}
(\sqrt{G})_{|A} = \partial_A(\sqrt{G}) - N^B_A \bar{\partial}_B (\sqrt{G}) - \sqrt{G} H^B_{AB}.
\end{equation}
Christoffel symbols of the second kind for the Levi-Civita connection are $\gamma^A_{BC}$,
Cartan's tensor is $C^A_{BC}$, and horizontal coefficients of the Chern-Rund and Cartan connections are $\Gamma^A_{BC}$. All are torsion-free (i.e., symmetric):
\begin{equation}
\label{eq:LC1}
\gamma^A_{BC}={\textstyle{\frac{1}{2}}} G^{AD} (\partial_C G_{BD} + \partial_B G_{CD} - \partial_D G_{BC})
=G^{AD} \gamma_{BCD},
\end{equation}
\begin{equation}
\label{eq:Cartan1}
C^A_{BC}={\textstyle{\frac{1}{2}}} G^{AD} (\bar{\partial}_C G_{BD} + \bar{\partial}_B G_{CD} - \bar{\partial}_D G_{BC})
=G^{AD} C_{BCD},
\end{equation}
\begin{equation}
\label{eq:CR1}
\Gamma^A_{BC}={\textstyle{\frac{1}{2}}} G^{AD} (\delta_C G_{BD} + \delta_B G_{CD} - \delta_D G_{BC})
=G^{AD} \Gamma_{BCD}.
\end{equation}
\begin{rmk}
Chern-Rund-Cartan coefficients are
metric-compatible for horizontal covariant differentiation of $\vec{G} = G_{AB} \, d X^A \otimes d X^B $ since $H^A_{BC} = \Gamma^A_{BC} \Rightarrow
G_{AB|C} = 0$ in \eqref{eq:horizG1}.
Similarly, Cartan's tensor is metric-compatible for vertical covariant differentiation
of ${\vec{G}}$: $Y^A_{BC} = C^A_{BC} \Rightarrow {G}_{AB}|_C = 0$.
\end{rmk}
\noindent From direct calculations with respective \eqref{eq:LC1}, \eqref{eq:Cartan1}, and \eqref{eq:CR1}, traces of linear connections are related to partial gradients of $G = \det (\vec{G})$:
\begin{equation}
\label{eq:Gids2}
\partial_A ( \ln \sqrt{G}) =   \gamma^B_{AB}, \qquad
\bar{\partial}_A ( \ln \sqrt{G}) 
=  C^B_{AB}, 
\qquad \delta_A (\ln \sqrt{G}) = 
 \textstyle{\frac{1}{2}} G^{BC}  {\delta}_A G_{CB} = \Gamma^B_{AB}.
\end{equation}
\begin{rmk} $H^A_{BC} = \Gamma^A_{BC} \Rightarrow
G_{|A} = 2 G (\ln \sqrt{G})_{|A} = 0$ and $Y^A_{BC} = C^A_{BC} \Rightarrow G|_A = 2 G (\ln \sqrt{G})|_A = 0$.
\end{rmk}
\noindent Nonlinear connection coefficients $N^A_B(X,D)$ admissible under \eqref{eq:trans1} and \eqref{eq:Ntrans} can be 
obtained in several ways.  When $T \mathcal{Z}$ is restricted to locally flat sections \cite{bao2000,minguzzi2014}, $N^A_B = 0$ in a preferred coordinate chart $\{ X,D \}$, but $\tilde{N}^A_B$ in \eqref{eq:Ntrans} do not vanish for heterogeneous transformations under which $\partial_B Q^J_K$ is nonzero.
A differentiable real Lagrangian function $\mathsf{L}(X,D)$ 
can be introduced, from which $N^A_B = \mathsf{G}^A_B$, where \cite{bejancu1990}
\begin{equation}
\label{eq:NLag}
\mathsf{G}^A_B = \bar{\partial}_B \mathsf{G}^A = \bar{\partial}_B [G^{AE} ( D^C \bar{\partial}_E \partial_C \mathsf{L} - \partial_E \mathsf{L} )].
\end{equation}
\begin{rmk} Let $G_{AB}(X,D)$ be positively homogeneous of degree zero in $D$. Then $G^A$ below are components of a spray \cite{minguzzi2014,bao2000}, and canonical nonlinear connection coefficients $N^A_B = G^A_B$ that obey \eqref{eq:Ntrans} are
\begin{equation}
\label{eq:spray1}
G^A = {\textstyle{\frac{1}{2}}} \gamma^A_{BC} D^B D^C, \qquad G^A_B = \bar{\partial}_B G^A.
\end{equation}
\end{rmk}
For classification, let $K^A_{BC} = H^A_{BC}$ and $Y^A_{BC} = V^A_{BC}$.  Then
a complete generalized Finsler connection is the set $(N^A_B,H^A_{BC},V^A_{BC})$.
The Chern-Rund connection is $(G^A_B,\Gamma^A_{BC},0)$.
The Cartan connection is $(G^A_B,\Gamma^A_{BC},C^A_{BC})$.
The Berwald connection is $(G^A_B,G^A_{BC},0)$, where $G^A_{BC} = N^A_{BC} = \bar{\partial}_B N^A_C = 
\bar{\partial}_B \bar{\partial}_C G^A$.

\subsubsection{A divergence theorem}
Let $\mathcal{M}$ be a manifold of dimension $n$ having $(n-1)$-dimensional boundary $\partial \! \mathcal{M}$ of class $C^1$,
a positively oriented hypersurface.
Stokes' theorem for a $C^1$ differentiable $(n-1)$ form $\pmb{\alpha}$ on $\mathcal{M}$ is
\begin{equation}
\label{eq:stokesforms}
\int_{\mathcal{M}}  d \pmb{\alpha} = \int_{\partial \! \mathcal{M}} \pmb{\alpha}.
\end{equation}
\begin{thm}
\label{thm:stokes}
Let $\mathcal{M}$, $\dim \mathcal{M} = n$, be the base manifold of a generalized Finsler bundle of total space $\mathcal{Z}$ with
positively oriented boundary $\partial \! \mathcal{M}$ of class $C^1$ and $\dim \partial \! \mathcal{M} = n-1$. 
Let $\pmb{\alpha}(X,D) = V^A(X,D) N_A(X,D) \Omega(X,D)$ be a differentiable $(n-1)$-form, and let $V^A$ be contravariant
components of vector field $\vec{V} = V^A \frac{\delta}{\delta X^A} \in H \mathcal{Z}$.  
Denote the field of positive-definite metric tensor components for the horizontal subspace by $G_{AB} (X,D)$ with $G= \det (G_{AB}) > 0$.
Assign a symmetric horizontal linear connection $H^A_{BC} = H^A_{CB}$ such
that $(\sqrt{G})_{|A} = 0$, and assume that $C^1$ functional relations $D = D(X)$ exist for representation of the vertical fiber coordinates
at each $X \in \mathcal{M}$.
Then in a coordinate chart $\{ X^A \}$, \eqref{eq:stokesforms} is explicitly, with volume and area forms given in the second of \eqref{eq:volforms} and \eqref{eq:areaform},
\begin{equation}
\label{eq:stokes}
\int_{\mathcal{M}} [V^A_{|A} + (V^A C^C_{BC} + \bar{\partial}_B V^A) D^B_{; \, A}]  \, d\Omega = \int_{\partial \! \mathcal{M}} V^A N_A \,\Omega ,
\end{equation}
where the horizontal covariant derivative is $V^A_{|A} = \delta_A V^A + H^B_{BA} V^A $, the definition $D^B_{; \, A} = \partial_A D^B + N^B_A$ with $\partial_A D^B  = \partial D^B / \partial X^A$, and $N_A$ is the unit outward normal component
of $\vec{N} = N_A \, dX^A$ to $\partial \! \mathcal{M}$.
\end{thm}
\begin{pf}
The proof, not repeated here, is given in the review article \cite{claytonMMS22}, implied but not derived explicitly in an earlier work \cite{clayton2017g}. The proof of \eqref{eq:stokes} \cite{claytonMMS22} extends that of Rund \cite{rund1975}--who specified a Finsler space $F_n$ with Cartan connection $(G^A_B,\Gamma^A_{BC}, C^A_{BC})$ and metric acquired from a Finsler (Lagrangian) function $\mathcal{F}$ (\S2.1.5)--to a generalized Finsler space with arbitrary positive-definite metric $G_{AB}(X,D)$ and arbitrary nonlinear connection $N^A_B(X,D)$.$\quad \square$
\end{pf}
\begin{rmk}
Under the stipulations of Stokes' theorem, \eqref{eq:stokes} holds when $\mathcal{M}$ and $\partial \! \mathcal{M}$ are replaced with any compact region of $\mathcal{M}' \subset \mathcal{M}$ and positively oriented boundary of that region.
\end{rmk}
\begin{rmk}
The Chern-Rund-Cartan horizontal connection coefficients, $H^A_{BC} = \Gamma^A_{BC}$, uniquely fulfill symmetry and metric-compatibility requirements.
\end{rmk}
\begin{rmk}
A different basis and its dual over $\mathcal{M}$ could be prescribed for $\vec{V}$ and $\vec{N}$ given certain stipulations \cite{claytonMMS22}.
However, geometric interpretation of covariant differentiation on the left side of \eqref{eq:stokes}
suggests $\{ \frac{\delta}{\delta X^A} \} $ should be used for $\vec{V}$, by which dual basis $\{ d X^B \} $ should be used for 
$\vec{N}$ to ensure invariance:
$\langle \vec{V}, \vec{N} \rangle \rightarrow V^A N_B \langle \frac{\delta}{\delta X^A}, dX^B  \rangle$.
If instead $\vec{V}$ is referred to the holonomic basis $ \{ \frac{\partial}{\partial X^A} \} $,
then $N^A_B = 0$ should be imposed for invariance with $N_B d X^B$.
As noted prior to \eqref{eq:NLag}, this choice would restrict \eqref{eq:stokes} to homogeneous transformations of coordinates $\{ X,D \}$. 
\end{rmk}
\noindent As assumed in Theorem 2.1.1 \cite{rund1975,clayton2017g,claytonMMS22}, $C^1$ functions $D = D(X)$ must
exist over all $ X \in \mathcal{M}$. 
Relations of generalized Finsler geometry \cite{bejancu1990} still apply, but additional relations emerge
naturally when metric $G_{AB}$ is interpreted as an osculating Riemannian metric \cite{rund1959,amari1962}.
Specifically, an alternative representation of \eqref{eq:stokes} is newly proven in the following. 

\begin{cor}
Given $C^1$ functions $D = D(X)$, let $\tilde{G}_{AB}(X) = G_{AB}(X,D(X))$ be components of the osculating Riemannian metric derived from $\vec{G} = G_{AB} \, dX^A \otimes dX^B$. Then \eqref{eq:stokes} is equivalent to 
\begin{equation}
\label{eq:stokeso}
\int_{\mathcal{M}} \tilde{V}^A_{: \, A} \, d\Omega = \int_{\partial \! \mathcal{M}} \tilde{V}^A \tilde{N}_A \,\Omega,
\end{equation}
where the vector $\tilde{V}^A(X) = V^A(X,D(X))$, unit normal $\tilde{N}_A(X) = N_A(X,D(X))$, and covariant derivative 
$\tilde{V}^A_{: \, A} = \partial_A \tilde{V}^A + \tilde{\gamma}^B_{BA} \tilde{V}^A$ with connection 
 $\tilde{\gamma}^B_{BA} (X) = \partial_A (\ln \sqrt {\tilde{G}(X)}) =  \tilde{\gamma}^B_{AB} (X)$
and $\tilde{G} = \det (\tilde{G}_{AB})$.
\end{cor}
\begin{pf}
The right of \eqref{eq:stokeso} is identical to the right of \eqref{eq:stokes} given the change of 
 variables.
In the left of \eqref{eq:stokeso}, from chain-rule differentiation, vanishing \eqref{eq:Gids}, and \eqref{eq:Gids2},
\begin{equation}
\label{eq:pfnew}
\partial_A \tilde{V}^A = \partial_A V^A + \bar{\partial}_B V^A \partial_A D^B,  
\end{equation}
\begin{equation}
\label{eq:pfnew2}
\begin{split}
\tilde{V}^A \tilde{\gamma}^B_{BA} & = \tilde{V}^A \partial_A ( \ln \sqrt{\tilde{G}} ) = 
V^A [ \partial_A ( \ln \sqrt{G}) + \bar{\partial}_B ( \ln \sqrt{G}) \partial_A D^B ] \\ & 
= V^A [ \delta_A ( \ln \sqrt{G}) + C^C_{BC} (N^B_A +  \partial_A D^B)] 
 =V^A [H^B_{AB} + C^C_{BC} D^B_{; \, A} ] = V^A [ H^B_{BA} + C^C_{BC} D^B_{; \, A} ].
\end{split}
\end{equation}
Adding \eqref{eq:pfnew} to \eqref{eq:pfnew2} with canceling $\pm N^B_A  \bar{\partial}_B V^A$ terms then produces
\begin{equation}
\label{eq:pfnew3}
\begin{split}
\tilde{V}^A_{: \, A} & = 
\{ \partial_A V^A + \bar{\partial}_B V^A \partial_A D^B - N^B_A  \bar{\partial}_B V^A \} + 
\{N^B_A  \bar{\partial}_B V^A + V^A [ H^B_{BA} + C^C_{BC} D^B_{; \, A} ] \} \\
& = \delta_A V^A +  V^A H^B_{BA}  +  \bar{\partial}_B V^A (\partial_A D^B + N^B_A) 
+ V^A C^C_{BC} D^B_{; \, A} = V^A_{|A} + (\bar{\partial}_B V^A +  V^A C^C_{BC})  D^B_{; \, A}.
\end{split}
\end{equation}
Integrands on the left sides of \eqref{eq:stokes} and \eqref{eq:stokeso} are thus verified to match,
completing the proof.$\quad \square$
\end{pf}
\begin{rmk}
Coefficients of the Levi-Civita connection of $\tilde{G}_{AB}$ satisfy the symmetry and metric-compatibility requirements used to prove \eqref{eq:stokeso}:
 \begin{equation}
\label{eq:LC1o}
\tilde{\gamma}^A_{BC}={\textstyle{\frac{1}{2}}} \tilde{G}^{AD} (\partial_C \tilde{G}_{BD} + \partial_B \tilde{G}_{CD} - \partial_D \tilde{G}_{BC})
=\tilde{G}^{AD} \tilde{\gamma}_{BCD}.
\end{equation}
\end{rmk}
\begin{rmk}
Given \eqref{eq:LC1o}, the form of the divergence theorem in \eqref{eq:stokeso} appears analogous to that
of a Riemannian manifold with boundary.
It is not identical, however, since the non-holonomic basis  $ \{ \frac{\delta}{\delta X^A} \} $ is used for $\vec{V}$. As in Remark 2.1.3,
the holonomic basis $\{ \frac{\partial}{\partial X^A} \} $ could be used in a preferred chart $\{ X, D(X) \}$ wherein $N^A_B = 0$;
under such special conditions the distinction vanishes.
\end{rmk}

\subsubsection{Pseudo-Finsler and Finsler spaces}
Preceding developments hold for generalized Finsler geometry, by which
the metric tensor components need not be derived from a Lagrangian \cite{watanabe1982,miron1983,bejancu1990}.
Subclasses of generalized Finsler geometry do require such a Lagrangian function, denoted by $\mathcal{L}$.
Let $\mathcal{Z} = T \mathcal{M} \backslash 0$ (i.e., the tangent bundle of $\mathcal{M}$ excluding zero section $D=0$). 
Let $\mathcal{L}(X,D): \mathcal{Z} \rightarrow \mathbb{R}$ be positive homogeneous of degree two in $D$, and as many times differentiable as needed with respect to $\{ X^A \}$ and $\{ D^A \}$ ($C^\infty$ is often assumed \cite{bao2000}, but $C^5$ is usually sufficient \cite{minguzzi2014}).  Then $(\mathcal{M},\mathcal{L})$ is a pseudo-Finsler space when the $n \times n$ matrix of components $G_{AB}$ is both non-singular over $\mathcal{Z}$ and obtained from Lagrangian 
$\mathcal{L}$:
\begin{equation}
\label{eq:pseudoF}
G_{AB}(X,D) = \bar{\partial}_A \bar{\partial}_B \mathcal{L}(X,D), \qquad \mathcal{L} = \textstyle{\frac{1}{2}} G_{AB} D^A D^B.
\end{equation}
 A Finsler space $(\mathcal{M},\mathcal{F})$, also denoted by $F_n$ where $n = \dim \mathcal{M}$, is a pseudo-Finsler space for which $G_{AB}(X,D)$ is always positive definite over
$\mathcal{Z}$.  For a Finsler space $F_n$ \cite{rund1959,bao2000}, the fundamental scalar Finsler function $\mathcal{F}(X,D)$ is introduced, positive homogeneous of degree one in $D$:
\begin{equation}
\label{eq:FFund}
\begin{split}
\mathcal{F}(X,D) = \sqrt{ 2 \mathcal{L}(X,D) }& = |G_{AB}(X,D) D^A D^B|^{1/2} \\ &
\quad \leftrightarrow \quad \mathcal{L}(X,D) =  {\textstyle{\frac{1}{2}}} \mathcal{F}^2(X,D);
\qquad \mathcal{F}(X,D) > 0 \, \forall D \, \neq 0.
\end{split}
\end{equation}
 In Finsler geometry \cite{bao2000,bejancu1990,rund1959}, it follows that $\mathsf{L} = \mathcal{L}$ and
 $\mathsf{G}^A = G^A$ in \eqref{eq:NLag} and \eqref{eq:spray1}, and that
\begin{equation}
\label{eq:strict}
G_{AB} = {\textstyle{\frac{1}{2}}} \bar{\partial}_A \bar{\partial}_B (\mathcal{F}^2), \quad
G^A_B = \gamma^A_{BC} D^C - C^A_{BC} \gamma^C_{DE} D^D D^E = \Gamma^A_{BC} D^C, 
\quad C_{ABC} = {\textstyle{\frac{1}{4}}}\bar{\partial}_A \bar{\partial}_B \bar{\partial}_C 
(\mathcal{F}^2).
\end{equation}
Reductions and embeddings for Finsler spaces are discussed elsewhere  \cite{rund1959,minguzzi2014,bao2000,deicke1955,yano1954,claytonMMS22}.

\subsection{Spatial configuration}
A description on a fiber bundle analogous to that of \S2.1 is used for the spatial configuration (i.e., current configuration) of a body.
A differential manifold $\mathfrak{m}$ of 
dimension $n$ represents a (deformed) physical body,
with base space embedded in ambient Euclidean space of dimension $N \geq n$.  
\begin{rmk} Definitions in \S2.2 parallel those of \S2.1, where lower-case indices and symbols, with the exception of connections, distinguish current-configurational quantities.
\end{rmk}
Let $x \in \mathfrak{m}$ denote the spatial image of a body
particle or point with $\{x^a\}(a=1,2,\ldots,n)$ being a coordinate chart
on $\mathfrak{m}$.  
At each spatial point is a vector $\vec{d}$, and 
chart(s) of secondary coordinates $\{d^k\}(k=1,2,\ldots,m)$ are assigned over $\mathfrak{m}$.  
Define 
$\mathsf{z} = (\mathfrak{z},\pi,\mathfrak{m},\mathfrak{u})$ as a fiber bundle of total
space $\mathfrak{z}$ (dimension $n+m$), 
where $\pi:\mathfrak{z} \rightarrow \mathfrak{m}$
is the projection and $\mathfrak{u} = \mathfrak{z}_x = \pi^{-1} (x) $ is the fiber at $x$.  
A chart covering a region of
$\mathfrak{z}$ is $\{x^a,d^k\}$. 
Each fiber is an $m$-dimensional vector space, so ($\mathfrak{z},\pi,\mathfrak{m})$ constitutes 
a vector bundle.

The global mapping from referential to spatial base manifolds is $\varphi$, referred to herein as the motion.
The global mapping from referential to current total spaces is the set $\Xi = (\varphi,\theta)$,
where in general $\varphi(X,D):\mathcal{M} \rightarrow \mathfrak{m}$ and $\Xi(X,D):\mathcal{Z} \rightarrow \mathfrak{z}$.
Functional forms of $\varphi(X,D)$ and $\Xi(X,D)$ vary in the literature \cite{claytonMMS22}; details are discussed in \S3.1. 
Mappings and field variables can be made time ($t$) dependent via introduction of independent parameter $t$ \cite{stumpf2000,clayton2017f,clayton2018c}.
Explicit time dependence is excluded from the current theoretical presentation that focuses on equilibrium configurations \cite{clayton2017g,clayton2017z}.  
The following diagram commutes \cite{bejancu1990}:
\newline \\
\begin{tikzcd}
 \mathcal{Z} \arrow[d, "\Pi"]  \arrow[r, "\Xi"] & \mathfrak{z}
  \arrow[d,"\pi" ] \\
  \mathcal{M} \arrow[r,"\varphi"] & \mathfrak{m}
\end{tikzcd} \\

\subsubsection{Basis vectors and nonlinear connections} 
Coordinate transformations
from $\{x,d\}$ to $\{\tilde{x},\tilde{d}\}$ on $\mathfrak{z}$ are of the general Finsler form 
\begin{equation}
\label{eq:trans1c}
\tilde{x}^a = \tilde{x}^a(x),
\qquad \tilde{d}^j(x,d) = q^j_k (x) d^k,
\end{equation}
where $q^j_k$ is non-singular and differentiable, with 
inverse obeying $\tilde{q}^i_k  q^k_j = \delta^i_j$. The holonomic basis for the tangent bundle $ T \mathfrak{z}$ is 
$\{\frac{\partial}{\partial x^a},\frac{\partial}{\partial d^k}\}$, and  
the holonomic basis for the cotangent bundle $ T^* \! \mathfrak{z}$
is $\{d x^a, d d^k \}$. 
Non-holonomic basis vectors $\{ \frac{\delta}{\delta x^a} \}$ 
and $\{ \delta d^k \}$ that transform traditionally under $x \rightarrow \tilde{x}$ are 
\begin{equation}
\label{eq:nonholc}
\frac{\delta}{\delta x^a} = \frac{\partial}{\partial x^a} - N^k_a \frac{\partial}{\partial d^k},
\qquad \delta d^k = dd^k + N^k_b dx^b.
\end{equation} 
The set $ \{ \frac{\delta}{\delta x^a}, \frac{\partial}{\partial d^k} \}$ is used as a local basis on $T \mathfrak{z}$, and
 $ \{ dx^a, \delta d^k \}$ on $T^* \mathfrak{z}$.
 Tangent bundle $T \mathfrak{z}$ with nonlinear connection admits an orthogonal decomposition into vertical vector bundle and horizontal distribution: $T \mathfrak{z} = V \mathfrak{z} \oplus H \mathfrak{z}$.
 The transformation law of the spatial nonlinear connection is 
 \begin{equation}
\label{eq:Ntransc}
\tilde{N}^j_a = \left(q^j_k  N^k_b - \frac{\partial {q}^j_k}{\partial {x}^b } d^k \right) \frac{\partial x^b} {\partial \tilde{x^a}}.
\end{equation}
Subsequently, take $m=n$.
Indices $j,k,\ldots \rightarrow a,b,\ldots$, summation over duplicate indices is from 1 to $n$,
and in \eqref{eq:trans1c}, the $d^a$ transform like components of a contravariant vector field over $\mathfrak{m}$:
\begin{equation}
\label{eq:subseqc}
q^a_b = \frac{\partial \tilde{d}^a}{\partial d^b} =  \frac{\partial \tilde{x}^a}{\partial x^b} .
\end{equation}
Spatial coordinate differentiation is described by the compact notation
\begin{equation}
\label{eq:diffnotc}
\partial_a f(x,d) = \frac{\partial f(x,d)}{\partial x^a},  \quad
\bar{\partial}_a f(x,d) = \frac{\partial f(x,d)}{\partial d^a};
\qquad \delta_a (\cdot) = \frac{\delta (\cdot)}{\delta x^a}
= \partial_a (\cdot) - N_a^b \bar{\partial}_b (\cdot);
\end{equation}
\begin{equation}
\label{eq:partialDc}
\partial_b x^a  = \frac{\partial x^a }{ \partial x^b} = \delta^a_b, \quad \bar{\partial}_b x^a  = 0; 
\qquad \partial_b d^a  = \frac{\partial d^a }{ \partial x^b}, \quad \bar{\partial}_b d^a  = \delta^a_b.
\end{equation}

\subsubsection{Length, area, and volume} 
The Sasaki metric tensor that produces an inner product of vectors over $\mathfrak{z}$ is \cite{yano1973}
\begin{equation}
\label{eq:Sasakic}
\pmb{\mathfrak{g}}(x,d) = {\vec{g}}(x,d) + \check{\vec{g}}(x,d) = g_{ab}(x,d) dx^a \otimes dx^b + \check{g}_{ab}(x,d) \delta d^a \otimes \delta d^b;
\end{equation}
\begin{equation}
\label{eq:Sasaki2c}
\mathfrak{g}_{ab} = g_{ab}  =
\vec{g} \left( \frac{\delta}{\delta x^a} , \frac{\delta}{\delta x^b} \right) = \check{g}_{ab} = 
\check{\vec{g}} \left( \frac{\partial}{\partial d^a} , \frac{\partial}{\partial d^b} \right) = \check{g}_{ba} = g_{ba} = \mathfrak{g}_{ba}.
\end{equation}
Denote by $\d \vec{x}$ a differential line element of $\mathfrak{m}$ referred to non-holonomic horizontal basis $ \{ \frac{\delta}{\delta x^a} \}$
and $\d \vec{d}$ a differentiable line element of $\mathfrak{u}$ referred to vertical basis $ \{ \frac{\partial}{\partial d^a} \}$.  Their squared lengths are
\begin{equation}
\label{eq:lengthsc}
|\d \vec{x}|^2 = \langle \d \vec{x}, \pmb{\mathfrak{g}} \d \vec{x} \rangle = g_{ab} \d x^a \d x^b,
\qquad
|\d \vec{d}|^2 = \langle \d \vec{d}, \pmb{\mathfrak{g}} \d \vec{d} \rangle = g_{ab} \d d^a \d d^b.
\end{equation}
The scalar volume element and volume form of $\mathfrak{m}$, where $\dim \mathfrak{m} = n$, 
and the area form of $\partial \! \mathfrak{m}$, the $(n-1)$-dimensional boundary of a compact region of $\mathfrak{m}$, are respectively
\begin{equation}
\label{eq:volformsc}
\d v = \sqrt{g} \, \d x^1 \d x^2 \ldots \d x^n, \quad d \omega = \sqrt{g} \, d x^1 \wedge d x^2 \wedge \ldots \wedge d x^n, \quad
\omega = \sqrt{b} \, d u^1 \wedge \ldots \wedge d u^{n-1}.
\end{equation}
The surface embedding in $\mathfrak{m}$ is $x^a = x^a(u^\alpha) \, (\alpha = 1,\ldots,n-1)$, $b^a_\alpha =
\frac{\partial x^a}{\partial u^\alpha}$, and $b = \det (b^a_\alpha g_{ab} b^b_\beta)$. 

\subsubsection{Covariant derivatives}
Denote by $\nabla$ the covariant derivative. Horizontal gradients of basis vectors are
determined by coefficients $H^a_{bc}$ and $K^a_{bc}$, and vertical gradients by
$V^a_{bc}$ and $Y^a_{bc}$:
\begin{equation}
\label{eq:horizgradc}
\nabla_{\delta / \delta x^b} \frac{\delta}{\delta x^c} = H^a_{bc} \frac{\delta}{\delta x^a}, \qquad
\nabla_{\delta / \delta x^b} \frac{\partial}{\partial d^c} = K^a_{bc} \frac{\partial}{\partial d^a};
\end{equation}
\begin{equation}
\label{eq:vertgradc}
\nabla_{\partial / \partial d^b} \frac{\partial}{\partial d^c} = V^a_{bc} \frac{\partial}{\partial d^a}, \qquad
\nabla_{\partial / \partial d^b} \frac{\delta}{\delta x^c} = Y^a_{bc} \frac{\delta}{\delta x^a}.
\end{equation}
By example, covariant derivative operations over $\mathfrak{z}$ are invoked like \eqref{eq:excov} for $\vec{V} = V^a \frac{\delta}{\delta x^a} \in H \mathfrak{z}$:
\begin{equation}
\label{eq:excovc}
\begin{split}
\nabla \vec{V} & = \nabla_{\delta / \delta x^b} \vec{V} \otimes d x^b + \nabla_{\partial / \partial d^b} \vec{V} \otimes \delta d^b
= V^a_{\, |b} \frac{\delta}{\delta x^a} \otimes d x^b + V^a|_b \frac{\delta}{\delta x^a} \otimes \delta d^b.
\end{split}
\end{equation}
Herein $(\cdot)_{|a}$ and $(\cdot)|_b$ denote horizontal and vertical covariant differentiation with respect to 
coordinates $x^a$ and $d^b$. 
Let $\gamma^a_{bc}$ be coefficients of the Levi-Civita connection on $\mathfrak{z}$,
 $C^a_{bc}$ coefficients of the Cartan tensor on $\mathfrak{z}$, and 
$\Gamma^a_{bc}$ horizontal coefficients of the Chern-Rund and Cartan connections on $\mathfrak{z}$:
 \begin{equation}
\label{eq:LC1c}
\gamma^a_{bc}={\textstyle{\frac{1}{2}}} g^{ad} (\partial_c g_{bd} + \partial_b g_{cd} - \partial_d g_{bc})
=g^{ad} \gamma_{bcd},
\end{equation}
\begin{equation}
\label{eq:Cartan1c}
C^a_{bc}={\textstyle{\frac{1}{2}}} g^{ad} (\bar{\partial}_c g_{bd} + \bar{\partial}_b g_{cd} - \bar{\partial}_d g_{bc})=g^{ad} C_{bcd},
\end{equation}
\begin{equation}
\label{eq:CR1c}
\Gamma^a_{bc}={\textstyle{\frac{1}{2}}} g^{ad} (\delta_c g_{bd} + \delta_b g_{cd} - \delta_d g_{bc})
=g^{ad} \Gamma_{bcd}.
\end{equation}

\subsubsection{A divergence theorem}
Let $\mathfrak{m}$, $\dim \mathfrak{m} = n$, be the base manifold of a generalized Finsler bundle of total space $\mathfrak{z}$
with positively oriented $(n-1)$-dimensional $C^1$ boundary $\partial \! \mathfrak{m}$. 
Let $\pmb{\alpha}(x,d) = V^a(x,d) n_a(x,d) \omega(x,d)$ be a differentiable $(n-1)$-form, and let $V^a$ be contravariant
components of vector field $\vec{V} = V^a \frac{\delta}{\delta x^a} \in H \mathfrak{z}$.  
Denote the field of components for positive-definite metric tensor on the horizontal subspace by $g_{ab} (x,d)$ with $g= \det (g_{ab})>0 $.
Assign horizontal connection $H^a_{bc} = H^a_{cb}$ such
that $(\sqrt{g})_{|a} = 0$ (e.g., $H^a_{bc} = \Gamma^a_{bc}$) , and assume that $C^1$ functional relations $d = d(x)$ exist for representation of the vertical fiber coordinates
$\forall x \in \mathfrak{m}$.
Then in a chart $\{ x^a \}$, with volume and area forms given in \eqref{eq:volformsc}, \eqref{eq:stokesforms} is
\begin{equation}
\label{eq:stokesc}
\int_{\mathfrak{m}} [V^a_{|a} + (V^a C^c_{bc} + \bar{\partial}_b V^a) d^b_{; \, a}] \, d \omega = \int_{\partial \! \mathfrak{m}} V^a n_a 
\, \omega,
\end{equation}
with $n_a$ the unit outward normal on $\partial \! \mathfrak{m}$, $V^a_{|a} = \delta_a V^a + V^a H^b_{ba}$, and $d^b_{; \, a} = \partial_a d^b + N^b_a$. Proof matches that of Theorem 2.1.1 upon changes of variables; a corollary akin to Corollary 2.1.1 also holds.


\section{Finsler-geometric continuum mechanics}
The original theory of Finsler-geometric continuum mechanics \cite{clayton2016,clayton2017g} accounts for finite deformations under conditions of static equilibrium for forces conjugate to material particle motion and state vector evolution.  
Subtle differences exist among certain assumptions for
different instantiations, incrementally revised in sucessive works.
Most differences are explained in a review \cite{claytonMMS22}. 

\subsection{Motion and deformation} 
\begin{figure}[hbp!]
\centering
    \includegraphics[width=1\textwidth]{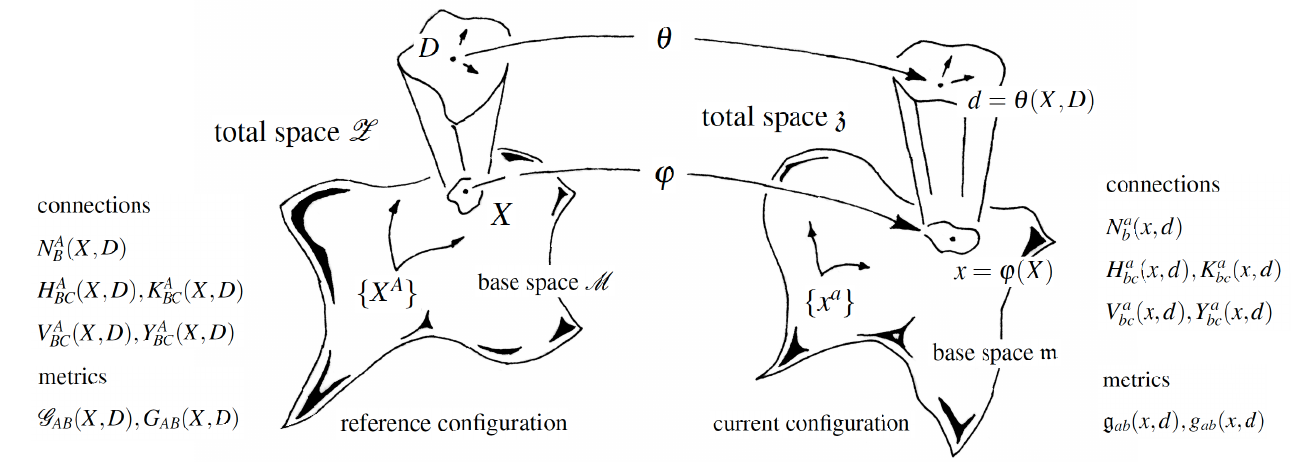}
    \caption{Total deformation $\Xi  = (\varphi,\theta): \mathcal{Z} \rightarrow \mathfrak{z}$ of material manifold $\mathcal{M}$ (dim $\mathcal{M} = n=m=2$) with base-space coordinates $\{X^A\}$ to spatial representation $\mathfrak{m}$ with base-space coordinates $\{x^a\}$. Internal structure fields are $(D,d)$ on total spaces $(\mathcal{Z},\mathfrak{z})$; arrows depict local components of state vectors $\vec{D}$ and $\vec{d}$ for neighborhoods centered at $X$ and $x$.}
    \label{fig1}
\end{figure}

Particle motion $\varphi: \mathcal{M} \rightarrow \mathfrak{m}$ and its inverse 
$\Phi: \mathfrak{m} \rightarrow \mathcal{M}$ are the one-to-one and $C^3$-differentiable functions
\begin{equation}
\label{eq:varphiC}
x^a = \varphi^a (X), \qquad X^A = \Phi^A (x),  \qquad (a,A = 1,\ldots, n)
\end{equation}
with $(\Phi \circ \varphi)(X) = X$.
Total motion is $\Xi: \mathcal Z \rightarrow \mathfrak{z}$, where $\Xi = (\varphi,\theta)$.  
Refer to Fig.~\ref{fig1}.
\begin{rmk}
Vector field $\vec{D}$ and its spatial counterpart $\vec{d}$ are referred to as internal state vector fields or director vector fields, but neither vector must be of unit length.
These are assigned physical interpretations pertinent to the specific class of mechanics problem under consideration \cite{claytonMMS22}.  
\end{rmk}
\noindent Motions of state vectors are defined as $C^3$ functions:
\begin{equation}
\label{eq:thetaC}
d^a = \theta^a (X,D), \qquad D^A = \Theta^A (x,d),  \qquad (a,A = 1, \ldots, n).
\end{equation}
\begin{rmk}
Fiber dimensions are $m = \dim \mathcal{U} = \dim \mathfrak{u} = \dim \mathfrak{m} = \dim \mathcal{M} = n$.
Extension for $m \neq n$ is conceivable \cite{ikeda1973,bejancu1990}. However, setting $m = n$
enables a more transparent physical interpretation of the vertical vector bundle, and it allows use of \eqref{eq:subseq} and \eqref{eq:subseqc}
that simplify notation and calculations.
For usual three-dimensional solid bodies, $n = 3$ as implied in parts of prior work \cite{claytonMMS22}, but other dimensions are permissible (e.g., two-dimensional membranes ($n = 2$) and one-dimensional rods ($n = 1$)).
\end{rmk}
From \eqref{eq:varphiC} and \eqref{eq:thetaC}, transformation formulae for partial differentiation operations between configurations of a differentiable function $ h(x,d): \mathfrak{z} \rightarrow \mathbb{R}$ are
\begin{equation}
\label{eq:diffC}
\frac{\partial ( h \circ \Xi )}{\partial X^A} = \frac{\partial  h}{\partial x^a} \frac{\partial \varphi^a}{\partial X^A}
+ \frac{\partial h}{\partial d^a} \frac{\partial \theta^a }{\partial X^A},
\qquad
\frac{\partial ( h \circ \Xi )}{\partial D^A} =   \frac{\partial h}{\partial d^a } \frac{\partial \theta^a }{\partial D^A}.
\end{equation}
\begin{rmk}
Unlike Chapter 8 of Bejancu \cite{bejancu1990}, basis vectors need not convect from $T \mathcal{Z}$ to
$T \mathfrak{z}$ with the motion $\Xi$. Rather, as in classical continuum field theories of mechanics \cite{truesdell1960,ericksen1960}, basis vectors---as well as metric tensors and connection coefficients---can
be assigned independently for configuration spaces $\mathcal{Z}$ and $\mathfrak{z}$.  As such, 
$(\frac{\delta}{\delta x^a},\frac{\partial}{\partial d^a},
g_{ab}, H^a_{bc},K^a_{bc},V^a_{bc},Y^a_{bc}, N^a_b)$ need not be obtained from 
$(\frac{\delta}{\delta X^A},\frac{\partial}{\partial D^A},
G_{AB}, H^A_{BC},K^A_{BC},V^A_{BC},Y^A_{BC}, N^A_B)$ via push-forward operations by $\Xi$.
But choosing $N^a_b$ as the push-forward of $N^A_B$ \cite{bejancu1990} is beneficial since
\begin{equation}
\label{eq:NrelateC}
N^b_a \frac{\partial \varphi^a}{\partial X^A} = N^B_A \frac{\partial \theta^b}{\partial D^B} - \frac{\partial \theta^b}{ \partial X^A }
\quad \Rightarrow \quad 
\frac{\delta (h \circ \Xi) }{\delta X^A} =  \frac{\delta h  }{\delta x^a}  \frac{\partial \varphi^a  }{ \partial X^A} =  \frac{\delta h  }{\delta x^a}  
\frac{\delta \varphi^a  }{ \delta X^A} =  \frac{\delta h  }{\delta x^a}  F^a_A,
\end{equation}
by which $\delta_A(\cdot) = F^a_A \delta_a(\cdot)$ simply relates the delta derivative across configurations.
\end{rmk}
As implied in \eqref{eq:NrelateC}, deformation gradient field $\vec{F}: H \mathcal{Z} \rightarrow H \mathfrak{z}$ is defined as
the two-point tensor field
\begin{equation}
\label{eq:defgradC}
\vec{F} = \frac{ \delta \pmb{\varphi}}{\delta \vec{X}} = \frac{\delta \varphi^a}{\delta X^A} \frac{\delta}{\delta x^a} \otimes
d X^A = \frac{\partial \varphi^a}{\partial X^A} \frac{\delta}{\delta x^a} \otimes d X^A,
\end{equation}
with \eqref{eq:varphiC} used in the rightmost equality. The inverse deformation gradient $\vec{f}: H \mathfrak{z} \rightarrow H \mathcal{Z}$ is defined as the following:
\begin{equation}
\label{eq:defgradiC}
\vec{f} = \frac{ \delta \pmb{\Phi}}{\delta \vec{x}} = \frac{\delta \Phi^A}{\delta x^a} \frac{\delta}{\delta X^A} \otimes
d x^a = \frac{\partial \Phi^A}{\partial x^a} \frac{\delta}{\delta X^A} \otimes d x^a.
\end{equation}
\begin{rmk}Accordingly, $F^a_A(X) f^A_b(x(X)) = \delta^a_b$ and $F^a_A(X) f^B_a(x(X)) = \delta^B_A$.
Usual stipulations on regularity \cite{marsden1983} of motions \eqref{eq:varphiC} apply such that $\det (F^a_A) > 0 $ and
$\det (f^A_a) >0 $. 
\end{rmk}
\noindent Transformation equations relating differential line elements of \eqref{eq:lengths} and \eqref{eq:lengthsc} follow:
\begin{equation}
\label{eq:linetrans}
\d \vec{x} = \d x^a \frac{\delta}{\delta x^a} =  F^a_A \d X^A \frac{\delta}{\delta x^a} = \vec{F} \d \vec{X},
\qquad
\d \vec{X} = \d X^A \frac{\delta }{\delta X^A }  = f^A_a \d x^a  \frac{\delta}{\delta X^A} = \vec{f} \d \vec{x}.
\end{equation}
Advancing \eqref{eq:linetrans}, with definition of the determinant, 
 \eqref{eq:volforms}, and \eqref{eq:volformsc}, 
volume elements and forms, respectively, transform between reference and spatial representations 
on $\mathcal{M}$ and $\mathfrak{m}$, with 
$J = \det(F^a_A) \sqrt{g/G} > 0 $ and $j = 1/J = J^{-1} > 0$, via (e.g., \cite{marsden1983,clayton2014})
\begin{equation}
\label{eq:voltrans}
\d v = J \d V = [ \det(F^a_A) \sqrt{g/G} ] \d V,  \qquad \d V = j \d v = [\det(f^A_a) \sqrt{G/g}] \d v,
\end{equation}
\begin{equation}
\label{eq:voltrans2}
\varphi^* d \omega = J d \Omega, \qquad  \Phi^* d \Omega = j d \omega.
\end{equation}
Strain can be quantified using symmetric
 Lagrangian deformation tensor $\vec{C} = C_{AB} dX^A \otimes dX^B$:
\begin{equation}
\label{eq:lengthC}
|\d \vec{x}|^2 =  F^a_A g_{ab} F ^b_B  \d X^A \d X^B = C_{AB} \d X^A \d X^B
= \langle \d \vec{X}, \vec{C} \d \vec{X} \rangle,
\quad
C_{AB} = F^a_A g_{ab} F^b_B = G_{AC} C^C_B = C_{BA}.
\end{equation}
From \eqref{eq:voltrans}, $\det(C_{AB}) =  \det (C^C_A G_{CB}) = J^2 G$.
Then from the first of \eqref{eq:horizgradc} and \eqref{eq:NrelateC} \cite{clayton2016,clayton2017z},
\begin{equation}
\label{eq:horiztranspf}
\nabla_{\delta / \delta X^A} \frac{\delta}{\delta x^c}  
=\frac{\delta x^a}{\delta X^A} \nabla_{\delta / \delta x^a} \frac{\delta}{\delta x^c}
=\delta_A \varphi^a H^b_{ac} \frac{\delta}{\delta x^b}
=F^a_A H^b_{ac} \frac{\delta}{\delta x^b}.
\end{equation}
Similarly, the second of \eqref{eq:horizgradc} gives
$\nabla_{\delta / \delta X^A} \frac{\partial}{\partial d^c} = F^a_A K^b_{ac}  \frac{\partial}{\partial d^b} $, though this is not needed later.

\subsection{Particular assumptions} 
\subsubsection{Director fields}
The divergence theorem \eqref{eq:stokes} is used to derive Euler-Lagrange equations for equilibrium of stress and state vector fields in \S3.3.3.  
Its derivation \cite{claytonMMS22,rund1975} requires existence of functional relations
\begin{equation}
\label{eq:funcC}
D^A = D^A (X), \qquad d^a = d^a(x),
\end{equation}
where the second of \eqref{eq:funcC} is implied by the first under a consistent change of variables per \eqref{eq:stokesc} and \eqref{eq:varphiC}. Existence of the 
following functional forms emerges from \eqref{eq:varphiC}, \eqref{eq:thetaC}, and \eqref{eq:funcC}:
\begin{equation}
\label{eq:thetaC2}
d^a = \theta^a (X,D) = \hat{\theta}^a(X,D(X)) = \bar{\theta}^a(X), 
\qquad D^A = \Theta^A (x,d) = \hat{\Theta}^A (x,d(x)) = \bar{\Theta}^A (x).
\end{equation}
\begin{rmk}
In some prior work \cite{clayton2016,clayton2017g}, alternative representations of particle motions incorporating state vector fields as arguments have been posited. These likely more complex alternatives are admissible but inessential \cite{claytonMMS22}. 
The current theory, like some others \cite{amari1962,saczuk1996,stumpf2000}, does not always require ${\theta}$ or ${\Theta}$ be
specified explicitly, though use of the former is implied later in \S5.  
\end{rmk}
\noindent The canonical, and pragmatic, choice for ${\theta}(X,D)$, given field $D^A(X)$, is \cite{clayton2017z}
\begin{equation}
\label{eq:thetasimp}
d = D \circ \Phi \quad \Leftrightarrow \quad d(x) = D (\Phi (x)) \quad \Rightarrow \quad {\theta}^a(D(X)) = D^A(X) \langle \delta d^a, \frac{\partial}{\partial D^A} \rangle =   D^A(X) \delta^a_A,
\end{equation}
where $\delta^a_A$ is viewed as a shifter between $V \mathfrak{z}$ and $ V \mathcal{Z}$.  
Accordingly,  $\delta^a_A = 1 \, \forall \, a=A, \delta^a_A = 0 \, \forall \, a \neq A$.
\begin{rmk}
Invoking \eqref{eq:thetasimp}, $\partial_A \theta^a (D(X)) = 0$
by definitions of $\theta^a = \theta^a(D(X))$ and $\partial_A(\cdot) = (\partial(\cdot)/\partial X^A)|_{D=\text{const}}$ in \eqref{eq:diffnot}. Also,
$\bar{\partial}_A \theta^a (D(X)) = \delta^a_A$ by \eqref{eq:partialD} and \eqref{eq:thetasimp}.
Then \eqref{eq:NrelateC} reduces to $N^a_b = N^A_B f^B_b \delta^a_A$, and conveniently for the degenerate case: $N^A_B = 0 \Leftrightarrow N^a_b = 0$.
\end{rmk}
 
 \subsubsection{Connections and metrics}
 Use of \eqref{eq:stokes} for any admissible $G_{AB}(X,D)$
necessitates a symmetric linear connection horizontally compatible with $G_{AB}$, meaning $H^A_{BC}= \Gamma^A_{BC}$, with
$\Gamma^A_{BC}$ Chern-Rund-Cartan coefficients of \eqref{eq:CR1}.  The simplest admissible choice of vertical coefficients is $V^A_{BC}=0$, corresponding to the Chern-Rund connection \cite{chern1948,rund1959,bao2000}.  The canonical choice $N^A_B = G^A_B$ of \eqref{eq:spray1} also corresponds to the Chern-Rund connection, but it is inessential for generalized Finsler geometry.
Choices $K^A_{BC} = H^A_{BC}$ \cite{clayton2017g,clayton2016} and $Y^A_{BC} = V^A_{BC}$ are logical given \eqref{eq:subseq}, but these are not mandatory.
Setting $K^A_{BC} = 0$, providing compatibility with Cartesian metric $\delta_{AB}$, may also be of utility \cite{claytonMMS22}.

Given a Sasaki metric $\pmb{\mathcal{G}}$ of \eqref{eq:Sasaki} with $G_{AB}$ of \eqref{eq:Sasaki2}, pragmatic  connection coefficients over $\mathcal{Z}$ are summarized 
in \eqref{eq:connrec}; complementary connections over $\mathfrak{z}$ given Sasaki metric $\pmb{\mathfrak{g}}$ with  $g_{ab}(x,d)$ of \eqref{eq:Sasaki2c} follow thereafter:
\begin{equation}
\label{eq:connrec}
H^A_{BC} =  \Gamma^A_{BC}, \quad V^A_{BC} = Y^A_{BC} = 0;
 \quad H^a_{bc} = \Gamma^a_{bc}, \quad V^a_{bc} = Y^a_{bc} = 0; \quad N^a_b = N^A_B f^B_b \delta^a_A.
\end{equation}
\begin{rmk}
Note $K^A_{BC}$ and $K^a_{bc}$ are left arbitrary to admit mathematical descriptions of different physics, in contrast to $Y^A_{BC}$ and $Y^a_{bc}$ set equal to their purely vertical counterparts for simplicity. Since nonlinear connection $N^A_B$ is also not explicitly chosen in \eqref{eq:connrec} but is left general to admit more physics than considered previously \cite{claytonMMS22}, \eqref{eq:Ntrans} is not necessarily ensured for arbitrary changes of coordinates, so transformation properties of $N^A_B$ should be verified. Once the former $N^A_B$ is chosen, $N^a_b$ in \eqref{eq:connrec} presumes \eqref{eq:thetasimp} is invoked with \eqref{eq:NrelateC}.  
\end{rmk}
\begin{rmk}
If the fields $G_{AB}(X,D)$ 
and $g_{ab}(x,d)$ are known, linear connection coefficients 
in \eqref{eq:connrec} can be calculated from definitions in \S2.1 and \S2.2. 
Metric $G_{AB}$ need not be homogeneous of degree zero with respect to $D$, but it can be.  Components 
$G_{AB}$ need not be derived, as in \S2.1.5, from a Lagrangian $\mathcal{L}$ or more specifically a fundamental Finsler function $\mathcal{F}$, but they can be.  Dependence of $\pmb{\mathcal{G}}$ on $X$ and $D$ 
is based on symmetry and physics pertinent to the particular problem of study.  Similar statements describe the spatial metric $\pmb{\mathfrak{g}}$ and components $g_{ab}$.
\end{rmk}

A decomposition of $G_{AB}$ into a Riemannian part $\bar{G}_{AC}$ and
a director-dependent part $\hat{G}^C_B$ is useful for describing fundamental physics and for solving boundary value problems \cite{clayton2016,clayton2017g,clayton2017z,clayton2018p}:
\begin{equation}
\label{eq:metricdecomp}
\begin{split}
\vec{G} & = \bar{\vec{G}} \hat{\vec{G}}; \qquad G_{AB}(X,D) = \bar{G}_{AC}(X) \hat{G}^C_B (X,D); 
\\ \bar{\vec{G}} & = \bar{G}_{AB} \, d X^A \otimes d X^B; \qquad
\hat{\vec{G}} = \hat{G}^A_B \frac{\delta}{\delta X^A} \otimes d X^B.
\end{split}
\end{equation}
More specific functional forms in \eqref{eq:metricdecomp} are advocated herein, as implied by past applications
\cite{claytonMMS22}:
\begin{equation}
\label{eq:mdec}
\begin{split}
G_{AB}(X,D) & = \bar{G}_{AC}(X) \hat{G}^C_B (D(X))
= \hat{G}^C_A (D(X))  \bar{G}_{CB}(X) ; 
\\  \bar{G}_{AB} & = \bar{G}_{BA},
\qquad \hat{G}^C_A G_{BC} = \hat{G}^C_B G_{CA}.
\end{split}
\end{equation}
\begin{rmk}
Components of $\bar{G}_{AB}$ are chosen to best represent symmetry of the physical body; in elasticity, often a Riemannian metric for rectilinear, cylindrical, or spherical coordinates on $\mathcal{M}$.
Components of $\hat{G}^C_B$ are assigned based on how microstructure $D$ affects measured lengths of material elements with respect to an observer in total generalized Finsler space $\mathcal{Z}$ (i.e., the space of the physical body enriched with microstructure geometry) \cite{clayton2017g,clayton2017z,claytonMMS22}.
\end{rmk}
Ideas apply analogously to spatial metric $g_{ab}(x,d)$ with $X$ replaced by $x$, and with $D$ replaced by $d$. For example, the spatial analog of \eqref{eq:mdec} is
\begin{equation}
\label{eq:mdecc}
\begin{split}
g_{ab}(x,d)  = \bar{g}_{ac}(x) \hat{g}^c_b (d(x))
= \hat{g}^c_a (d(x))  \bar{g}_{cb}(x) ; \qquad
 \bar{g}_{ab}  = \bar{g}_{ba}, \quad
\qquad \hat{g}^c_a g_{bc} = \hat{g}^c_b g_{ca}.
\end{split}
\end{equation}
All metrics in \eqref{eq:mdec} and \eqref{eq:mdecc} are assumed invertible with positive determinants. A symmetric  tensor $\bar{\vec{C}}$ 
\cite{clayton2017z} and volume ratio $\bar{J} > 0$ are defined to
exclude internal state-dependence of strain:
\begin{equation}
\label{eq:Cbar}
\bar{\vec{C}}(X) = \bar{C}_{AB}(X) \, d X^A \otimes d X^B, \qquad 
\bar{C}_{AB} = F^a_A \bar{g}_{ab} F^b_B, \quad \bar{C}^A_B = \bar{G}^{AC} \bar{C}_{CB};
\end{equation}
\begin{equation}
\label{eq:Jbar}
\bar{J}(X) = \sqrt{ \det (\bar{C}^A_B (X)) }; \qquad \bar{J}  = J \sqrt{ \hat{G} / \hat{g}},
\quad \hat{G} = \det (\hat{G}^A_B), \quad
\hat{g} = \det (\hat{g}^a_b).
\end{equation}

\subsection{Energy and equilibrium}

\subsubsection{Variational principle}

 A variational principle \cite{clayton2017g,clayton2016,claytonMMS22} is implemented.
Let $\Psi$ denote the total energy functional 
for a compact domain $\mathcal{M}' \subset \mathcal{M}$ with positively oriented boundary
$\partial \! \mathcal{M}'$, and let $\psi$ be the local free energy density per unit reference volume of material:
\begin{equation}
\label{eq:Psitot}
\Psi  [\pmb{\varphi},\vec{D}] = \int_{\mathcal{M}'} \psi (F^a_A, D^A, D^A_{|B},X^A) \, d \Omega. 
\end{equation}
Denote surface forces as $\vec{p} = p_a d x^a$, a mechanical
load vector (force per unit reference area), and $\vec{z} = z_A \delta D^A$, a thermodynamic
force conjugate to the internal state vector. Denote
a generic local, vector-valued volumetric source term conjugate to structure variations by $\vec{R} = R_A \delta D^A$ , extending prior theory \cite{clayton2017g,clayton2016,claytonMMS22} to accommodate more
possible physics \cite{lubarda2002,yavari2010} (Appendix B). A variational principle for Finsler-geometric continuum mechanics, holding
$X$ fixed but with $x = \varphi(X)$ and $D$ variable, is 
\begin{equation}
\label{eq:variation1}
\delta \Psi [\pmb{\varphi},\vec{D}] = \oint_{\partial \! \mathcal{M}'} ( \langle \vec{p}, 
\delta \pmb{\varphi} \rangle + \langle \vec{z}, \delta \vec{D} \rangle ) \Omega
+ \int_{\mathcal{M}'}  \langle \vec{R}, \delta \vec{D} \rangle  \, d \Omega.
\end{equation}
In coordinates, with variation of $\vec{D}$ in parentheses to distinguish from non-holonomic basis $\{ \delta D^A \}$,
\begin{equation}
\label{eq:variation2}
\delta \int_{\mathcal{M}'} \psi \, d \Omega = \oint_{\partial \! \mathcal{M}'} \{ p_a \delta \varphi^a
\} \Omega 
+  \oint_{\partial \! \mathcal{M}'} \{ z_C \delta (D^C) \} \Omega
+  \int_{\mathcal{M}'}  \{ R_C \delta (D^C) \} d \Omega.
\end{equation}
Results used in \S3.3.3 are now noted, with $\alpha = 1$ or $\alpha = 2$  (derived in Appendix A  using \eqref{eq:connrec}):
\begin{equation}
\label{eq:variations}
\delta F^a_A = \delta_A (\delta \varphi^a), 
\qquad
\delta D^A_{|B} = [\delta(D^A)]_{|B} - (\bar{\partial}_C N^A_B - \bar{\partial}_C K^A_{BD} D^D)\delta(D^C),
\end{equation}
\begin{equation}
\label{eq:varG}
\delta (d \Omega) = \textstyle{\frac{1}{2}} \alpha G^{AB} \bar{\partial}_C G_{AB} \delta (D^C) d \Omega 
= \alpha \bar{\partial}_C (\ln \sqrt{G}) \delta (D^C) d \Omega
= \alpha C^A_{CA}  \delta (D^C) d \Omega.
\end{equation}

\subsubsection{General energy density}

As evident in \eqref{eq:variation1}, independent variables entering total free energy density, per unit reference volume, function $\psi$ are the deformation gradient, the internal state vector, the horizontal gradient of the internal state vector, and the reference position of the material particle:
\begin{equation}
\label{eq:psivars}
\psi = \psi (\vec{F},\vec{D}, \nabla \vec{D}, \vec{X} ) = \psi (F^a_A, D^A, D^A_{|B},X^A).
\end{equation}
Dependence on $\vec{F}$ accounts for bulk elastic strain energy.
Dependence on $\vec{D}$ generally accounts for effects of microstructure
on stored energy.  Energy from heterogeneity
of microstructure (e.g., internal material surfaces) is captured by dependence on the internal state gradient:
\begin{equation}
\label{eq:gradD}
\nabla \vec{D}  = D^A_{|B} \frac{\partial}{\partial D^A} \otimes d X^B + D^A|_B 
\frac{\partial}{\partial D^A}  \otimes \delta D^B;
\end{equation}
\begin{equation}
\label{eq:gradDi}
 D^A_{|B}  = \delta_B D^A + K^A_{BC} D^C = \partial_B D^A - N_B^A  + K^A_{BC} D^C,
 \qquad 
 D^A|_B = \bar{\partial}_B D^A + V ^A_{BC} D^C = \delta^A_B.
 \end{equation}
 Dependence on $\vec{X}$ permits heterogeneous properties.
Prior work \cite{clayton2017g,claytonMMS22} adds motivation for \eqref{eq:psivars}. 
\begin{rmk} Vertical gradient $D^A|_B = \delta^A_B$, calculated from $V^A_{BC} = 0$ by \eqref{eq:connrec}, provides no information, so it 
is excluded from the arguments of energy density in \eqref{eq:psivars}.
\end{rmk}
Expansion of the integrand on the left in \eqref{eq:variation2}, with $\delta X^A = 0$ by definition, is
\begin{equation}
\label{eq:deltapsi}
\begin{split}
\delta \psi & = \frac{\partial \psi}{\partial F^a_A} \delta F^a_A + \frac{\partial \psi}{\partial D^A} \delta (D^A)
+ \frac{\partial \psi}{\partial D^A_{|B}} \delta D^A_{|B} 
= P^A_a \delta F^a_A + Q_A \delta (D^A)
+ Z^B_A \delta D^A_{|B};
\\
P^A_a & = \frac{\partial \psi}{\partial F^a_A}, \qquad
Q_A =  \frac{\partial \psi}{\partial D^A} , \qquad
Z^A_B = \frac{\partial \psi}{\partial D^B_{|A}}.
\end{split}
\end{equation}
Denoted by $\vec{P}$ is the mechanical stress tensor (i.e., the first Piola-Kirchhoff stress, a two-point tensor, generally non-symmetric), $\vec{Q}$ an internal force vector conjugate to $\vec{D}$, and $\vec{Z}$ a micro-stress tensor conjugate to the horizontal gradient of $\vec{D}$. 

\subsubsection{Euler-Lagrange equations}
Connection coefficients in \eqref{eq:connrec} are employed along with \eqref{eq:varphiC},
\eqref{eq:horiztranspf}, \eqref{eq:funcC}, \eqref{eq:variations}, and \eqref{eq:varG}.  Insertion of \eqref{eq:deltapsi} into the left side of \eqref{eq:variation2}, followed by
 integration by parts and use of \eqref{eq:stokes} of Theorem \ref{thm:stokes}, produces 
 \begin{equation}
 \label{eq:intpart}
 \begin{split}
 \delta \int_{\mathcal{M}'} & \psi \, d \Omega  =
  \int_{\mathcal{M}'} 
\{ P^A_a \delta F^a_A + Q_A \delta (D^A)
+ Z^B_A \delta D^A_{|B} \} d \Omega 
 +
 \int_{\mathcal{M}'} 
 \psi \delta( d \Omega) \\ = & 
 - \int_{\mathcal{M}'} 
 \{ \partial_A P^A_a  +  \bar{\partial}_B P^A_a \partial_A D^B + P^B_a \Gamma^A_{AB} - P^A_c \Gamma^c_{ba} F^b_A + P^A_a C^C_{BC} (\partial_A D^B + N^B_A) \} \delta \varphi^a
 d \Omega 
 \\ & \quad  - \int_{\mathcal{M}'} 
 \{
  \partial_A Z^A_C +   \bar{\partial}_B Z^A_C \partial_A D^B   + Z^B_C \Gamma^A_{AB}  - Z^A_B K^B_{AC} - Q_C 
   \\  
& \qquad \qquad  + Z^B_A [\bar{\partial}_C N^A_B - \bar{\partial}_C K^A_{BD} D^D + \delta^A_C C^D_{ED} (\partial_B D^E + N^E_B )] - \alpha  \psi C^A_{CA}
 \} \delta(D^C) d \Omega 
  \\ & \quad  + \oint_{\partial \! \mathcal{M}'} \{ P^A_a \delta \varphi^a \} N_A \Omega
  + \oint_{\partial \! \mathcal{M}'} \{ Z^A_C  \delta(D^C) \} N_A \Omega.
 \end{split}
 \end{equation}
  Euler-Lagrange equations consistent with any admissible variations $\delta \pmb{\varphi}$ and $\delta \vec{D}$ locally at each $X \in \mathcal{M}'$, as well as natural boundary conditions on $\partial \! \mathcal{M}'$ are obtained as follows.  
Steps follow those outlined in the original works \cite{clayton2017g,clayton2016} with minor departures \cite{claytonMMS22}.

The first of these culminating Euler-Lagrange equations is the macroscopic balance of linear momentum, derived by setting the first integral on the right-hand side of \eqref{eq:intpart} equal to zero, consistent with the right side of \eqref{eq:variation2}. Localizing the outcome and presuming the result must hold for any admissible variation $\delta \varphi^a$,
\begin{equation}
\label{eq:linmom}
\partial_A P^A_a  +  \bar{\partial}_B P^A_a \partial_A D^B + P^B_a \Gamma^A_{AB} - P^A_c \Gamma^c_{ba} F^b_A =
 - P^A_a C^C_{BC} (\partial_A D^B + N^B_A).
\end{equation}
The second Euler-Lagrange equation is the balance of micro-momentum (i.e., director momentum or internal state equilibrium).
It is derived by setting the second integral on the right side of \eqref{eq:intpart}
equal to the rightmost term in \eqref{eq:variation2} and then localizing,
giving for any admissible
variation $\delta (D^C)$,
\begin{equation}
\label{eq:dirmom}
\begin{split}
  \partial_A Z^A_C +   \bar{\partial}_B Z^A_C \partial_A D^B  & + Z^B_C \Gamma^A_{AB}  - Z^A_B K^B_{AC} - 
  (Q_C - R_C)
   \\ & = 
 \alpha  \psi C^A_{CA} - Z^B_A [\bar{\partial}_C N^A_B - \bar{\partial}_C K^A_{BD} D^D + \delta^A_C C^D_{ED} (\partial_B D^E + N^E_B )].
\end{split} 
\end{equation}
Natural boundary conditions on $\partial \! \mathcal{M}'$ are 
derived by setting the second-to-last and last boundary integrals
in \eqref{eq:intpart} equal to the remaining, respective first and second boundary integrals on the
right side of \eqref{eq:variation2} and localizing the results, yielding for any admissible
variations $\delta \varphi^a$ and $\delta (D^C)$,
\begin{equation}
\label{eq:neumann}
p_a = P^A_a N_A, \qquad z_A = Z^B_A N_B.
\end{equation}
\begin{rmk}
With natural boundary conditions \eqref{eq:neumann} or essential boundary conditions (i.e., prescribed $\pmb{\varphi}(X)$ and $\vec{D}(X)$ for $X \in \partial \! \mathcal{M}'$) and local force density vector $\vec{R}(X)$ for each $X \in \mathcal{M}'$, \eqref{eq:linmom} and \eqref{eq:dirmom} comprise 2$n$ coupled PDEs in 2$n$ degrees-of-freedom
$x^a = \varphi^a(X) $ and $D^A(X)$ at any $X \in \mathcal{M}'$, and by extension, any $X \in \mathcal{M}$.
\end{rmk}

\begin{rmk}
Consider the simplified case when Riemannian metrics are used: no $D$-dependence of $\vec{G}$ and
no $d$-dependence of $\vec{g}$. Then $\Gamma^A_{BC} = \gamma^A_{BC}$, $\Gamma^a_{bc} = \gamma^a_{bc}$,  and $C^A_{BC} = 0$.  The right side of \eqref{eq:linmom} 
vanishes, and \eqref{eq:linmom} is of the form of the static momentum balance of classical continuum mechanics with null 
body force \cite{toupin1960,marsden1983,clayton2011}. Further taking
$N^A_B$ and $K^A_{BC}$ independent of $D$, \eqref{eq:dirmom} is similar to equilibrium equations for gradient materials \cite{capriz1989} as in phase-field mechanics \cite{gurtin1996,clayton2011b}.
\end{rmk}

\begin{rmk}
In some prior work \cite{clayton2017g}, $G_{AB}(X,D)$ was an argument of $\psi$, extending \eqref{eq:psivars}, and $D$-dependence of the metric manifested in a distinct thermodynamic force, rather than entering implicitly in $Q_A$.
The present approach is favored for brevity \cite{claytonMMS22}, but the former is admissible.
\end{rmk}

\begin{prop}
Euler-Lagrange equations can be expressed in the following alternative way:
\begin{equation}
\label{eq:altmac}
\partial_A P^A_a  +  \bar{\partial}_B P^A_a \partial_A D^B + P^B_a \gamma^A_{AB} - P^A_c \Gamma^c_{ba} F^b_A =
 - P^A_a C^C_{BC} \partial_A D^B,
\end{equation}
\begin{equation}
\label{eq:altmic}
\begin{split}
  \partial_A Z^A_C +   \bar{\partial}_B Z^A_C \partial_A D^B  & + Z^B_C \gamma^A_{AB}  - Z^A_B K^B_{AC} - 
  (Q_C - R_C)
   \\ & = 
 \alpha  \psi C^A_{CA} - Z^B_A ( \bar{\partial}_C N^A_B - \bar{\partial}_C K^A_{BD} D^D + \delta^A_C C^D_{ED} \partial_B D^E ).
\end{split} 
\end{equation}
\end{prop}
\begin{pf}
From \eqref{eq:diffnot} and \eqref{eq:Gids2},
\begin{equation}
\label{eq:trGamma}
\Gamma^A_{AB} = \Gamma^A_{BA} = \partial_B (\ln \sqrt{G}) - N^A_B \bar{\partial}_A (\ln \sqrt{G})
= \gamma^A_{BA} - N^A_B C^C_{AC}.
\end{equation}
Substitution of \eqref{eq:trGamma} with symmetry $\gamma^A_{BC} = \gamma^A_{CB}$
into \eqref{eq:linmom} and \eqref{eq:dirmom} yields \eqref{eq:altmac} and \eqref{eq:altmic}.$\quad \square$
\end{pf}
\begin{rmk} Notably, \eqref{eq:altmac} and \eqref{eq:altmic} show how the nonlinear connection terms $N^A_B$
cancel, simplifying calculations.
Nonlinear connection $N^A_B$ still ultimately affects governing equations via
influence on $D^A_{|B} = \partial_B D^A - N_B^A + K^A_{BC} D^C$, thus affecting $Z^B_A = \partial \psi / \partial D^A_{|B}$, and
through $\bar{\partial}_C N^A_B$ in \eqref{eq:altmic}.
Spatial $N^a_b$ can enter $\Gamma^a_{bc}$ in \eqref{eq:altmac}.
The linear connection $K^A_{BC}$ and
its gradient $\bar{\partial}_D K^A_{BC}$ in \eqref{eq:altmic} are somewhat unique to Finsler-geometric continuum mechanics. The trace of Cartan's tensor, $C^A_{BA}$, in all forms of the Euler-Lagrange equations is also 
a distinctive feature. This term, of course, vanishes when $G_{AB}$ is independent of $D$ (i.e., a Riemannian rather than Finslerian metric).
\end{rmk}

\subsubsection{Spatial invariance and material symmetry}
First consider rotations of the spatial frame of reference, given by orthonormal transformation
$q^a_b$ in \eqref{eq:trans1c} whereby 
$\det (q^a_b) = 1$ and $\tilde{q}^a_b = g^{ac} q^d_c g_{bd}$ (i.e., $\vec{q}^{-1} = \vec{q}^\text{T}$ \cite{marsden1983}).
Since $\vec{F} \rightarrow \vec{q} \vec{F}$ under such coordinate changes,
$\psi$ in \eqref{eq:psivars} should obey more restricted forms to maintain proper observer independence. 
Two possibilities are
\begin{equation}
\label{eq:psi2}
\psi = \hat{\psi}[\vec{C}(\vec{F},\vec{g}),\vec{D},\nabla \vec{D},\vec{X}] = \hat{\psi}(C_{AB}, D^A, D^A_{|B},X^A),
\end{equation}
\begin{equation}
\label{eq:psi3}
\psi = \bar{\psi}[\bar{\vec{C}}(\vec{F},\bar{\vec{g}}),\vec{D},\nabla \vec{D},\vec{X}] = \bar{\psi}(\bar{C}_{AB}, D^A, D^A_{|B}, X^A),
\end{equation}
noting that \eqref{eq:psivars} can be consistently expressed 
from \eqref{eq:varphiC}, \eqref{eq:thetaC}, \eqref{eq:mdec}, and \eqref{eq:mdecc} as
\begin{equation}
\label{eq:psivars2}
 \psi (\vec{F},\vec{D}, \nabla \vec{D}, \vec{X} ) = \check{\psi} 
 (\vec{F},\vec{D}, \bar{\vec{G}}(\vec{X}), \hat{\vec{G}}(\vec{D}), 
 \bar{\vec{g}}({\varphi} (\vec{X})), \hat{\vec{g}}({\theta} (\vec{X},\vec{D})), \nabla \vec{D}, \vec{X} ).
\end{equation}
From \eqref{eq:lengthC}, \eqref{eq:Cbar}, \eqref{eq:psi2} and \eqref{eq:psi3}, first Piola-Kirchhoff stress $P^A_a$ of \eqref{eq:deltapsi} is
calculated using the chain rule:
\begin{equation}
\label{eq:Pk1}
P^A_a = \frac{\partial \psi} {\partial F^a_A} = 2 g_{ab} F^b_B \frac{\partial \hat{\psi}}{\partial C_{AB}} = 2 \bar{g}_{ab} F^b_B \frac{\partial \bar{\psi}}{\partial \bar{C}_{AB}}.
\end{equation}
The resulting Cauchy stress tensors with spatial components $\sigma^{ab}$ and  $\bar{\sigma}^{ab}$ obey symmetry rules consistent with the classical local balance of angular momentum \cite{truesdell1960,marsden1983,clayton2011}:
\begin{equation}
\label{eq:cauchy}
\sigma^{ab} = \frac{1}{J} g^{ac} P_c^A F^b_A = 
\frac{2}{J} F^a_A F^b_B \frac{\partial \hat{\psi}}{\partial C_{AB}}  = \sigma^{ba}, \quad
\bar{\sigma}^{ab} =
\frac{1}{\bar{J}} \bar{g}^{ac} P_c^A F^b_A =
\frac{2}{\bar{J}} F^a_A F^b_B \frac{\partial \bar{\psi}}{\partial \bar{C}_{AB}} 
= \bar{\sigma}^{ba}.
\end{equation}

Now consider changes of the material frame of reference, given by transformation $Q^A_B$ of
\eqref{eq:trans1} and \eqref{eq:subseq} with inverse $\tilde{Q}^B_A$.
Under affine changes of coordinates $X^A \rightarrow Q^C_A X^A$, it follows that $d X^A \rightarrow  Q^C_A d X^A$, 
$F^a_A \rightarrow \tilde{Q}^A_C F^a_A$, $G_{AB} \rightarrow \tilde{Q}^A_C \tilde{Q}^B_D G_{AB}$,
$C_{AB} \rightarrow \tilde{Q}^A_C \tilde{Q}^B_D C_{AB}$,
$\bar{G}_{AB} \rightarrow \tilde{Q}^A_C \tilde{Q}^B_D \bar{G}_{AB}$, 
$\bar{C}_{AB} \rightarrow \tilde{Q}^A_C \tilde{Q}^B_D \bar{C}_{AB}$,
$D^A \rightarrow Q^C_A D^A$, $\delta  D^A \rightarrow  Q^C_A \delta D^A$, and 
$D^A_{|B} \rightarrow Q^C_A \tilde{Q}^B_D D^A_{|B}$.
Energy densities $\psi$, $\hat{\psi}$, and $\bar{\psi}$ should be 
invariant under all transformations $\tilde{Q}^A_B$ (e.g., rotations, reflections, inversions) belonging to the symmetry group $\mathbb{Q}$ of the material \cite{maugin1972b,clayton2011,spencer1971,balzani2006} (e.g., $\psi \rightarrow \psi$).
The present focus is on polynomial invariants \cite{spencer1971,balzani2006} with basis $\mathcal{P}$ of invariant functions with respect to $\tilde{\vec{Q}} \in \mathbb{Q}$ and energy offsets $\hat{\psi}_0 = \text{constant}$,  $\bar{\psi}_0 = \text{constant}$: 
\begin{equation}
\label{eq:pbasis1}
\hat{\mathcal{P}} = \{ I_1, I_2, \ldots, I_\upsilon \}; \qquad I_\alpha = I_\alpha (\vec{C},\vec{D},\nabla \vec{D}), \qquad
\hat{\psi} = \hat{\psi} (I_1, I_2, \ldots, I_\upsilon, \vec{X}) + \hat{\psi}_0;
\end{equation}
\begin{equation}
\label{eq:pbasis2}
\bar{\mathcal{P}} = \{ \bar{I}_1, \bar{I}_2, \ldots, \bar{I}_\zeta \}; \qquad \bar{I}_\alpha = \bar{I}_\alpha (\bar{\vec{C}},\vec{D},\nabla \vec{D}), \qquad
\bar{\psi} = \bar{\psi} (\bar{I}_1, \bar{I}_2, \ldots, \bar{I}_\zeta, \vec{X}) + \bar{\psi}_0.
\end{equation}
The total number of applicable invariants is $\upsilon$ or $\zeta$ for \eqref{eq:psi2} or \eqref{eq:psi3}.  Stress of \eqref{eq:Pk1} becomes 
\begin{equation}
\label{eq:Pk1b}
P^A_a = 2 g_{ab} F^b_B \sum_{\alpha = 1}^\upsilon \hat{\psi}_\alpha \frac{\partial I_\alpha }{\partial C_{AB}}
= 2 \bar{g}_{ab} F^b_B \sum_{\alpha = 1}^\zeta \bar{\psi}_\alpha \frac{\partial \bar{I}_\alpha} {\partial \bar{C}_{AB}};
\quad \hat{\psi}_\alpha = \frac{ \partial \hat{\psi}} {\partial I_\alpha}, 
\quad \bar{\psi}_\alpha = \frac{ \partial \bar{\psi}} {\partial \bar{I}_\alpha}.
\end{equation}
\begin{rmk} A thorough and modern geometric treatment of material symmetry, uniformity, and homogeneity in continuous media is included in a recent monograph \cite{leon2021}.
\end{rmk}

\section{One-dimensional base manifold}
The framework of \S2 and \S3 is applied for $n = 1$: a 1-D base manifold $\mathcal{M}$.
In \S4.1, geometry and kinematics are presented, including assumptions that
enable tractable solutions to several classes of boundary value problems while at the same time
maintaining sufficient generality to address broad physical behaviors.
Resulting 1-D governing equations are derived in \S4.2.
General solutions are obtained for two problem classes in \S4.3.
Constitutive functions for a soft biological tissue, namely a 1-D strip of skin under axial extension, are given in \S4.4. Model parameters and analytical solutions for 1-D skin stretching and tearing are reported in \S4.5.

\subsection{Geometry and kinematics}
Let $X = X^1$. Considered is a reference domain $\{\mathcal{M}: X \in [-L_0,L_0] \}$, where the total length relative to a Euclidean metric
is $2 L_0$, and boundary $\partial \! \mathcal{M}$ is the endpoints $X = \pm L_0$.
The referential internal state vector reduces to the single component $D = D^1$, which is assumed to have physical units, like $X$, of length. 
The spatial coordinate is $x = x^1$, and the spatial state component is $d = d^1$.
A normalization constant (i.e., regularization length) $l$ is introduced, and the physically meaningful domain for internal state
is assumed as $D \in [0,l]$. The associated order parameter is 
\begin{equation}
\label{eq:OP}
\xi(X)  = \frac{D(X)}{l} = \frac{d(\varphi(X))}{l}, \qquad l > 0,
\end{equation}
with meaningful domain $\xi \in [0,1]$, and where \eqref{eq:funcC} and \eqref{eq:thetasimp} are invoked.
For generic $f$ and $h$ differentiable in their arguments, let
\begin{equation}
\label{eq:not1D}
f ' (X) = \frac{ \d f(X)}{ \d X}, \quad f ''(X) = \frac{ \d^2 f (X)}{ \d X^2}; \qquad
\dot{h}(\xi) = \frac{ \d h( \xi )}{ \d \xi},  \quad \ddot{h}(\xi) = \frac{ \d^2 h(\xi)}{ \d \xi^2}.
\end{equation}
For 1-D manifolds, the following metrics apply from \eqref{eq:mdec} and \eqref{eq:mdecc}:
\begin{equation}
\label{eq:1D}
G_{11}(X,D) = G(X,D) = \bar{G}(X) \hat{G}(D) = \hat{G}(D),
\quad
g_{11}(x,d) = g(x,d) = \bar{g}(x) \hat{g}(d) = \hat{g}(d).
\end{equation}
Since $\bar{g} = \bar{G} = 1$ for isometric 1-D Riemannian spaces, setting 
\begin{equation}
\label{eq:isometric}
\hat{g}( d (\varphi(X))) = \hat{G}( D(X)) \leftrightarrow g(\xi) = G(\xi)
\end{equation}
renders $\mathfrak{m}$ and $\mathcal{M}$ isometric 
when $\phi(X) = X + c_0 \Leftrightarrow F(X) = 1$, regardless of local values of $D$, $d$, or $\xi$ at corresponding points $x = \varphi(X)$.
\begin{rmk}
This assumption \eqref{eq:isometric}, used henceforth in \S4, may be relaxed in future applications to address residual stress (e.g., from growth \cite{yavari2010}; see Appendix B), especially for $n = \dim \mathcal{M} > 1$.
\end{rmk}
Henceforth in \S4, functional dependence on $D$ or $d$ is replaced with that on $\xi$. Then
\begin{equation}
\label{eq:xiderivs}
D' = \frac{\xi'}{l}, \qquad \frac{\partial f(X,D)}{ \partial D } = \frac{1}{l} \frac{\partial f(X,\xi(D))}{ \partial \xi }.
\end{equation}
The following functional forms are assumed for 
referential nonlinear connection $N^A_B$ and linear connection $K^A_{BC}$, with $N_0 = \text{constant}$ and $\hat{K}(X)$  both dimensionless:
\begin{equation}
\label{eq:N1D}
N^A_B \rightarrow N^1_1 = N = -N_0 l \xi', \qquad K^A_{BC} \rightarrow K^1_{11} (X,\xi) = K(X,\xi)= \frac{\hat{K}(X)}{l \xi}
\, \Rightarrow \, \bar{\partial}_1 K^1_{11} D = - K^1_{11}. 
\end{equation}
Spatial coefficients $K^a_{bc}$ do not affect the governing equations and thus are left unspecified. Conditions \eqref{eq:connrec} apply in 1-D, leading to, with \eqref{eq:OP}--\eqref{eq:N1D}, 
\begin{equation}
\label{eq:conn1D}
\Gamma^A_{BC} \rightarrow \Gamma^1_{11} = {{\frac{1}{2G}}} \delta_1 G  
=  {{\frac{1}{2G}}} ( \partial_1 G - N^1_1 \bar{\partial}_1 G) = -N \bar{\partial}_1 ( \ln \sqrt{G}) = -N C^1_{11} 
= - N \frac{\chi}{l},
\end{equation}
\begin{equation}
\label{eq:chi1D}
\chi(\xi) = \frac { \dot{G}(\xi)}{2 G(\xi) } = \frac { \dot{g}(\xi)}{2 g(\xi) } = l C^1_{11} (\xi),
\end{equation}
\begin{equation}
\label{eq:N1Ds}
N^a_b \rightarrow \frac{N}{F}  = -\frac{N_0 l \xi'}{F}   = - {N_0 l}\frac{ \d \xi}{\d x},
\qquad
\Gamma^a_{bc} \rightarrow -\frac{N}{F} \frac{ \dot{g}} {2 g}  = -\frac{N}{F} \frac{\chi}{l}.
\end{equation}
The deformation gradient, deformation tensor, Jacobian determinant, and director gradient are
\begin{equation}
\label{eq:def1D}
F^a_A \rightarrow F^1_1 = F =  \frac{\d \varphi}{\d X} =  \varphi', \quad
C^A_B \rightarrow C^1_1 = C= G^{11} g_{11} F^1_1 F^1_1 = F^2 = (\varphi')^2, \quad J = F = \sqrt{C},
\end{equation}
\begin{equation}
\label{eq:dgrad1D}
D^A_{|B} \rightarrow D^1_{|1}= \frac{\d D}{\d X} - N + K D = 
(1+ N_0) l \xi' + \hat{K}.
\end{equation}
From \eqref{eq:isometric}, $\bar{C} = \bar{C}^1_1 = C^1_1 = C$  and $\bar{J} = J$ in 
1-D reductions of \eqref{eq:Cbar} and \eqref{eq:Jbar}.

\subsection{Governing equations}
A generic energy density is assigned and equilibrium equations are derived for the 1-D case given prescriptions of \S4.1.

\subsubsection{Energy density}
In 1-D, $C_{AB}$ consists of a single invariant $C$, and $D^A$ and $D^A_{|B}$ likewise.
Dependencies in \eqref{eq:psivars} are suitably represented by $F$, $\xi$, and $(\xi',X)$ with \eqref{eq:OP} and \eqref{eq:dgrad1D}. 
Since $\bar{C} = C = F^2$, all energy densities $\psi$ of \eqref{eq:psivars} in \eqref{eq:psi2} -- \eqref{eq:psivars2}
are expressed simply as
\begin{equation}
\label{eq:psi1D}
\psi = \psi(C,\xi,\xi',X).
\end{equation}
Denote by $\mu_0$ a constant, later associated to an elastic modulus, with units of energy density.
\begin{rmk}
For comparison with data from experiments in ambient Euclidean 3-space, $\mu_0$ can be given units of energy per unit (3-D) volume, such that $\Psi = \int_\mathcal{M} \psi d \Omega$ is energy per unit cross-sectional area normal to $X$.
For a 1-D $\mathcal{M}$, this cross-sectional area is, by definition, constant.  
\end{rmk}
\noindent Denote by $\Upsilon_0$ a constant, related to surface energy, with units of energy per unit (2-D fixed cross-sectional) area.
Let $W$ be strain energy density and $\Lambda$ energy density associated with microstructure.
Denote by $w$ a dimensionless strain energy function, $y$ a dimensionless interaction function (e.g., later representing elastic degradation from microstructure changes), $\lambda$ a dimensionless phase energy function, and $\iota$ a dimensionless gradient energy function assigned a quadratic form.
Free energy density \eqref{eq:psi1D} is then prescribed in intermediate functional form as follows:
\begin{equation}
\label{eq:psi1d1}
\psi(C,\xi,\xi',X) = W(C,\xi) + \Lambda(\xi,\xi',X) = {\frac{\mu_0}{2}} w(C) y(\xi) + \frac{\Upsilon_0}{l} [ \lambda(\xi) + \iota(\xi',X)],
\end{equation}
\begin{equation}
\label{eq:iota}
\iota = |D^1_{|1}|^2-\hat{K}^2 = D^1_{|1} G_{11} G^{11} D^1_{|1} -\hat{K}^2 = [(1+ N_0) l \xi' + \hat{K}]^2 -\hat{K}^2, \quad (N_0 = \text{const}, \hat{K} =  \hat{K}(X)).
\end{equation}
Note $\iota(0,X) = 0$. For null ground-state energy and stress, $\psi(1,0,0,X) = 0$ and $\frac{\partial \psi}{ \partial {C}}(1,0,0,X)= 0$:
\begin{equation}
\label{eq:ground1D}
w(1) = 0, \qquad \frac{\d w }{\d C}(1) = 0, \qquad \frac{ \d^2w}{ \d C^2} \geq 0; \qquad \lambda(0) = 0.
\end{equation}
The third of \eqref{eq:ground1D} ensures convexity of $w$. Thermodynamic forces originating in \eqref{eq:deltapsi} are derived as
\begin{equation}
\label{eq:P1d}
P = P^1_1 = \frac{\partial \psi}{\partial F} =  2 \frac{g}{G} F \frac{\partial \psi}{\partial C}  = 2 \sqrt{C} \frac{\partial \psi}{\partial C} 
= \mu_0 y \sqrt{C}  \frac{\d w}{\d C} ,
\end{equation}
\begin{equation}
\label{eq:Q1d}
Q = Q_1 = \frac{\partial \psi}{\partial D} =  \frac{1}{l} \frac{\partial \psi}{\partial \xi} = \frac{\mu_0}{2l} w \frac{\d y}{\d \xi}
+ \frac{\Upsilon_0}{l^2} \frac{\d \lambda}{\d \xi}
= \frac{\Upsilon_0}{l^2} \left( A_0 w \dot{y} + \dot{\lambda} \right), \quad A_0 = \frac{\mu_0 l}{2 \Upsilon_0},
\end{equation}
\begin{equation}
\label{eq:Z1d}
Z = Z^1_1 = \frac{\partial \psi}{\partial D^1_{|1}} =  \frac{\Upsilon_0}{l} \frac{\partial \iota}{\partial D^1_{|1}} 
= 2 \frac{\Upsilon_0}{l} D^1_{|1} = 2 \frac{\Upsilon_0}{l} [(1+ N_0) l \xi' + \hat{K}].
\end{equation}
The volumetric source term in \eqref{eq:variation1} is prescribed as manifesting from changes in energy density proportional to changes of the local referential volume form (e.g.,  physically representative of local volume changes from damage/tearing, similar to effects of tissue growth on energy (Appendix B)):
\begin{equation}
\label{eq:R1d}
R = R_1 = \beta \psi \bar{\partial}_1 ( \ln \sqrt{G}) = \frac{\beta}{l} \psi \chi, \qquad (\beta = \text{constant}). 
\end{equation}

\subsubsection{Linear momentum}
The macroscopic momentum balance, \eqref{eq:linmom} or \eqref{eq:altmac} is, upon use of relations in \S4.1 and \S4.2.1,
\begin{equation}
\label{eq:linmom1d}
\frac{\d P}{\d X} 
  = 
 P  (N_0 - 1) \chi \frac{\d \xi}{\d X} = -  (1 - N_0 ) \frac{P}{2G}  \frac{\d G}{\d X}.
\end{equation}
This is a separable first-order ordinary differential equation (ODE) that can be integrated directly:
\begin{equation}
\label{eq:linmom1d2}
\int_{P_0}^P{\d  (\ln \mathsf{P}}) = - (1 -N_0) \int_{G_0}^G \d ({ \ln \sqrt{\mathsf{G}} } )
\quad \Rightarrow \quad P = P_0 \left( \sqrt{{G_0}/{G}} \right)^{1-N_0}.
\end{equation}
The integration limit on $G(\xi(X))$ is $G_0 = G(0)$, and $P_0$ is a constant
stress corresponding to $\xi = 0$. 
\begin{rmk}
If $G$ is Riemannian, then $G = G_0$ and $P = P_0 = \text{constant}$.
In the Finslerian setting, $P$ can vary with $X$ if $\xi$ varies with $X$ and $N_0$ differs from unity.
However, if $P$ vanishes on $\partial \! \mathcal{M}$ (i.e., at $X = \pm L_0$),
then $P_0 = 0$ necessarily, so $P(X) = 0 \, \forall \, X \in \mathcal{M}$, meaning this 1-D
domain cannot support residual stress. The same assertion applies when $\eqref{eq:isometric}$ is relaxed and $N_0$ vanishes.
\end{rmk}
\noindent From \eqref{eq:P1d} and \eqref{eq:linmom1d2}, when $\mu_0$ is nonzero,
\begin{equation}
\label{eq:linmom1d3}
\sqrt{C(X)}   \frac{\d w(C(X))}{\d C}y(\xi(X)) \left[ \frac{G (\xi(X)) }{G_0} \right]^{(1-N_0)/2} = \frac{ P_0}{\mu_0} = \text{constant},
\end{equation}
where the value of $P_0$, constant for a given static problem, depends on the boundary conditions.

\subsubsection{Micro-momentum}
Define $\bar{K}(X) = l \hat{K}(X)$. Then the microscopic momentum balance, \eqref{eq:dirmom} or \eqref{eq:altmic}, is, upon use of relations in \S4.1 and \S4.2.1
and dividing by $2 \Upsilon_0 ( 1 + N_0)$,
\begin{equation}
\label{eq:micmom1d}
\begin{split}
\frac{\d ^2 \xi}{\d X^2} & + \chi(\xi) \left[ 1  - \frac{(1+N_0)(\alpha-\beta)}{2} \right] \left( \frac{ \d \xi } {\d X} \right)^2 
+ \frac{\bar{K}(X)}{l^2} \chi(\xi) \left[ \frac{1}{1+N_0} - (\alpha - \beta) \right] \frac{ \d \xi } {\d X} 
\\ & 
+ \frac{ \d \bar{K}(X)}{\d X} \frac {1}{l^2(1+ N_0)} 
 - \frac{1}{2 l^2(1+N_0)} \left[ \frac{\d \lambda (\xi)}{\d \xi} 
+   (\alpha - \beta) \chi(\xi) \lambda(\xi) \right]
 \\ &
 = \frac{A_0  w(C(X)) }{2l^2(1 + N_0)} \left[ \frac{ \d y(\xi)}{\d \xi}  
+ (\alpha - \beta) \chi(\xi) y(\xi) \right].
\end{split}
\end{equation}
This is a nonlinear and non-homogeneous second-order ODE with variable coefficients.
General analytical solutions are not feasible. However, the following assumption is made henceforth in \S4 to
reduce the nonlinearity (second term on left side) and render some special solutions possible:
\begin{equation}
\label{eq:ab}
\beta = \alpha - {2}/{(1+ N_0)}.
\end{equation}
\begin{rmk}
Assumption \eqref{eq:ab} generalizes, yet is consistent with, physically realistic choices for fracture, shear bands, cavitation, and phase transitions
 \cite{clayton2017g,clayton2016,clayton2017z}: $\alpha=2, \beta = 0, N_0 = 0$.
 \end{rmk}
 \noindent Applying \eqref{eq:ab} with notation of \eqref{eq:not1D}, \eqref{eq:micmom1d} reduces to the form studied in the remainder of \S4:
 \begin{equation}
 \label{eq:1dred}
 \begin{split}
 l^2 (1 + N_0) \xi''  - \frac{\dot{\lambda} }{2} 
-  \frac{ \chi \lambda} {1+N_0} 
- {\bar{K} \chi}  \xi' 
+  {\bar{K}'} 
 = \frac{A_0 w }{2} \left[ \dot{y} + \frac{2 \chi y}{1 + N_0} \right]  .
 \end{split}
 \end{equation}
 This is a linear second-order ODE, albeit generally non-homogeneous with variable coefficients.
For the special case that $\Upsilon_0 ( 1+N_0) = 0$, terms on the left of \eqref{eq:micmom1d} all vanish, and equilibrium demands 
\begin{equation}
\label{eq:special1d}
\mu_0 w(C(X)) \left[ \frac{ \d y(\xi)}{\d \xi}  
+ \frac{2 \chi(\xi) y(\xi)}{1+N_0} \right]= 0.
\end{equation}











\subsection{General solutions}

\subsubsection{Homogeneous fields}
Consider cases wherein $\xi(X) \rightarrow \xi_\text{H} = \text{constant} \,  \forall \, X \in [-L_0,L_0]$. Assign the notation
$ f_\text{H} (X) = f(X,\xi_\text{H})$. Then stress and momentum conservation in \eqref{eq:P1d} and \eqref{eq:linmom1d2} combine to
\begin{equation}
\label{eq:linmomH}
 P_\text{H} = \mu_0 \sqrt{C}  \frac{\d w}{\d C} y_\text{H} =
 P_0 \left( \frac{G_0}{G_\text{H}} \right)^{(1-N_0)/2}  = \text{constant}. 
\end{equation}
If $\mu_0$, $y_\text{H}$, and $\d w / \d C$ are nonzero, convexity of $w$ suggests $C = C_\text{H} = F^2_\text{H} = \text{constant}$.
Accordingly, $\varphi_\text{H} (X) = F_\text{H} X + c_0$. If $\mu_0 = 0$, $y_\text{H} = 0$, or $\d w / \d C = 0$, then $P_\text{H} = 0$ and $\varphi_\text{H}(X)$ is arbitrary. 
Assume now that none of the former are zero, such that $F = F_\text{H}$, $C = \text{C}_\text{H}$, $w = w_\text{H} = w(C_\text{H})$ are constants. Then equilibrium equation \eqref{eq:1dred} becomes, with  $\bar{K}'_\text{H} = K'_0$ a dimensionless constant,
\begin{equation}
 \label{eq:1dH}
 \begin{split}
  - \frac{\dot{\lambda}_\text{H} }{2 } 
-  \frac{ \chi_\text{H} \lambda_\text{H}} {1+N_0} 
+ {K}'_0
 = \frac{A_0 w_\text{H} }{2} \left[ \dot{y}_\text{H} + \frac{2 \chi_\text{H} y_\text{H}}{1 + N_0} \right]  .
 \end{split}
 \end{equation}
 \begin{rmk}
 If $\varphi_\text{H}$ is imposed by displacement boundary conditions, then $C_\text{H}$ is known, as is $w_\text{H}$. In that case,
 \eqref{eq:1dH} is an algebraic equation that can be solved implicitly for $\xi_\text{H}$, the value of which is substituted into 
 \eqref{eq:linmomH} for stress $P_\text{H}$. If $P_\text{H}$ is imposed by traction boundary conditions, then
 \eqref{eq:linmomH} and \eqref{eq:1dH} are to be solved simultaneously for $C_\text{H}$ and $\xi_\text{H}$. 
 \end{rmk}

\subsubsection{Stress-free states}
Now consider cases wherein  $P = 0 \, \forall \, X \in [-L_0,L_0]$. Relation \eqref{eq:linmom1d} is trivially satisfied.
Assume $\mu_0$ is nonzero. Then \eqref{eq:linmom1d3} requires, since $C > 0$, $G > 0$, 
\begin{equation}
\label{eq:linmomzero}
\frac{\d w (C(X))}{ \d C} y( \xi(X)) = 0. 
\end{equation}
This is obeyed for any $y(\xi)$ at $C = 1$ (i.e., rigid-body motion) via \eqref{eq:ground1D}. 
Assume further that $w = 0$, again satisfied at $C = 1$ via \eqref{eq:ground1D}.
Then the right side of \eqref{eq:1dred} vanishes, leaving 
 \begin{equation}
 \label{eq:1d0}
 \begin{split}
 \xi''
  - \frac{\bar{K} \chi}{l^2 (1 + N_0) }   \xi' 
   - \frac{\dot{\lambda} }{2 l^2(1+N_0)} 
-  \frac{ \chi \lambda} {l ^2(1+N_0)^2} 
+ \frac {\bar{K}'}{l^2(1+ N_0)} 
 = 0, 
 \end{split}
 \end{equation}
with functional dependencies $\xi(X)$, $\chi(\xi)$, $\bar{K}(X)$, and $\lambda(\xi)$.
The ODE is linear or nonlinear depending on forms of $\lambda$ and $\chi$; analytical solutions can be derived for special cases.
If $\bar{K} = \text{constant}$, \eqref{eq:1d0} is autonomous. If $\bar{K} = 0$, then
\eqref{eq:1d0} is
\begin{equation}
\label{eq:1dA}
\frac{ \d^2 \xi}{ \d X^2}  = \zeta \frac{ \d \zeta} { \d \xi} = \frac{1}{2 l^2 ( 1 + N_0) } \left[ \frac{\d \lambda }{ \d \xi} + \frac{2 \chi(\xi) \lambda(\xi)}{1 + N_0} \right], 
\end{equation}
where $\zeta = \xi' \Rightarrow  \xi'' = \zeta \d \zeta / \d \xi $. The right equation can be separated and integrated as
\begin{equation}
\label{eq:auto1}
\begin{split}
\frac{1}{2} \zeta^2&  =  \frac{1}{2 l^2 ( 1 + N_0) }\int  \left[ \frac{\d \lambda }{ \d \xi} + \frac{2 \chi(\xi) \lambda(\xi)}{1 + N_0} \right] \d \xi + c_1 \\ & \Rightarrow
\frac{ \d \xi }{\d X} = \pm \frac{1}{ l \sqrt{ 1 + N_0} } \left( \int  \left[ \frac{\d \lambda }{ \d \xi} + \frac{2 \chi(\xi) \lambda(\xi)}{1 + N_0} \right] \d \xi + c_1 \right)^{1/2}.
\end{split}
\end{equation}
This first-order ODE can be separated and solved for $\xi = \arg [X(\xi)]$, where
\begin{equation}
\label{eq:auto2}
\begin{split}
X(\xi)  = \pm  l \sqrt{ 1 + N_0}  \int \frac{ \d \xi}{ \{ {\textstyle{\int}}  \left[ {\d \lambda }/{ \d \xi} + {2 \chi(\xi) \lambda(\xi)}/{(1 + N_0)} \right] \d \xi + c_1 \}^{1/2}} + c_2.
\end{split}
\end{equation}
Integration constants are $c_1$ and $c_2$, determined by boundary conditions.

Now allow arbitrary $\bar{K}(X)$ but restrict $\chi = 0$ (e.g., $G = G_0$). 
Assume $\lambda(\xi)$ is quadratic such that $\dot{\lambda} = 2 \omega_0 + 2 \omega_1 \xi$.
Now \eqref{eq:1d0} is linear:
 \begin{equation}
 \label{eq:1dhet}
 \frac{ \d^2 \xi} { \d X^2}  - \frac{ \omega_1 }{ l^2(1+N_0)} \xi  
=  \frac {1}{l^2(1+ N_0)} \left( { \omega_0 } - \frac{ \d \bar{K}}{ \d X} \right).
 \end{equation}
 This ODE is non-homogeneous but has constant coefficients. Assume
 $\omega_1 >0$ and $N_0 > -1$.  Then 
 \begin{equation}
 \label{eq:gensol}
 \xi(X) = c_1 \text{exp} \left[ ({X}/{l}) \sqrt{ {\omega_1}/{(1 + N_0)} } \right] + 
 c_2 \text{exp} \left[ -({X}/{l}) \sqrt{ {\omega_1}/({1 + N_0)} } \right]
  + \xi_\text{p}(X),
 \end{equation}
 where  $c_1$ and $c_2$ are new constants and $\xi_\text{p}$ is the particular solution from $\omega_0$ and  
 $\bar{K}(X) = l \hat{K}(X)$.

\subsection{Constitutive model}

The framework is applied to a strip of skin loaded in tension along the $X$-direction.
\begin{rmk}
A 1-D theory cannot distinguish between uniaxial strain or uniaxial stress conditions, nor can it account for anisotropy.
Thus, parameters entering the model (e.g., $\mu_0$, $\Upsilon_0$) are particular to those loading conditions and material
orientations from experiments to which they are calibrated (e.g., uniaxial stress along a preferred fiber direction).
\end{rmk}
\noindent The nonlinear elastic potential of \S4.4.2 specializes a 3-D model \cite{gasser2006,holzapfel2009,ni2012,nolan2014} to 1-D.
The internal structure variable $\xi = D/l$ accounts for local rearrangements that lead to softening and degradation
under tensile load \cite{munoz2008,gasser2011,yang2015,tubon2022}: fiber sliding, pull-out, and breakage of collagen fibers, as well as rupture of the elastin fibers and ground matrix.
\begin{rmk}
Specifically, $D$ is a representative microscopic sliding or separation distance among microstructure constituents,
and $l$ is the value of this distance at which the material can no longer support tensile load. In the context of cohesive theories of fracture \cite{clayton2005b,tubon2022,sree2022}, 
$D$ can be interpreted as a crack opening displacement.
\end{rmk}
\begin{rmk}
Some physics represented by the present novel theory, not addressed by nonlinear elastic-continuum damage \cite{balzani2012,tubon2022} or phase-field \cite{clayton2014i,gultekin2019} approaches, are summarized as follows. The Finslerian metrics $G(\xi) = g(\xi)$ account for local rescaling of material and spatial manifolds $\mathcal{M}$ and $\mathfrak{m}$ due to
 microstructure changes (e.g, expansion due to tearing or cavitation).  Nonlinear connection $N_0$ rescales the quadratic contribution
of the gradient of $\xi$ to surface energy by a constant, and linear connection $\hat{K}$ rescales the linear contribution of the gradient of $\xi$ to surface energy by a continuous and differentiable function of $X$, enabling a certain material heterogeneity.  
\end{rmk}
 
 \subsubsection{Metrics}
From \eqref{eq:lengths}, \eqref{eq:lengthsc}, \eqref{eq:lengthC}, \eqref{eq:1D},  \eqref{eq:isometric},  and \eqref{eq:def1D},
the difference in squared lengths of line elements $\d \vec{x}$ and $\d \vec{X}$ is 
\begin{equation}
\label{eq:dX1d}
(|\d \vec{x}|^2 - | \d \vec{X}|^2)(C,\xi) = G(\xi) (C - 1) \d X \, \d X.
\end{equation}
\noindent Herein, the metric is assigned an exponential form frequent in generalized Finsler geometry \cite{watanabe1983,clayton2017g} and Riemannian geometry \cite{yavari2010,ozakin2010}:
\begin{equation}
\label{eq:G1D}
G(\xi) = \text{exp} \left( \frac{2k}{r} \xi^r \right) \quad \Rightarrow \quad \chi(\xi) = \frac{ \dot{G}} { 2 G} = \frac{ \dot{g}}{2 g} =  k \xi^{r-1}.
\end{equation}
For $\xi \in [0,1]$, two constants are $k$, which is positive for expansion, and $r > 0$.
\begin{rmk}
Local regions of $\mathcal{M}$ at $X$ and $\mathfrak{m}$ at $x = \varphi(X)$ are rescaled isometrically by $G(\xi(X))$. Physically, this rescaling arises from changes in structure associated with degradation, to which measure $  \frac{1}{2} \ln {G ( \xi) }$ is interpreted as a contributor to remnant strain.
For Riemannian metrics, $G = \bar{G} = \bar{g} = g = 1$, in which case \eqref{eq:dX1d} is independent of $\xi$ and this remnant strain always vanishes. 
\end{rmk}
\noindent The ratio of constants is determined by the remnant strain contribution at failure: $\hat{\epsilon} =  \frac{k}{r} = \frac{1}{2} \ln {G ( 1) }$. 
Since $\xi \in [0,1]$, smaller $r$ at fixed $\frac{k}{r}$ gives a sharper increase in $ \frac{1}{2} \ln G$ versus $\xi$. Values of $k$ and $r$ are calibrated to data in \S4.5;
choices of $N_0$ and $\bar{K}$ are explored parametrically therein. 
 Nonlinear connection $N_0 = \text{constant}$ and linear connection
$\hat{K}(X) = \bar{K}(X)/l$ affect the contribution of state gradient $\xi'$ to surface energy $\iota$ via \eqref{eq:psi1d1} and \eqref{eq:iota}.
Constraint $N_0 > -1$ is applied to avoid model singularities and encompass trivial choice $N_0 = 0$.
The value of $N_0$ uniformly scales the contribution of $(\xi')^2$ to $\iota$ and $\psi$.
Function $\hat{K}$ scales, in a possibly heterogeneous way, the contribution of $\xi'$ to $\iota$ and $\psi$. Even when $\xi'$ vanishes, $N_0$ and $\bar{K}$ can affect solutions.
 
\subsubsection{Nonlinear elasticity}
Strain energy density $W$ in \eqref{eq:psi1d1} is dictated by the normalized (dimensionless) function $w(C)$:
\begin{equation}
\label{eq:w1d}
w(C) = (\sqrt{C} - 1)^2 + \frac{a_1}{2 b_1} \left[ \text{exp} \{ b_1 (C -1)^2 \}  - 1\right],
\end{equation}
where dimensionless constants are $a_1  \geq 0$ and $b_1 > 0$, and $\mu_0 > 0$ is enforced along with $\Upsilon_0 > 0$ in \eqref{eq:psi1d1}. This adapts prior models for collagenous tissues \cite{gasser2006,holzapfel2009,ni2012,nolan2014} to the 1-D case. The
first term on the right, linear in $C$, accounts for the ground matrix and elastin.
The second (exponential) term accounts for the collagen fibers, which, in the absence of damage processes, stiffen significantly at large $C$. Such stiffening is dominated by the parameter $b_1$, whereas
$a_1$ controls the fiber stiffness at small stretch $\sqrt{C} \approx 1$ \cite{ni2012}. 

The elastic degradation function $y(\xi)$ and independent energy contribution $\lambda(\xi)$
in \eqref{eq:psi1d1} are standard from phase-field theories \cite{clayton2014i,gultekin2019}, where $\vartheta \in [0,\infty)$ is a constant with 
$\vartheta = 2$ typical for brittle fracture and $\vartheta = 0 \mapsto y = 1$ for purely elastic response:
\begin{equation}
\label{eq:ylam1d}
y(\xi) = (1-\xi)^\vartheta, \quad \dot{y}(\xi) = - \vartheta ( 1 - \xi)^{\vartheta - 1}; \qquad \lambda(\xi) = \xi^2, \quad \dot{\lambda}(\xi) = 2 \xi.
\end{equation}
When $\vartheta > 0$, $y(1) = 0$: no strain energy $W$ or tensile load $P$ are supported at $X$ when $D(X) = l$.
Verification of \eqref{eq:ground1D} for prescriptions \eqref{eq:w1d} and \eqref{eq:ylam1d} is straightforward \cite{balzani2006,holzapfel2009}. 
Stress $P$ conjugate to $F = \sqrt{C}$ and force $Q$ conjugate to $D = l \xi$ are, from \eqref{eq:P1d},
\eqref{eq:Q1d}, \eqref{eq:w1d}, and \eqref{eq:ylam1d}: 
\begin{equation}
\label{eq:P1ds}
P(C,\xi) = \mu_0 (1 - \xi)^\vartheta \left[  (\sqrt{C} - 1) + a_1 \sqrt{C}(C-1)  \text{exp} \{ b_1 (C -1)^2 \} \right], 
\end{equation}
\begin{equation}
\label{eq:Q1ds}
Q(C,\xi) =  \frac{2 \Upsilon_0}{l^2} \left[  \xi -  \frac{A_0  \vartheta}{2}(1-\xi)^{\vartheta - 1} \left( (\sqrt{C} - 1)^2 + \frac{a_1}{2 b_1} \left[ \text{exp} \{ b_1 (C -1)^2 \}  - 1\right] \right) \right]. 
\end{equation}
\begin{rmk}
Ideal elasticity (i.e., no structure-mediated metric variation or degradation), is obtained when $k = 0 \Rightarrow G = 1 \Rightarrow \chi = 0$, $\vartheta = 0 \leftrightarrow y = 1 \Rightarrow \dot{y} = 0$, and $\bar{K}' = 0$.
In this case, as $\dot{\lambda}(0) = 0$ by \eqref{eq:ylam1d}, trivial solutions to \eqref{eq:linmom1d2} and \eqref{eq:micmom1d} are
 $P(X) = P_0 = \text{constant}, \, \xi(X) = 0 \, \forall \, X \in \mathcal{M}$.
 \end{rmk}

\subsection{Specific solutions}
Inputs to the model are nine constants $l > 0 $, $k$, $r > 0 $, $N_0 > -1$,
$\mu_0 > 0$, $a_1 \geq 0$, $b_1 > 0$, $\vartheta \geq 0$, $\Upsilon_0 > 0$, and the function $\bar{K}(X)$.
These are evaluated for stretching and tearing of skin \cite{yang2015,tubon2022,sree2022} by applying the constitutive model of \S4.4 to the general solutions derived in \S4.3.

\subsubsection{Homogeneous fields}
Here the skin specimen is assumed to degrade homogeneously in a gauge section of initial length $2 L_0$ (i.e., diffuse damage), an idealization fairly characteristic of certain experiments \cite{fung1993,munoz2008,yang2015,mitsuhashi2018,ani2012}.
Per \S4.3.1, assume deformation control, with $F = F_\text{H} = \sqrt{C_\text{H}} \geq 1$ increased incrementally from unity. The analytical solution for $\xi = \xi_\text{H}$ is then the implicit solution of \eqref{eq:1dH} upon substitution of \eqref{eq:G1D}, \eqref{eq:w1d} and \eqref{eq:ylam1d}, here for
$\vartheta > 0$:
\begin{equation}
 \label{eq:homsol}
 \begin{split}
   \xi_\text{H}  +  [{k}/({1+N_0}) ]   & \xi_\text{H}^{1+ r} 
   =  \textstyle{\frac{1}{2}}  {A_0 \vartheta} (1 - \xi_\text{H})^{\vartheta - 1}  
\{  (\sqrt{C_\text{H}} - 1)^2 + [{a_1}/({2 b_1) ] }
\\ & \times \left[ \text{exp} \{ b_1 (C_\text{H} -1)^2 \}  - 1\right] \}
    \{ 1 - { 2 k \xi^{r-1}_\text{H} 
  (1 - \xi_\text{H})
 }/{[(1 + N_0)}\vartheta] \} + {K}'_0.
 \end{split}
 \end{equation}
This dimensionless solution does not depend on $\mu_0$, $\Upsilon_0$, or $l$ individually, but only on dimensionless ratio $A_0 = \frac{\mu_0 l}{2 \Upsilon_0}$. However, stress $P_\text{H} = P (C_\text{H}, \xi_\text{H})$ is found from \eqref{eq:P1ds}, which does depend on $\mu_0$.

\begin{figure}
	\centering
	\subfigure[axial stress]{\includegraphics[width=0.44\textwidth]{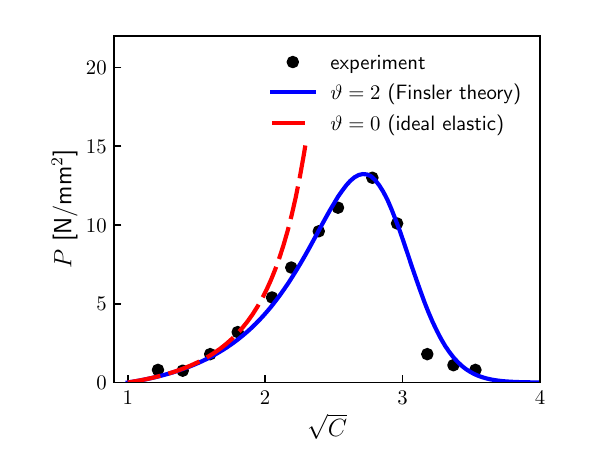}\label{fig2a}} \quad
	\subfigure[axial stress]{\includegraphics[width=0.44\textwidth]{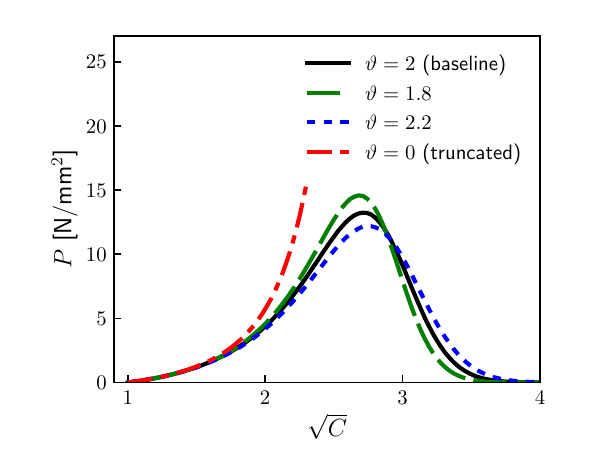}\label{fig2b}} \\
	\vspace{-3mm}
	\subfigure[axial stress]{\includegraphics[width=0.44\textwidth]{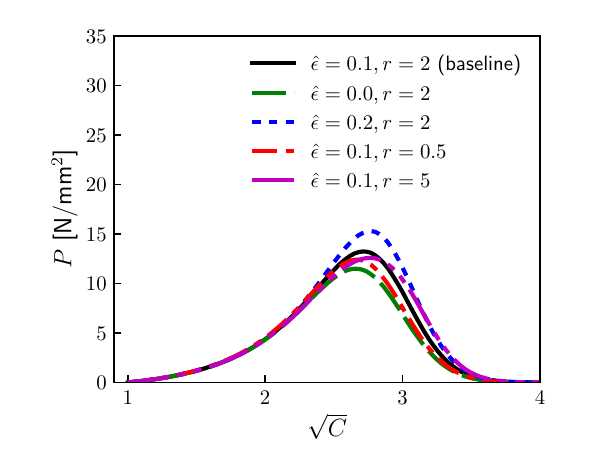}\label{fig2c}} \quad
	\subfigure[axial stress]{\includegraphics[width=0.44\textwidth]{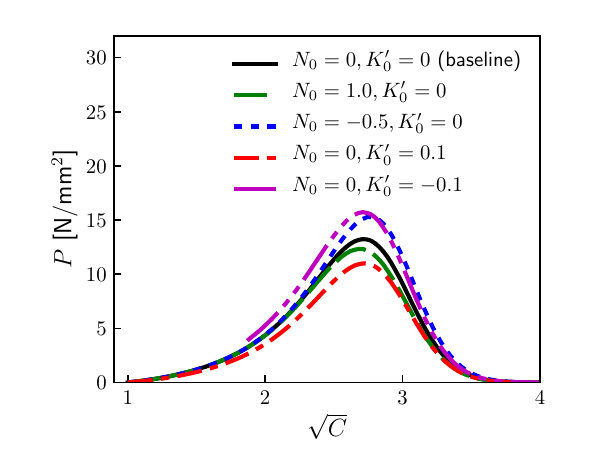}\label{fig2d}} \\
	\vspace{-3mm}
	\subfigure[structure variable]{\includegraphics[width=0.44\textwidth]{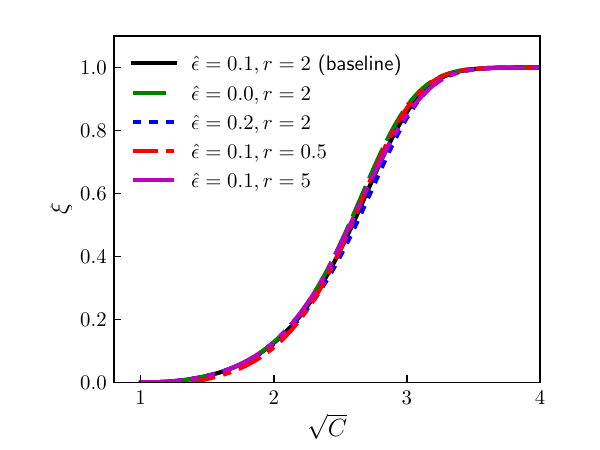}\label{fig2e}} \quad
	\subfigure[structure variable]{\includegraphics[width=0.44\textwidth]{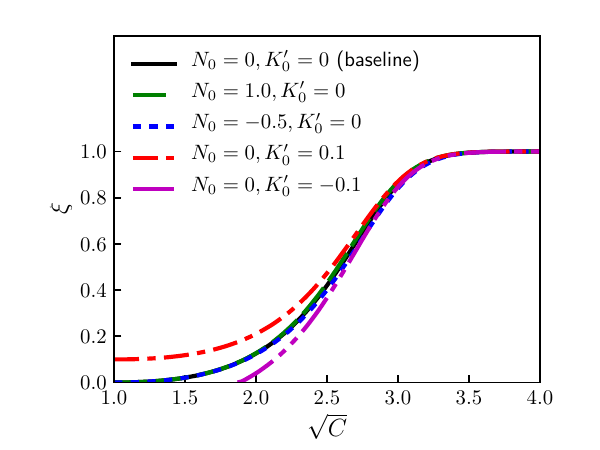}\label{fig2f}} 
	\vspace{-2mm}
\caption{Extension and tearing of skin for imposed axial stretch ratio $\sqrt{C}$, 1-D model: 
(a) stress $P$ comparison with data \cite{yang2015} (see text \S4.5.1 for definition of experimental stretch ratio) of Finsler model (baseline) and ideal nonlinear elasticity (null structure change)
(b) effect on stress $P$ of energy degradation exponent $\vartheta$ with $\hat{\epsilon} = 0.1$, $r=2$, $N_0 = 0$, and $K'_0 = 0$
(c) effect on stress $P$ of Finsler metric scaling $\hat{\epsilon} = \frac{k}{r}$ and $r$
with $\vartheta = 2$, $N_0 = 0$, and $K'_0 = 0$
(d) effect on stress $P$ of nonlinear connection $N_0$ and linear connection $K'_0$ with $\vartheta = 2$, $\hat{\epsilon} = 0.1$, and $r=2$
(e) effect on internal structure $\xi = D/l$ of Finsler metric scaling $\hat{\epsilon} = \frac{k}{r}$ and $r$
with $\vartheta = 2$, $N_0 = 0$, and $K'_0 = 0$
(f) effect on internal structure $\xi = D/l$ of nonlinear connection $N_0$ and linear connection $K'_0$ 
with $\vartheta = 2$, $\hat{\epsilon} = 0.1$, and $r=2$
\label{fig2}}
\end{figure}

Stress $P$ is shown in Fig.~\ref{fig2a}, first assuming $N_0 = 0$ and $K'_0 = 0$ for simplicity. The Finsler model, with $A_0 = 8.5 \times 10^{-2}$ corresponding to baseline parameters given in Table~\ref{table1}, successfully matches experimental\footnote{Stretch corresponding to Fig.~5(e) in experimental work \cite{yang2015} is defined as engineering strain plus 1.2 in Fig.~\ref{fig2a} of \S4.5.1 and Fig.~\ref{fig4a} of \S5.5.1 to account for pre-stress ($\approx 0.7$ MPa) and pre-strain ($\approx 0.2$), so $\sqrt{C} = 1$ consistently for stress-free reference states among models and experiments. Stress-free states at null strain are consistent with data in Fig.~3(a) of that work \cite{yang2015}. Alternatively, $2 \frac{\sigma_0}{\mu_0} (\sqrt{C} - 1)$, with $\sigma_0= \text{constant}$, can be added to $w$ of \eqref{eq:w1d} giving a pre-stress of $P_{(C=1,\xi=0)} = \sigma_0$ to fit data with pre-stress; this, however, would require relaxation of the second of \eqref{eq:ground1D}.} data \cite{yang2015}. 
The value of $\mu_0$ is comparable to the low-stretch tensile modulus in some experiments \cite{ni2012,jood2018}, acknowledging significant variability in the literature. 
\begin{rmk}
The ideal elastic solution ($\xi = 0$) is shown for comparison.
Excluding structure evolution corresponding to collagen fiber rearrangements, sliding, and breakage,
the model is too stiff relative to this data for which such microscopic mechanisms have been observed \cite{yang2015}. The ideal elastic model is unable to replicate the linearizing, softening, and failure mechanisms
with increasing stretch $\sqrt{C}$ reported in experiments on skin and other soft tissues \cite{fung1993,ani2012,yang2015,takano2017,mitsuhashi2018}.
\end{rmk}
In Fig.~\ref{fig2b}, effects of $\vartheta$ on $P$ are revealed for $\hat{\epsilon} = 0.1$, $r=2$, $N_0 = 0$, and $K'_0 = 0$, noting 
 $\vartheta = 0$ produces the ideal nonlinear elastic solution $\xi_\text{H} = 0$ in \eqref{eq:homsol}. 
 Peak stress increases with decreasing $\vartheta$; the usual choice from phase-field theory $\vartheta = 2$
 provides the close agreement with data in Fig.~\ref{fig2a}.
 In Fig.~\ref{fig2c}, effects of Finsler metric scaling factors
  $\hat{\epsilon} = \frac{k}{r} $ and $r$ on stress $P$ are demonstrated,
  where at fixed $r$, peak stress increases (decreases) with increasing (decreasing) $\hat{\epsilon}$ and $k$.
Baseline choices  $\hat{\epsilon} = 0.1$ and $r=2$ furnish agreement with experiment in
 Fig.~\ref{fig2a}. A remnant strain of 0.1 is the same order of magnitude observed in cyclic loading experiments \cite{rubin2002,munoz2008}.
 Complementary effects on evolution of structure versus stretch are shown in Fig.~\ref{fig2e}: modest changes in $\xi$ produce significant changes in $P$.
In Fig.~\ref{fig2d}, effects of connection coefficients $N_0$ and $K'_0$ are revealed, holding material parameters at their baseline values of Table~\ref{table1}.
For this homogeneous problem, maximum $P$ decreases with increasing $N_0$ and $K'_0$.
Corresponding evolution of $\xi$ is shown in Fig.~\ref{fig2f}.
When $K'_0 < 0$, a viable solution $\xi_\text{H} \in [0,1]$ exists only for $\sqrt{C} > 1$.

The total energy per unit cross-sectional area of the specimen is $\bar{\Psi}$, found upon integration of $\psi(C_\text{H},\xi_\text{H})$ in \eqref{eq:psi1d1} over $\mathcal{M}$ with local volume element $\d V = \sqrt{G(\xi_\text{H}) } \, \d X$: 
\begin{equation}
\label{eq:totensol}
\frac{\bar{\Psi}}{L_0}  = {\mu_0}  \left[
 (1 - \xi_\text{H})^\vartheta 
 \{  (\sqrt{C_\text{H}} - 1)^2 + \frac{a_1}{2 b_1}   \left[ \text{exp} \{ b_1 (C_\text{H} -1)^2 \}  - 1\right] \}
  + \frac{ \xi_\text{H}^2}{A_0}
\right] \text{exp} \left( \frac{k}{r} \xi_\text{H}^r \right) .
\end{equation}

\begin{table}
\footnotesize
\caption{Baseline model parameters for rabbit skin tissue: 1-D and 2-D theories}
\label{table1}       
\centering
\begin{tabular}{lllrr}
\hline\noalign{\smallskip}
Parameter & Units & Definition & Value (1-D) & Value (2-D) \\
\hline\noalign{\smallskip}
$l $ & mm & length scale & 0.04 & 0.04 \\
$k$ & $\cdots$ & metric scaling factor & 0.2 & 0.2 \\
$m$ & $\cdots$ & metric scaling factor & $\cdots$ & 0.3 \\
$r$ & $\cdots$ & metric scaling exponent & 2 & 2 \\ 
 $\mu_0$ & N/mm$^2$ & shear modulus (axial 1-D)& 0.2 & 0.2 \\
 $\kappa_0$ & N/mm$^2$ & bulk modulus ($\kappa_0 = k_0 \mu_0$) & $\cdots$ & 1.2 \\
 $ a_1 $ & $\cdots$ & nonlinear elastic constant & 2.8  & 2.8 \\ 
 $ a_2 $ & $\cdots$ & nonlinear elastic constant & $\cdots$ & 6 \\
 $ b_1 $ & $\cdots$ & nonlinear elastic constant & 0.055  & 0.055 \\
 $ b_2 $ & $\cdots$ & nonlinear elastic constant & $\cdots$  & 0.17 \\
 $ \vartheta $ & $\cdots$ & degradation exponent & 2  & 2 \\
 $ \varsigma $ & $\cdots$ & degradation exponent & $\cdots$  & 2  \\
$ \Upsilon_0$ & mJ/mm$^2$ & isotropic surface energy & 0.47 & 0.47 \\
$ \gamma_\xi $ & $\cdots$ & anisotropic energy factor & $\cdots$  & 1 \\
$ \gamma_\eta $ & $\cdots$ & anisotropic energy factor & $\cdots$  & 0.84 \\
\noalign{\smallskip}\hline\noalign{\smallskip}
 \\
\end{tabular}
\end{table}

\subsubsection{Stress-free states}
The stress-free solutions of \S4.3.2 are applied to evaluate the remaining unknown parameters $l$ and $\Upsilon_0$, given $\mu_0$ and $A_0$ from \S4.5.2. 
Assume the specimen tears completely at its midpoint at $X = 0$, such that $\xi(0) = 1$. No load is supported anywhere, and only 
rigid body motion is possible at other locations $X$ where $\xi(X) > 0$.
Assume the specimen is clamped at its ends where it is gripped, such that $\xi(-L_0) = \xi(L_0) =0$.
Symmetry conditions $\xi(-X) = \xi(X)$ are imposed, with $\xi'(0)$ discontinuous, such
that a solution need be calculated only for the half-space $X \in [0,L_0]$.

\begin{figure}
	\centering
	\subfigure[structure profiles]{\includegraphics[width=0.44\textwidth]{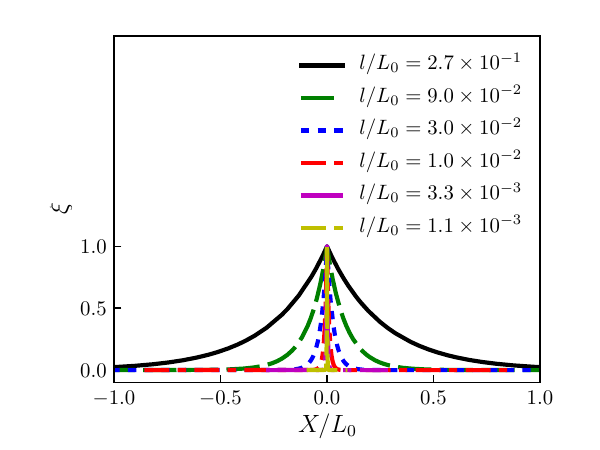}\label{fig3a}} \quad
	\subfigure[surface energy]{\includegraphics[width=0.44\textwidth]{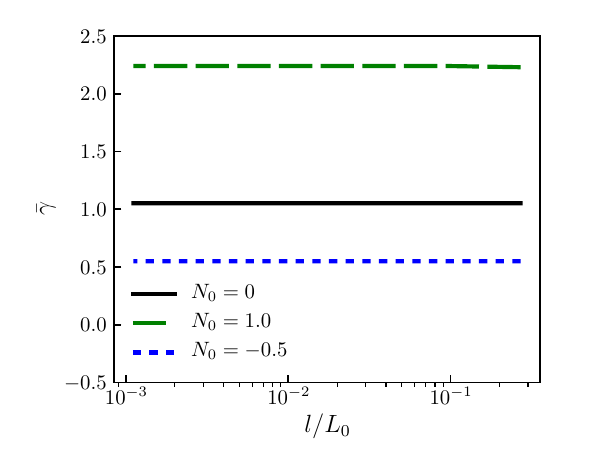}\label{fig3b}} \\
		\vspace{-3mm}
	\subfigure[energy ratio]{\includegraphics[width=0.44\textwidth]{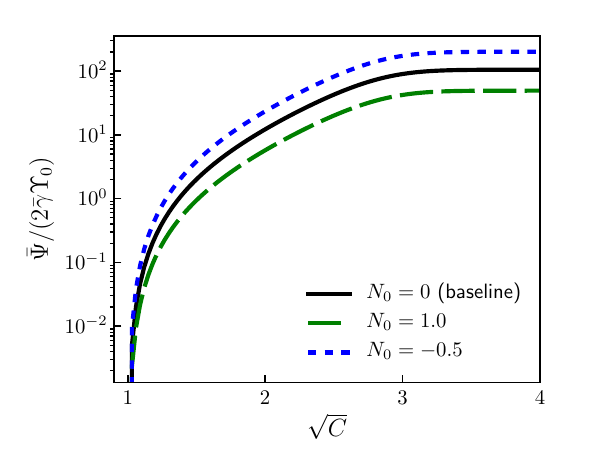}\label{fig3c}} \quad
	\subfigure[structure profiles]{\includegraphics[width=0.44\textwidth]{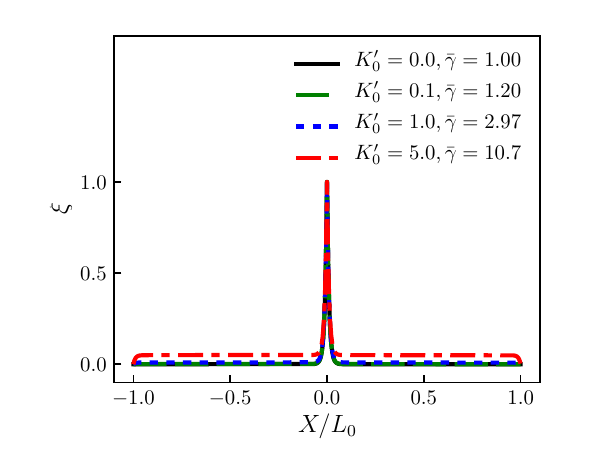}\label{fig3d}}
	\vspace{-2mm}
\caption{Extension and tearing of skin, 1-D model: 
(a) stress-free solution for internal state profile (baseline parameters, $N_0 = 0$)
(b) normalized surface energy for rupture versus regularization length
(c) ratio of homogeneous energy to energy for stress-free localized rupture
(d) stress-free solution, $\hat{\epsilon} = 0$, $l/L_0 = 10^{-2}$, heterogeneous connection $\bar{K}(X)$  \label{fig3}}
\end{figure}

First take $\bar{K} = l \hat{K} = 0$ so that \eqref{eq:auto2} holds.
Assume $c_1 = 0$ corresponding to $\xi ' = 0$ where $\xi = 0$ since the anti-derivative in \eqref{eq:auto1} vanishes at $\xi = 0$ when
$\lambda = \xi^2, \chi = k \xi^{r-1}, r > 0$. It is verified a posteriori \cite{clayton2017g,clayton2017z,clayton2018c} that this closely approximates
true boundary conditions $\xi(\pm L_0) = 0$ as well as $\xi'(\pm L_0) = 0$ for $L_0 \gg l$.
Then the physically valid (negative) root for the half-domain giving $X \geq 0 $ in \eqref{eq:auto2} becomes, with \eqref{eq:G1D} and \eqref{eq:ylam1d}, 
\begin{equation}
\label{eq:autosol}
\begin{split}
\frac{X(\xi)}{L_0} = - \frac{l}{L_0} \sqrt{ 1 + N_0}
    \int_{z = 1}^{z = \xi} \frac{ \d z }
 { z \sqrt{ 1 +  2 k z^r / [(1+N_0)(2+r)]} }.
\end{split}
\end{equation}
The lower limit follows from $X(1) = 0$, obviating $c_2$ in \eqref{eq:auto2}.
Analytical solution $ \xi = \arg{X}(\xi)$ is exact, but it is most easily evaluated by quadrature when $k$ is nonzero, decrementing $z$ from 1 to 0 in small negative steps. 
The profile of $\xi(X)$ depends on $X / L_0$ and $l / L_0$, but not $l$ or $L_0$ individually.
\begin{rmk} 
This new 1-D solution, \eqref{eq:autosol}, agrees with more specific solutions derived in past work: $N_0 = 0$ and $r = 1$ \cite{clayton2017g,clayton2016} with slight correction \cite{clayton2018c}
and $N_0 = 0$ and $r=2$ \cite{clayton2018c}. 
\end{rmk}
Normalized surface energy per two-sided cross-sectional area, $\bar{\gamma}$, is obtained by integration 
of $\psi = \Lambda$ in \eqref{eq:psi1d1} over $\mathcal{M}$: 
\begin{equation}
\label{eq:surfsol}
\bar{\gamma} = 
 \frac{1}{ 2  \Upsilon_0}  \int_{-L_0}^{L_0} \psi  \sqrt{G} \, \d X 
= 
\frac{1}{ 2 l }  \int_{-L_0}^{L_0}
\{  \xi^2 + (1+N_0) l \xi' [2 \hat{K} + (1+N_0) l \xi' ] \} 
\, \text{exp} [ ({k}/{r}) \xi^r]  \, \d X.
\end{equation}
This energy likewise depends on $l/L_0$ but not $l$ or $L_0$ individually. Baseline values of $k$ and $r$ are now taken from Table~\ref{table1}.  The solution \eqref{eq:autosol} is shown for $N_0 = 0$ and different $l/L_0$ in Fig.~\ref{fig3a}. The smaller (larger) the regularization length ratio $l/L_0$, the
sharper (more diffuse) the zone centered at the midpoint of the domain over which prominent structure changes occur. 

Normalized energy density \eqref{eq:surfsol} is shown in Fig.~\ref{fig3b} versus $l/L_0$ for several $N_0$. Increasing $N_0$ increases this energy, as might be anticipated from \eqref{eq:psi1d1} with \eqref{eq:iota} when $\hat{K} = 0$. A stress-free ruptured state is energetically favorable to a stressed homogeneous state (\S4.5.1) from applied deformation $C_\text{H}$ when $\bar{\Psi}  > 2  \bar{\gamma} \Upsilon_0$, with $\bar{\Psi}$ given by \eqref{eq:totensol}. Ratio $\bar{\Psi} / ( 2 \bar{\gamma} \Upsilon_0)$ is shown in Fig.~\ref{fig3c} versus $\sqrt{C} = \sqrt{C_\text{H}}$ with $l/L_0 = 10^{-2}$ and several $N_0$, recalling $K_0' = 0$. 
Increasing $N_0$ increases $\bar{\gamma}$, reducing $\bar{\Psi} / ( 2 \bar{\gamma} \Upsilon_0)$.
For cases in Fig.~\ref{fig3a}, Fig.~\ref{fig3b}, and Fig.~\ref{fig3c}, $\xi(\pm L_0) < 10^{-8}$ and  $| l \xi' (\pm L_0) | < 10^{-8}$ are observed for $l/L_0 \leq 0.03$, verifying $c_1 = 0$ in \eqref{eq:auto2} and \eqref{eq:autosol} under this length constraint.

The remaining parameters $l$ and $\Upsilon_0$ are now quantified. To match the measured energy release rate $\mathsf{J}_\text{C}$ (i.e., toughness) of skin,  
$2 \bar{\gamma} \Upsilon_0  \approx \mathsf{J}_\text{C}$.   
Take $L_0 = 4$ mm, the span of specimens  \cite{yang2015} whose data are represented in Fig.~\ref{fig2a}. Then $l/L_0 = 10^{-2} \Rightarrow l = 40 \, \mu$m is more than sufficiently small to adhere to the aforementioned boundary constraints (i.e., $c_1 = 0$) while providing a damage profile of intermediate diffusivity in Fig.~\ref{fig3a}. This value of $l$ then gives $\Upsilon_0 =  \frac{\mu_0 l}{2 A_0} = 0.47$ kJ/m$^2$ (Table~\ref{table1}).
\begin{rmk}
Along with the choice $N_0 = 0$, the Finsler model with full set of baseline parameters in Table~\ref{table1} produces $\bar{\gamma} \approx 1$ in Fig.~\ref{fig3b} 
and $2 \bar{\gamma} \Upsilon_0 = 1.0$ kJ/m$^2$, in concurrence with experimental data: $0.5 \lesssim \mathsf{J}_\text{C} \lesssim 2.5$ kJ/m$^2$ \cite{per1997,tubon2022,sree2022}.
Value $l = 40\, \mu$m is between $4 \times$ and $40 \times$ the collagen fiber diameter \cite{fung1993,cowin2007,yang2015}.
Though not shown in Fig.~\ref{fig3b}, increasing $\hat{\epsilon} = \frac{k}{r}$ from 0.1 to 0.2 at $\vartheta = r  = 2$ with $N'_0 = K'_0 = 0$ and $l/L_0 = 10^{-2}$ increases effective toughness to $2 \bar{\gamma} \Upsilon_0 =  1.02$ kJ/m$^2$.
Under the same conditions, reducing $\hat{\epsilon}$ to 0 diminishes the predicted toughness to $2 \bar{\gamma} \Upsilon_0 = 0.94$ kJ/m$^2$.
\end{rmk}
Finally, take $k = 0$ but permit nonzero $\bar{K}(X) = l \hat{K}(X) $, such that \eqref{eq:gensol} applies.  
As an example, let $\bar{K} = - K'_0  l \cdot (1-X/L_0 )$ for $X \in [0,L_0]$ and
 $\bar{K} =  K'_0 l \cdot (1 + X/L_0)$ for $X \in [-L_0,0)$.
 Boundary conditions $\xi(0) =1$ and $\xi(\pm L_0) = 0$ still apply, as does symmetry
 relation $\xi(X) = \xi(-X)$.
From \eqref{eq:ylam1d}, $\omega_1 = 1$ and $\omega_0 = 0$. For the  whole domain $X \in [-L_0,L_0]$, 
$\bar{K}' = K_0'l/L_0 = \text{constant}$, and simply $\xi_\text{p} = K_0'l/L_0$. Then \eqref{eq:gensol} gives
\begin{equation}
 \label{eq:psol}
 \begin{split}
 & \xi(X)  = c_1 \text{exp} [ {X}/ \{ l  \sqrt{ 1 + N_0 } \} ] + 
 c_2 \text{exp} [ -{X}/ \{ l  \sqrt{ 1 + N_0} \} ]
   + K_0'l/L_0, \\
 &  c_1  = 1 - c_2 - K_0'l/L_0  = \frac{ - K_0'l/L_0 - [ 1 - K_0'l/L_0] \, \text{exp}[ - {L_0}/ \{ l  \sqrt{ 1 + N_0 } \} ] } 
   { \text{exp}[  {L_0}/ \{ l  \sqrt{ 1 + N_0 } \} ]  - \text{exp}[ - {L_0}/ \{ l  \sqrt{ 1 + N_0 } \} ] }. 
  \end{split}
 \end{equation}
Profiles of $\xi(X)$ are shown in Fig.~\ref{fig3d} for $K'_0 \geq 0 $ with baseline $l/L_0 = 10^{-2}$.
 Normalized surface energy $\bar{\gamma}$ from \eqref{eq:surfsol} is reported in Fig.~\ref{fig3d} for each case, recalling $\hat{\epsilon}=k = 0$ produces Riemannian (Euclidean) metric $G=1$.  Setting $K'_0 > 0$ increases $\bar{\gamma}$ for this problem. Setting $K'_0 < 0$ reduces $\bar{\gamma}$ and produces a physically invalid solution (not shown in Fig.~\ref{fig3d}) in \eqref{eq:psol}: $\xi < 0$ on part of $\mathcal{M}$.

\section{Two-dimensional base manifold}

The framework of \S2 and \S3 is applied for $n = 2$: a 2-D base manifold $\mathcal{M}$.
In \S5.1, geometry and kinematics are presented.
Governing equations are derived in \S5.2.
Solutions are considered for general problem classes in \S5.3.
Constitutive functions for an orthotropic 2-D patch of skin under planar deformations are assigned in \S5.4. Solutions for stretching and tearing follow in \S5.5.

\subsection{Geometry and kinematics}
Reference coordinates are Cartesian (orthogonal): $\{ X^1, X^2 \}$. Considered is a reference domain $\{\mathcal{M}: X^1 \in [-L_0,L_0],
X^2 \in [-W_0, W_0] \}$, where the total area relative to a Euclidean metric
is $4 L_0 W_0$, and boundary $\partial \! \mathcal{M}$ is the edges $(X ^1, X^2) = (\pm L_0, \pm W_0)$.
The referential internal state vector has coordinates $\{D^1,D^2\}$, both with physical units of length. Spatial coordinates are Cartesian $\{x^1,x^2\}$ and $\{d^1, d^2\}$.
A normalization constant (i.e., regularization length) is $l$, with physically meaningful domain
assumed as $D^A \in [0,l]$ ($A=1,2$). With notation $f(X,D) = f(X^A,D^B)$, dimensionless order parameters are,
with \eqref{eq:funcC} and \eqref{eq:thetasimp} invoked,
\begin{equation}
\label{eq:OP2D}
\xi(X)  = \frac{D^1(X)}{l} = \frac{d^1(\varphi(X))}{l}, \qquad
\eta(X)  = \frac{D^2(X)}{l} = \frac{d^2(\varphi(X))}{l},
\qquad l > 0.
\end{equation}
Physically meaningful domains are $\xi \in [0,1]$ and $\eta \in [0,1]$. 
For 2-D manifolds with Cartesian base coordinates $\{X^1, X^2\}$ and $\{x^1,x^2\}$, the following metrics apply from \eqref{eq:mdec} and \eqref{eq:mdecc}:
\begin{equation}
\label{eq:2D}
\bar{G}_{AB} = \delta_{AB}, \quad
\bar{g}_{ab} = \delta_{ab}; \quad
G_{AB}(X,D) = \hat{G}_{AB}(D),
\quad
g_{ab}(x,d) = \hat{g}_{ab}(d).
\end{equation}
Herein, the following constraint is imposed:
\begin{equation}
\label{eq:isometric2D}
\hat{g}_{ab}( d (\varphi(X))) = \delta^A_a \delta^B_b \hat{G}_{AB}( D(X)) \leftrightarrow 
g_{ab}(\xi,\eta) =  \delta^A_a \delta^B_b G_{AB} (\xi,\eta),
\end{equation}
making $\mathfrak{m}$ and $\mathcal{M}$ isometric 
when $\phi^a (X) = \delta^a_A X^A  + c^a_0 \Leftrightarrow F^a_A = \delta^a_A$ regardless of $\{\xi, \eta\}$ at $x = \varphi(X)$.
\begin{rmk} 
Equation \eqref{eq:isometric2D} may be removed in other settings to directly model residual stress (e.g., Appendix B), but all residual stresses are not necessarily eliminated with \eqref{eq:isometric2D} in place.
\end{rmk}
Though other non-trivial forms are admissible (e.g., \S4.1), assume nonlinear $N^A_B$ and linear $K^A_{BC}$ connections vanish:
\begin{equation}
\label{eq:N2D}
N^A_B =0 \, \Rightarrow \, N^a_b = \delta^a_A N^A_B (F^{-1})^B_b = 0 , \qquad K^A_{BC} = 0. 
\end{equation}
The $K^a_{bc}$ do not affect the governing equations to be solved later, so they are unspecified.  Applying
\eqref{eq:connrec} and \eqref{eq:OP2D}--\eqref{eq:N2D},
\begin{equation}
\label{eq:conn2D}
\delta_A G_{BC} = \partial_A G_{BC} - N^D_A \bar{\partial}_D G_{BC} = 0
\, \Rightarrow \, \Gamma^A_{BC} = 0,
\qquad \delta_a g_{bc} = 0 \, \Rightarrow \, \Gamma^a_{bc} = 0,
\end{equation}
\begin{equation}
\label{eq:chi2D}
\chi_A (\xi,\eta) = l C^B_{AB} (\xi,\eta) = l \bar{\partial}_A \{ \ln \sqrt{ G(\xi,\eta)} \};
\quad l \bar{\partial}_1(\cdot) = \partial (\cdot) / \partial \xi,
\quad l \bar{\partial}_2(\cdot) = \partial (\cdot) / \partial \eta.
\quad
\end{equation}
The deformation gradient, deformation tensor, Jacobian determinant, and director gradient are
\begin{equation}
\label{eq:def2D}
F^a_A  =  \frac{\partial \varphi^a}{\partial X^A}, \quad
C^A_B = G^{AC} g_{bc} F^b_B F^c_C = 
G^{AC}  F^c_C  \delta_c^F  G_{FE} \delta_b^E F^b_B
, \quad J = \det (F^a_A) = \sqrt{ \det (C^A_B)},
\end{equation}
\begin{equation}
\label{eq:dgrad2D}
D^A_{|B} = \delta_B D^A + K^A_{BC} D^C = \partial_B D^A;
\qquad D^1_{|A} = l \partial_A \xi, \quad
D^2_{|A} = l \partial_A \eta.
\end{equation}
Unless $F^a_A$ and $G_{AB}$ are diagonal, $\vec{C}$ and $\bar{\vec{C}}$ can differ.
From \eqref{eq:Cbar} and \eqref{eq:Jbar},
\begin{equation}
\label{eq:Cbar2D}
\bar{C}^A_B = \delta^{AC} \bar{C}_{CB} = \delta^{AC} \delta_{bc} F^b_B F^c_C,
 \qquad \bar{J} =  \sqrt{ \det (\bar{C}^A_B)} = J.
\end{equation}

\subsection{Governing equations}
A generic energy density is chosen and equilibrium equations are derived for the 2-D case of \S5.1.

\subsubsection{Energy density}
For the present case, dependencies on $D^A$ and $D^A_{|B}$ are suitably represented by 
($\xi,\eta$) and $(\partial_A \xi, \partial_A \eta)$ of \eqref{eq:OP2D} and \eqref{eq:dgrad2D}. 
The functional form of \eqref{eq:psi3} is invoked without explicit $X$ dependency, whereby
\begin{equation}
\label{eq:psi2D}
\psi = \bar{\psi}(\bar{C}_{AB},\xi,\eta, \partial_A \xi, \partial_A \eta).
\end{equation}
Henceforth in \S5, the over-bar is dropped from $\psi$ to lighten the notation. Denote by $\mu_0$ a constant, later associated to a shear modulus, with units of energy density.
\begin{rmk}
For comparison with experiments in ambient 3-space, $\mu_0$ has units of energy per unit 3-D volume, so $\Psi = \int_\mathcal{M} \psi \, d \Omega$ is energy per unit thickness normal to the $X^1$ and $X^2$.
\end{rmk}
Denote by $\Upsilon_0$ a constant related to surface energy with units of energy per unit (e.g., 2-D fixed cross-sectional) area, and by $\gamma_\xi$ and $\gamma_\eta$ two dimensionless constants.
Let $W$ be strain energy density and $\Lambda$ energy density associated with microstructure.
Denote by $w$ a dimensionless strain energy function (embedding possible degradation), $\lambda$ and $\nu$ dimensionless phase energy functions, $\iota$ a dimensionless gradient energy function assigned a sum of quadratic forms, and $\nabla_0 (\cdot) = \frac{\partial}{\partial \vec{X}} (\cdot)$ the partial material gradient. Free energy \eqref{eq:psi2D} is prescribed in intermediate functional form as
\begin{equation}
\label{eq:psi2d1}
\begin{split}
\psi(\bar{\vec{C}},\xi, \eta, \nabla_0 \xi, \nabla_0 \eta)
& = W(\bar{\vec{C}},\xi,\eta) + \Lambda(\xi,\eta, \nabla_0 \xi, \nabla_0 \eta) 
\\ & = {\frac{\mu_0}{2}} w(\bar{\vec{C}},\xi,\eta)+ \frac{\Upsilon_0}{l} [ \gamma_\xi \lambda(\xi) + 
\gamma_\eta \nu(\eta) + \iota(\nabla_0 \xi, \nabla_0 \eta)],
\end{split}
\end{equation}
\begin{equation}
\label{eq:iota2D}
\iota = \gamma_\xi  |l \nabla_0 \xi|^2 + \gamma_\eta l^2 | l \nabla_0 \eta|^2 = 
l^2  \delta^{AB} ( \gamma_\xi  \partial_A \xi \partial_B \xi +  \gamma_\eta   \partial_A \eta \partial_B \eta). 
\end{equation}
Note $\iota(0,0) = 0$. Therefore, for null ground-state energy density $\psi$ and stress $P^A_a$,
\begin{equation}
\label{eq:ground2D}
w(\delta_{AB},\xi,\eta) = 0, \qquad \frac{\partial w}{ \partial {\bar{C}_{AB}}}(\delta_{AB},\xi, \eta)= 0; 
\qquad \lambda(0) = \nu(0) = 0.
\end{equation}
Convexity and material symmetry are addressed in \S5.4.2. Thermodynamic forces of \eqref{eq:deltapsi} are, applying \eqref{eq:Pk1}, 
\begin{equation}
\label{eq:P2d}
 P^A_a = \frac{\partial \psi}{\partial F^a_A} =  2 \delta_{ab}  F^b_B  \frac{\partial \psi}{\partial \bar{C}_{AB}}  
 = \mu_0 \delta_{ab}  F^b_B \frac{\partial w}{\partial \bar{C}_{AB}} ,
\end{equation}
\begin{equation}
\label{eq:Q2d}
Q_1 =  \frac{1}{l} \frac{\partial \psi}{\partial \xi} 
= \frac{\Upsilon_0}{l^2} \left( A_0  \frac{\partial w}{\partial \xi}
 + \gamma_\xi \frac{\d \lambda}{\d \xi} \right ), 
 \quad
 Q_2 =  \frac{1}{l} \frac{\partial \psi}{\partial \eta} 
= \frac{\Upsilon_0}{l^2} \left( A_0  \frac{\partial w}{\partial \eta}
 + \gamma_\eta \frac{\d \nu}{\d \eta} \right );
 \quad A_0 = \frac{\mu_0 l}{2 \Upsilon_0},
 \end{equation}
\begin{equation}
\label{eq:Z2d}
Z^A_1 = \frac{\partial \psi}{\partial D^1_{|A}} =  
\frac{\Upsilon_0}{l^2}  \frac{\partial \iota}{\partial (\partial_A \xi) }
= 2 {\Upsilon_0} \gamma_\xi  \delta^{AB} \partial_ B \xi, 
\quad
Z^A_2 = \frac{\partial \psi}{\partial D^2_{|A}} =  
\frac{\Upsilon_0}{l^2}  \frac{\partial \iota}{\partial (\partial_A \eta) }
= 2 {\Upsilon_0} \gamma_\eta \delta^{AB} \partial_B \eta.
\end{equation}
The source term in \eqref{eq:variation1} manifests from changes in energy proportional to changes of the local referential volume form (e.g.,  local volume changes from damage, treated analogously to an energy source from tissue growth (Appendix B)):
\begin{equation}
\label{eq:R2d}
R_A =  \beta \psi \bar{\partial}_A ( \ln \sqrt{G}) = \frac{\beta}{l} \psi \chi_A, \qquad (\beta = \text{constant}; \, A =1,2). 
\end{equation}

\subsubsection{Linear momentum}
Linear momentum balance \eqref{eq:linmom} or \eqref{eq:altmac} is, invoking relations in \S5.1 and \S5.2.1,
\begin{equation}
\label{eq:linmom2d}
\begin{split}
\mu_0 \delta_{ab} & \left[
\frac{ \partial^2 \varphi^b}{\partial X^A \partial X^B} \frac{\partial w}{\partial \bar{C}_{AB}} +
\frac{ \partial \varphi^b}{\partial X^B } \left(  \frac{\partial^2 w}{\partial \bar{C}_{AB} \partial X^A } + 
  \frac{\partial^2 w}{ \partial \bar{C}_{AB} \partial \xi}   \frac{\partial \xi }{\partial X^A} + 
    \frac{\partial^2 w}{ \partial \bar{C}_{AB} \partial \eta}   \frac{\partial \eta }{\partial X^A}\right)
\right]
\\ & = 
- \mu_0 \delta_{ab} \frac{ \partial \varphi^b}{\partial X^B } \frac{\partial w}{\partial \bar{C}_{AB}}
\left[ \frac{ \partial} {\partial \xi} \left( \ln \sqrt{G} \right)  \frac{\partial \xi }{\partial X^A}
+  \frac{ \partial} {\partial \eta} \left( \ln \sqrt{G} \right) \frac{\partial \eta }{\partial X^A}
 \right].
\end{split}
\end{equation}
\begin{rmk}
For nonzero $\mu_0$, \eqref{eq:linmom2d} is two coupled nonlinear PDEs ($a=1,2$) in four field variables:
$\varphi^1(X)$, $\varphi^2(X)$, $\xi(X)$, and $\eta(X)$.
\end{rmk}

\subsubsection{Micro-momentum}
State-space equilibrium \eqref{eq:dirmom} or \eqref{eq:altmic} is, using relations of \S5.1 and \S5.2.1 and dividing by $2 \Upsilon_0$, the two equations
\begin{equation}
\label{eq:micro2da}
\begin{split}
\gamma_\xi & \delta^{AB} \frac{\partial^2 \xi}{\partial X^A \partial X^B} 
+  \left( 1 - \frac{\alpha - \beta}{2} \right) \gamma_\xi \delta^{AB} \frac{\partial}{\partial \xi} \left( \ln \sqrt{G} \right)
\frac{\partial \xi}{\partial X^A} \frac{\partial \xi}{\partial X^B} -
\frac{\gamma_\xi}{2 l^2} \frac{ \d \lambda}{\d \xi} 
\\ & +
\gamma_\xi \delta^{AB} \frac{\partial}{\partial \eta} \left( \ln \sqrt{G} \right) \frac{\partial \xi}{\partial X^A} \frac{\partial \eta}{\partial X^B} 
-
\left(\frac{\alpha - \beta}{2} \right) \gamma_\eta \delta^{AB} \frac{\partial}{\partial \xi} \left( \ln \sqrt{G} \right) \frac{\partial \eta}{\partial X^A} \frac{\partial \eta}{\partial X^B} 
\\ & 
-  \left(\frac{\alpha - \beta}{2 l^ 2} \right)  \frac{\partial}{\partial \xi} \left( \ln \sqrt{G} \right)  [ \gamma_\xi \lambda + \gamma_\eta \nu ]
= \frac{A_0}{2 l^2} \left[ \frac{\partial w}{ \partial \xi} + (\alpha-\beta)  \frac{\partial}{\partial \xi} \left( \ln \sqrt{G} \right)  w \right],
\end{split}
\end{equation}
\begin{equation}
\label{eq:micro2db}
\begin{split}
\gamma_\eta & \delta^{AB} \frac{\partial^2 \eta}{\partial X^A \partial X^B} 
+  \left( 1 - \frac{\alpha - \beta}{2} \right) \gamma_\eta \delta^{AB} \frac{\partial}{\partial \eta} \left( \ln \sqrt{G} \right)
\frac{\partial \eta}{\partial X^A} \frac{\partial \eta}{\partial X^B} -
\frac{\gamma_\eta}{2 l^2} \frac{ \d \nu}{\d \eta} 
\\ & +
\gamma_\eta \delta^{AB} \frac{\partial}{\partial \xi} \left( \ln \sqrt{G} \right) \frac{\partial \eta}{\partial X^A} \frac{\partial \xi}{\partial X^B} 
-
\left(\frac{\alpha - \beta}{2} \right) \gamma_\xi \delta^{AB} \frac{\partial}{\partial \eta} \left( \ln \sqrt{G} \right) \frac{\partial \xi}{\partial X^A} \frac{\partial \xi}{\partial X^B} 
\\ & 
-  \left(\frac{\alpha - \beta}{2 l^ 2} \right)  \frac{\partial}{\partial \eta} \left( \ln \sqrt{G} \right)  [ \gamma_\xi \lambda + \gamma_\eta \nu ]
= \frac{A_0}{2 l^2} \left[ \frac{\partial w}{ \partial \eta} + (\alpha-\beta)  \frac{\partial}{\partial \eta} \left( \ln \sqrt{G} \right)  w \right].
\end{split}
\end{equation}
\begin{rmk} For nonzero $\Upsilon_0$,  \eqref{eq:micro2da} and \eqref{eq:micro2db} are two coupled nonlinear PDEs in four field variables:
$\varphi^1(X)$, $\varphi^2(X)$, $\xi(X)$, and $\eta(X)$, where
derivatives of $\varphi^1(X)$ and  $\varphi^2(X)$ enter $w$ on the right sides via $\bar{C}_{AB} = \partial_A \varphi^a \delta_{ab} \partial_B \varphi^b$.
For the special case $\Upsilon_0 = 0$, the left sides of \eqref{eq:micro2da} and \eqref{eq:micro2db} vanish, whereas for $\mu_0 = 0$, the right sides vanish.
\end{rmk}

\subsection{General solutions}

\subsubsection{Homogeneous fields}
Examine cases for which $\xi(X) \rightarrow \xi_\text{H} = \text{constant}$
and $\eta(X) \rightarrow \eta_\text{H} = \text{constant}$ at all points $X \in \mathcal{M}$; the constants may differ: $ \xi_\text{H} \neq \eta_\text{H}$ in general. Apply the notation $f_\text{H}(X) = f(X,\xi_\text{H},\eta_\text{H})$. Restrict $\mu_0 > 0$. Then \eqref{eq:P2d} and \eqref{eq:linmom2d} reduce to
\begin{equation}
\label{eq:linmom2dh}
\begin{split}
\frac{\partial P^A_a} {\partial X^A} = \mu_0 \delta_{ab}  \left[ \frac{ \partial^2 \varphi^b}{\partial X^A \partial X^B} \frac{\partial w}{\partial \bar{C}_{AB}} + \frac{ \partial \varphi^b}{\partial X^B }  \frac{\partial^2 w}{\partial \bar{C}_{AB} \partial X^A } \right]
= 0 \, \Rightarrow \, (P_\text{H})^A_a = \frac{\mu_0}{2} \left(\frac{\partial w}{\partial F^a_A}\right)_\text{H} = \text{constant}.
\end{split}
\end{equation}
This should be satisfied for any homogeneous $F^a_A = (F_\text{H})^a_A$ for which
$\partial^2 \varphi^a / \partial X^A \partial X^B = 0$.
Micro-momentum conservation laws \eqref{eq:micro2da} and \eqref{eq:micro2db} become
\begin{equation}
\label{eq:micro2dag}
\begin{split}
 - {\gamma_\xi} \frac{ \d \lambda}{\d \xi} 
-  ( {\alpha - \beta})  \frac{\partial}{\partial \xi} \left( \ln \sqrt{G} \right)  [ \gamma_\xi \lambda + \gamma_\eta \nu ]
= {A_0} \left[ \frac{\partial w}{ \partial \xi} + (\alpha-\beta)  \frac{\partial}{\partial \xi} \left( \ln \sqrt{G} \right)  w \right],
\end{split}
\end{equation}
\begin{equation}
\label{eq:micro2dbg}
\begin{split}
-{\gamma_\eta} \frac{ \d \nu}{\d \eta} 
-  ({\alpha - \beta} )  \frac{\partial}{\partial \eta} \left( \ln \sqrt{G} \right)  [ \gamma_\xi \lambda + \gamma_\eta \nu ]
= {A_0} \left[ \frac{\partial w}{ \partial \eta} + (\alpha-\beta)  \frac{\partial}{\partial \eta} \left( \ln \sqrt{G} \right)  w \right],
\end{split}
\end{equation}
wherein $\lambda = \lambda_\text{H}$, $\nu = \nu_\text{H}$, $(\frac{\partial}{\partial \xi} \ln \sqrt{G})_\text{H}$, and $(\frac{\partial}{\partial \eta} \ln \sqrt{G})_\text{H}$
 are all algebraic functions of $(\xi_\text{H},\eta_\text{H})$,
while $w = w_\text{H}$, 
 $(\frac{\partial}{\partial \xi} w)_\text{H}$, and $(\frac{\partial}{\partial \eta} w)_\text{H}$ 
 are algebraic functions of of $(\xi_\text{H},\eta_\text{H},(F_\text{H})^a_A)$.
 \begin{rmk}
 Homogeneous equilibrium is satisfied by the six algebraic equations
 \eqref{eq:linmom2dh} ($a,A = 1,2$), \eqref{eq:micro2dag}, and \eqref{eq:micro2dbg}
 in ten unknowns $(P_\text{H})^A_a$, $(F_\text{H})^a_A$, $\xi_\text{H}$, $\eta_\text{H}$.
 Given $(P_\text{H})^A_a$ or $(F_\text{H})^a_A$ as mechanical loading,
 the remaining six unknowns can be obtained from a simultaneous solution.
If $(F_\text{H})^a_A$ is imposed, \eqref{eq:micro2dag} and \eqref{eq:micro2dbg}
are two equations in $\xi_\text{H}$, $\eta_\text{H}$. Then
\eqref{eq:linmom2dh} yields the remaining $(P_\text{H})^A_a$.
\end{rmk}
\begin{rmk}
Essential boundary conditions for homogeneous states are $\xi = \xi_\text{H}$ and
$\eta = \eta_\text{H}$, both $\forall \, X \, \in \partial \! \mathcal{M}$.
Since $\xi$ and $\eta$ are constants, $Z^B_A = 0$
by \eqref{eq:Z2d}, so corresponding natural boundary conditions for forces conjugate to internal structure parameters in \eqref{eq:neumann} are 
$z_A = Z^B_A N_B = 0$.
\end{rmk}
 
\subsubsection{Stress-free states}
Consider cases whereby $P^A_a = 0 \, \forall X \in {\mathcal{M}}$. 
Linear momentum conservation laws \eqref{eq:linmom}, \eqref{eq:altmac}, and \eqref{eq:linmom2d}  are trivially satisfied.
Restrict $\mu_0 > 0$. Since $F^a_A$ is non-singular, \eqref{eq:P2d} requires 
$\partial w / \partial \bar{C}_{AB} = 0$.
This is obeyed at $\bar{C}_{AB} = \delta_{AB}$ via \eqref{eq:ground2D}; thus assume rigid body motion
(i.e., $\varphi^a = Q^a_A X^A + c^a_0$,
with $Q^a_A$ constant and proper orthogonal and $c^a_0$ constant)
whereby $w = 0$ vanishes as well by \eqref{eq:ground2D}.
\begin{rmk}
General analytical solutions for stress-free states are not apparent without particular forms of functions $w(\bar{C}_{AB},\xi,\eta)$, $G(\xi,\eta)$, $\lambda(\xi)$, $\nu(\eta)$, and values of
$\gamma_\xi$, $\gamma_\eta$, $\alpha$, $\beta$, and $l$.
\end{rmk}
\begin{rmk}
If $\partial w / \partial \xi = \partial w / \partial \eta = 0$ for $\bar{C}_{AB} = \delta_{AB}$,
then right sides of \eqref{eq:micro2da} and \eqref{eq:micro2db} vanish.
Whether or not stress-free deformation states with  $\bar{C}_{AB} \neq \delta_{AB}$ (e.g., locally) exist depends on $w$.
\end{rmk}

\subsection{Constitutive model}

The framework is applied to a rectangular patch of skin loaded in the $X^1$-$X^2$ plane.
A 2-D theory (i.e., membrane theory)
cannot distinguish between plane stress and plane strain conditions \cite{hoz2009}, nor can it account for out-of-plane anisotropy.
Nonetheless, 2-D nonlinear elastic models are widely used
to represent soft tissues, including skin \cite{tong1976,fung1993}.
Thus, parameters entering the model (e.g., $\mu_0$, $\Upsilon_0$) are particular to loading conditions and material orientations from experiments to which they are calibrated (e.g., here, plane stress).
\begin{rmk}
In a purely 2-D theory, incompressibility often used for 3-D modeling of biological tissues \cite{fung1993,holzapfel2000,holzapfel2009,ni2012}, cannot be assumed since contraction under biaxial stretch is not quantified in a 2-D theory.
Incompressibility is also inappropriate if the material dilates due to damage.
\end{rmk}
The skin is treated as having orthotropic symmetry, with two constant orthogonal directions in the reference configuration denoted by unit vectors $\vec{n}_1$ and $\vec{n}_2$:
\begin{equation}
\label{eq:n12}
\vec{n}_1 = n_1^A \frac{ \delta}{\delta X^A}, \quad
\vec{n}_2 = n_2^A \frac{ \delta}{\delta X^A}; \quad
n^A_i \delta_{AB} n^B_j = \delta_{ij} \quad (i,j=1,2).
\end{equation}
\begin{rmk}
The collagen fibers in the plane of the skin need not all align with $\vec{n}_1$ or $\vec{n}_2$, so 
long as orthotropic symmetry is respected.  For example, each $\vec{n}_i$ can bisect the alignments of two equivalent primary families of fibers in the skin whose directions are not necessarily orthogonal \cite{gasser2006,ni2012}.  In such a case, $\vec{n}_1$ is still a unit vector orthogonal to $\vec{n}_2$; planar orthotropy is maintained with respect to reflections about both unit vectors $\vec{n}_i$.
\end{rmk}
\begin{rmk}
The internal structure variables $\xi = D^1/l$ and $\eta = D^2/l$ account for mechanisms that lead to softening and degradation
under tensile load: fiber sliding, pull-out, and breakage of collagen fibers, and rupture of the elastin fibers and ground matrix.
Each $D^A$ ($A=1,2$) is a representative microscopic sliding or separation distance in the $\vec{n}_i \delta^i_A$ direction,
with $l$ the distance at which the material can no longer support tensile load along
that direction. 
\end{rmk}
\begin{rmk}
In the cohesive zone interpretation, 
each $D^A$ is viewed as a crack opening displacement for separation on a material surface (line in 2-D) normal to $\vec{n}_i \delta^i_A$.
\end{rmk}
\noindent Finslerian metrics $G_{AB}(\xi,\eta) = \delta^a_A \delta^b_B g_{ab}(\xi,\eta)$ of 
\S5.4.1 anisotropically rescale material and spatial manifolds $\mathcal{M}$ and $\mathfrak{m}$ due to
 microstructure changes in different directions.
 In the absence of damage, the nonlinear elastic potential of \S5.4.2 specializes a 3-D model \cite{gasser2006,holzapfel2009,ni2012,nolan2014} to 2-D.
 
\subsubsection{Metrics}
From \eqref{eq:lengths}, \eqref{eq:lengthsc}, \eqref{eq:2D},  \eqref{eq:isometric2D},  and \eqref{eq:def2D},
the difference in squared lengths of line elements $\d \vec{x}$ and $\d \vec{X}$ is 
\begin{equation}
\label{eq:dX2d}
(|\d \vec{x}|^2 - | \d \vec{X}|^2)(\vec{F},\xi,\eta) = [\delta_a^E \delta_b^F G_{EF}(\xi,\eta) F^a_A F^b_B - G_{AB}(\xi,\eta)] \d X^A \, \d X^B.
\end{equation}
\begin{rmk} 
Local regions of $\mathcal{M}$ at $X$ and $\mathfrak{m}$ at $x = \varphi(X)$ are rescaled isometrically by components $G_{AB}(\xi(X),\eta(X))$. 
When $F^a_A = \delta^a_A$, $| \d \vec{x} | = | \d \vec{X} |$ regardless of $G_{AB}$, $\xi$, or $\eta$.
For degenerate Riemannian metrics $G_{AB} = \bar{G}_{AB} = \delta_{AB}$ and $g_{ab} =  \bar{g}_{ab} = \delta_{ab}$, \eqref{eq:dX2d} becomes independent of ($\xi,\eta$).  
\end{rmk}
\noindent The Cartesian coordinate chart $\{X^A\}$ is prescribed such that $n^A_i = \delta^A_i$ in \eqref{eq:n12}; thus $\vec{n}_1$ and $\vec{n}_2$ are parallel to respective $X^1$- and $X^2$-directions on $\mathcal{M}$.
 Rescaling arises from changes in structure associated with degradation and damage in orthogonal directions, to which remnant strain contributions $\frac{1}{2} \ln [ G_{11} ( \xi)] $ 
and $\frac{1}{2} \ln [G_{22} ( \eta) ]$ can be linked.
The metric tensor $G_{AB}$ is hereafter assigned specific exponential terms, generalizing the 1-D form of \S4.4.1 to an anisotropic 2-D form appropriate for orthotropic symmetry:
\begin{equation}
\label{eq:G2D}
[ G_{AB}(\xi,\eta) ] = 
\begin{bmatrix}
\text{exp} \left( \frac{2k}{r} \xi^r \right) & 0 \\
0 & \text{exp} \left( \frac{2m}{r} \eta^r \right) 
\end{bmatrix}
 \, \Rightarrow \, G(\xi,\eta) = \det [ G_{AB} (\xi,\eta) ] =
\text{exp} \left( { \textstyle{ \frac{2}{r} }} [k \xi^r + m \eta^r ]  \right).
\end{equation}
For $\xi \in [0,1]$ and $\eta \in [0,1]$, two constants in \eqref{eq:G2D} are $k$ and $m$, positive for expansion.
A third constant $r > 0$ modulates rates of change of $G_{11}(\xi)$ and $G_{22}(\eta)$ 
with respect to their arguments.
Ratios are determined by remnant strain contributions at failure: $ \hat{\epsilon}_\xi = \frac{k}{r}$
and $\hat{\epsilon}_\eta = \frac{m}{r}$.
Values of $k$, $m$, and $r$ are calibrated to data in \S5.5.1.
Isotropy arises in \eqref{eq:G2D} when $\eta = \xi$ and $m = k$.
\begin{rmk}
More general forms of $G_{AB}(\xi,\eta)$, likely with more parameters, are possible; \eqref{eq:G2D} is a simple
form sufficient to address experimental observations for extension and tearing of skin.
\end{rmk}
\noindent From \eqref{eq:G2D}, non-vanishing components of Cartan's tensor in \eqref{eq:Cartan1} and \eqref{eq:chi2D} are
\begin{equation}
\label{eq:cartan2d}
l C^1_{11}  =  \chi_1 = \frac{\partial}{\partial \xi} (\ln \sqrt{G}) =  k \xi^{r-1}, \qquad
l C^2_{22}  =  \chi_2  = \frac{\partial}{\partial \eta} (\ln \sqrt{G}) =  m \eta^{r-1}.
\end{equation}

\subsubsection{Nonlinear elasticity}

The nonlinear elasticity model generalizes that of \S4.4.2 to a 2-D base space $\mathcal{M}$
with anisotropic Finsler metric depending on two structure variable components, $\xi$ and $\eta$ in normalized dimensionless form.
For the 2-D case, material symmetry of \S3.3.4 requires careful consideration.
Here, the skin is treated as a planar orthotropic solid \cite{tong1976,fung1993,jood2018}. 

Viewing the $D^A$ as components of a material vector field,
orthotropic symmetry suggests invariants $\xi^2$ and $\eta^2$.
For physically admissible ranges $\xi \in [0,1]$ and $\eta \in [0,1]$,
these can be replaced with $\xi$ and $\eta$.
Viewing the $D^A_{|B}$ similarly, orthotropic symmetry permits
a more general functional dependence than the sum of quadratic forms
in $\iota$ of \eqref{eq:psi2d1} and \eqref{eq:iota2D}. However, the chosen form of $\iota$ in  \eqref{eq:iota2D} allows
for partial anisotropy, not inconsistent with orthotropy, when $\gamma_\xi$ and $\gamma_\eta$ differ.
Thus, the structure-dependent contribution to $\psi$, $\Lambda l = {\Upsilon_0}(\gamma_\xi \lambda + \gamma_\eta \nu + \iota)$, more specifically here
\begin{equation}
\label{eq:lam2d}
\lambda(\xi) = \xi^2, \quad \nu(\eta) = \eta^2; \qquad
\iota( \nabla_0 \xi, \nabla_0 \eta) = l^2  \delta^{AB} ( \gamma_\xi  \partial_A \xi \partial_B \xi +  \gamma_\eta   \partial_A \eta \partial_B \eta),
\end{equation}
is consistent with material symmetry requirements.
Strain energy density $W$ in \eqref{eq:psi2d1} is dictated by 
dimensionless function $w(\bar{C}_{AB},\xi,\eta)$.
Per the above discussion, $\xi$ and $\eta$ are treated as scalar invariant arguments.
A partial list of remaining invariants \cite{holzapfel2009,gultekin2016} of \eqref{eq:pbasis2} for orthotropic symmetry of a 2-D material entering $w$ (and thus $\psi = \bar{\psi}$) is then, applying $n^A_i = \delta^A_i$ in \eqref{eq:n12},
\begin{equation}
\label{eq:invar2d}
\bar{I}_1 = \text{tr} \bar{\vec{C}} = \delta^{AB} \bar{C}_{AB}, \quad
\bar{I}_2 = J^2 = \det \bar{\vec{C}}, \quad
\bar{I}_3 = \bar{C}_{AB} n^A_1 n^B_1 = \bar{C}_{11}, \quad
\bar{I}_4 = \bar{C}_{AB} n^A_2 n^B_2 = \bar{C}_{22}.
\end{equation}
\begin{rmk}
As $\vec{n}_1$ and $\vec{n}_2$ are orthonormal, 
$\bar{I}_1 = \bar{I}_3 + \bar{I}_4$, so one of $\bar{I}_1,\bar{I}_3,\bar{I}_4$ in \eqref{eq:invar2d} is redundant.
Since $J \geq 1$, dependence on $\bar{I}_2 = \bar{C}_{11} \bar{C}_{22} - (\bar{C}_{12})^2$ can be replaced by $J$ (or by $(\bar{C}_{12})^2$ given $\bar{I}_3,\bar{I}_4$).
\end{rmk}
The Euclidean metric $\bar{G}^{AB} = \delta^{AB}$, rather than Finsler metric $G^{AB}$, is used for scalar products in \eqref{eq:lam2d} and \eqref{eq:invar2d}, consistent with \eqref{eq:Cbar2D}.
In 2-space, $\bar{I}_1$ and $\bar{I}_2$ are the complete set of isotropic invariants of $\bar{\vec{C}}$.
Two orthotropic invariants are $\bar{I}_3$ and $\bar{I}_4$;
several higher-order invariants are admissible \cite{holzapfel2009,gultekin2016} but excluded here since \eqref{eq:invar2d} is sufficient for the present application.
The dimensionless strain energy function entering \eqref{eq:psi2d1} is prescribed specifically as
\begin{equation}
\label{eq:w2d}
\begin{split}
w(\bar{C}_{AB},\xi,\eta) & = \left[ \frac{1}{J} (\bar{C}_{11} + \bar{C}_{22}) + k_0 (J-1)^2 - 2 \right] 
y_\mu(\xi,\eta) \\
& \, + \left[ \frac{a_1}{2 b_1} \left( \text{exp} \{ b_1 (\bar{C}_{11} -1)^2 \}  - 1\right) \right] 
\text{H}(\bar{C}_{11} -1) 
 y_\xi(\xi) \\
& \, + \left[ \frac{a_2}{2 b_2} \left( \text{exp} \{ b_2 (\bar{C}_{22} -1)^2 \}  - 1\right) \right] 
\text{H}(\bar{C}_{22} -1) 
y_\eta(\eta).
\end{split}
\end{equation}
Dimensionless constants are $k_0 > 0$, $a_1 \geq 0$, $b_1 > 0$, $a_2 \geq 0$, and $b_2 > 0$.
Right-continuous Heaviside functions $\text{H}(f)  = 1 \, \forall f \geq 0, \text{H}(f)  = 0 \, \forall f < 0$.
Also, $\mu_0 > 0$ and $\Upsilon_0 > 0$ are enforced in \eqref{eq:psi2d1}. 
\begin{rmk}
Potential $w$ in \eqref{eq:w2d} extends prior models for collagenous tissues \cite{gasser2006,holzapfel2009,ni2012,nolan2014} to include anisotropic structure changes. The first term on the right, linear in $\bar{I}_1/J$ and independent of volume change, accounts for isotropic shearing resistance of ground matrix and elastin.
The second term on the right accounts for resistance to volume (area) change, $k_0$ being
 a dimensionless bulk (area) modulus finite for a 2-D model; the dimensional bulk modulus $\kappa_0 = k_0 \mu_0 $.
Exponential terms account for stiffening from collagen fibers
in orthogonal directions $\vec{n}_i$.
 Heaviside functions prevent fibers from supporting compressive load \cite{holzapfel2009,lat2016} since they would likely buckle.
\end{rmk}
Degradation functions are $y_\mu(\xi,\eta)$, $y_\xi(\xi)$, and $y_\eta(\eta)$, where for the anisotropic theory,
\begin{equation}
\label{eq:yaniso}
y_\mu = (1-\xi)^\vartheta(1-\eta)^\varsigma = y_\xi y_\eta, \qquad
y_\xi = (1-\xi)^\vartheta, \qquad
y_\eta = (1- \eta)^\varsigma .
\end{equation} 
Corresponding material constants are $\vartheta \in [0,\infty)$ and $\varsigma \in [0,\infty)$.
Notice that matrix strain energy degrades equivalently with increasing $\xi$ and $\eta$ via $y_\mu$, maintaining isotropy of the first term in \eqref{eq:w2d}. As collagen fibers debond, the ground matrix and elastin simultaneously weaken.
\begin{rmk}
Choices $\vartheta = \varsigma =2$ are typical for phase-field fracture \cite{clayton2014}, though other values are possible for soft biologic tissues \cite{gultekin2019}.
Setting $\vartheta = \varsigma = 0$ implies null degradation (i.e., ideal elastic stress-stretch response).
\end{rmk}
\begin{rmk}
When $\xi = \eta = 0$, $w$ of \eqref{eq:w2d} is polyconvex \cite{balzani2006, balzani2012}, facilitating existence and uniqueness of solutions.
Also, $\psi$ with \eqref{eq:lam2d}, \eqref{eq:w2d}, and \eqref{eq:yaniso} obeys \eqref{eq:ground2D}. 
\end{rmk}
Stress components $P^A_a$ conjugate to $F^a_A = \partial_A \varphi^a$ are found from 
\eqref{eq:Pk1b}, \eqref{eq:P2d}, \eqref{eq:w2d}, and \eqref{eq:yaniso}, while forces $Q_{1,2}$ conjugate to $\xi,\eta$ are found from 
\eqref{eq:Q2d}, \eqref{eq:lam2d}, \eqref{eq:w2d}, and \eqref{eq:yaniso}: 
\begin{equation}
\label{eq:P2ds}
\begin{split}
{P^A_a}/{\mu_0}  =J^{-1} &  [   \delta_{ab} \delta^{AB} F^b_B
- {\textstyle{\frac{1}{2}}}  \bar{C}_{BC} \delta^{BC} (F^{-1})^A_a
+ k_0 J^2 (J-1) (F^{-1})^A_a ]  (1-\xi)^\vartheta(1-\eta)^\varsigma
\\
& +   [a_1(\bar{C}_{11} -1) \text{exp} \{ b_1 (\bar{C}_{11} -1)^2 \} \delta_{ab} \delta^A_1 F^b_1 ]
(1-\xi)^\vartheta \text{H}(\bar{C}_{11} -1)
\\ 
& +   [a_2(\bar{C}_{22} -1) \text{exp} \{ b_2 (\bar{C}_{22} -1)^2 \} \delta_{ab} \delta^A_2 F^b_2 ]
(1- \eta)^\varsigma \text{H}(\bar{C}_{22} -1),
\end{split}
\end{equation}
\begin{equation}
\label{eq:Q2ds1}
\begin{split}
{Q_1 l^2}/{(2\Upsilon_0)}  & = 
 \gamma_\xi \xi -   A_0 \vartheta \left[
  (1-\xi)^{\vartheta -1} (1-\eta)^\varsigma  \{ J^{-1} (\bar{C}_{11} + \bar{C}_{22}) + k_0 (J-1)^2 - 2 \} \right] 
  \\ & 
   \quad - A_0  \vartheta (1-\xi)^{\vartheta -1} \left[  {\textstyle{\frac{1}{2}}}  ({a_1}/{ b_1}) \left( \text{exp} \{ b_1 (\bar{C}_{11} -1)^2 \}  - 1\right) \right] 
\text{H}(\bar{C}_{11} -1),
  \end{split}
 \end{equation}
\begin{equation}
\label{eq:Q2ds2}
\begin{split}
{Q_2 l^2}/({2\Upsilon_0})  & = 
 \gamma_\eta \eta -   A_0 \varsigma \left[
  (1-\eta)^{\varsigma - 1} (1-\xi)^\vartheta  \{ J^{-1} (\bar{C}_{11} + \bar{C}_{22}) + k_0 (J-1)^2 - 2 \} \right] 
  \\ & 
   \quad - A_0 \varsigma (1-\eta)^{\varsigma - 1} \left[  {\textstyle{\frac{1}{2}}}  ({a_2}/{ b_2}) \left( \text{exp} \{ b_2 (\bar{C}_{22} -1)^2 \}  - 1\right) \right] 
\text{H}(\bar{C}_{22} -1).
  \end{split}
\end{equation}
\begin{rmk}
An ideal elastic response is obtained when $k = m = 0 \Rightarrow G_{AB} = \delta_{AB} \Rightarrow \chi_A = 0$, and $\vartheta = \varsigma = 0 \Rightarrow  \frac{\partial w}{\partial \xi} = \frac{\partial w}{\partial \eta}  = 0$.
Then since $\frac{\d \lambda }{ \d \xi} (0) = 0$ 
and $\frac{ \d \nu }{ \d \eta}(0) = 0$ by \eqref{eq:lam2d}, 
the right side of \eqref{eq:linmom2d} vanishes identically, and the (trivial) solutions to \eqref{eq:micro2da} 
and \eqref{eq:micro2db} are $\xi(X) = \eta(X) = 0 \, \forall \, X \in \mathcal{M}$.
\end{rmk}
\begin{rmk}
An isotropic version of the theory can be obtained, if along with
$m = k$ in \eqref{eq:G2D}, the following choices are made instead of \eqref{eq:yaniso}:
\begin{equation}
\label{eq:yiso}
y_\mu = {\textstyle{\frac{1}{2}}} [ (1-\xi)^\vartheta + (1-\eta)^\vartheta ], \quad
\varsigma = \vartheta, 
\quad y_\xi = y_\eta = 0;
 \qquad \gamma_\xi = \gamma_\eta =  {\textstyle{\frac{1}{2}}} \gamma_\mu \geq 0.
\end{equation}
Collagen fiber contributions to strain energy are removed such that $w$ now only depends on
isotropic invariants of $\bar{\vec{C}}$. Equilibrium equations \eqref{eq:linmom2d}, \eqref{eq:micro2da}, and \eqref{eq:micro2db} are identical under the change of variables $\xi \leftrightarrow \eta$, implying $\eta(X) = \xi(X)$ if identical boundary conditions on $D^A$ or $z_A$ are applied for each field on $\partial \! \mathcal{M}$.
In this case, one of \eqref{eq:micro2da}, and \eqref{eq:micro2db} is redundant and replaced with $\eta = \xi$.
\end{rmk}

\subsection{Specific solutions}

Possible inputs to the 2-D model are seventeen constants $l > 0 $, $k$, $m$, $r > 0$, 
$\mu_0 > 0$, $k_0 > 0$, $a_1 \geq 0$, $b_1 > 0$, $a_2 \geq 0$, $b_2 > 0$, 
$\vartheta \geq 0$, $\varsigma \geq 0$, $\Upsilon_0 > 0$,
$\gamma_\xi$, $\gamma_\eta$, $\alpha$, and $\beta$. 
Values of $l$ and $\Upsilon_0$ are taken from the analysis in \S4.5.2 of complete tearing
of a 1-D specimen of skin to a stress-free state. This is appropriate given that 1-D and 2-D theories are applied to describe surface energy and material length scale pertinent to the same experiments \cite{yang2015,per1997,tubon2022,sree2022}, and since stress-free solutions in \S5.5.3 perfectly parallel those of \S4.5.2. The remaining parameters are evaluated, in \S5.5.1, by applying the constitutive model of \S5.4 to the general solutions for homogeneous fields derived in \S5.3.1 to uniaixal-stress extension of 2-D skin specimens along the material $X^1$- and $X^2$-directions, respectively aligned  perpendicular and parallel to Langer's lines. 
\begin{rmk}
Collagen fibers of the microstructure in the dermis are aligned predominantly along Langer's lines and are more often pre-stretched in vivo along these directions \cite{jood2018}. In vivo or in vitro, elastic stiffness at finite stretch tends to be larger in directions along Langer's lines (i.e., parallel to $X^2$ and $\vec{n}_2$) than in orthogonal directions (e.g., parallel to $\vec{n}_1$). Degradation and failure behaviors are also anisotropic:
rupture stress tends to be larger, and failure elongation lower, for stretching in the stiffer $\vec{n}_2$-direction \cite{ani2012,yang2015,jood2018}. 
\end{rmk}
In \S5.5.2, model outcomes are reported for planar biaxial extension \cite{lanir1974,fung1993,hoz2009} of 2-D specimens, highlighting simultaneous microstructure degradation perpendicular and parallel to Langer's lines. 
Lastly, in \S5.5.3, stress-free states analogous to those modeled in a 1-D context in \S4.5.2 are
evaluated for the 2-D theory.

In \S5.5.1 and \S5.5.2, equilibrium solutions of \S5.3.1 hold. Invoking \eqref{eq:cartan2d}, \eqref{eq:lam2d}, \eqref{eq:w2d}, \eqref{eq:yaniso}, and \eqref{eq:P2ds}, and dropping $(\cdot)_\text{H}$ notation for brevity, \eqref{eq:linmom2dh}, \eqref{eq:micro2dag}, and \eqref{eq:micro2dbg} comprise the algebraic system
\begin{equation}
\label{eq:LM2D}
\begin{split}
{P^A_a}  & =\mu_0 J^{-1}   [   \delta_{ab} \delta^{AB} F^b_B
- {\textstyle{\frac{1}{2}}}  \bar{C}_{BC} \delta^{BC} (F^{-1})^A_a
+ k_0 J^2 (J-1) (F^{-1})^A_a ]  (1-\xi)^\vartheta (1-\eta)^\varsigma 
\\
& \qquad +   [a_1(\bar{C}_{11} -1) \text{exp} \{ b_1 (\bar{C}_{11} -1)^2 \} \delta_{ab} \delta^A_1 F^b_1 ]
(1-\xi)^\vartheta \text{H}(\bar{C}_{11} -1)
\\ 
& \qquad +   [a_2(\bar{C}_{22} -1) \text{exp} \{ b_2 (\bar{C}_{22} -1)^2 \} \delta_{ab} \delta^A_2 F^b_2 ]
(1-\eta)^\varsigma \text{H}(\bar{C}_{22} -1)
\\
& = \text{constant},
\end{split}
\end{equation}
\begin{equation}
\label{eq:MM2DA}
\begin{split}
  {\gamma_\xi}  \xi 
+  k\xi^{r-1} [ \gamma_\xi \xi^2 + \gamma_\eta \eta^2 ]
= - \frac{A_0}{2} \left[ \frac{\partial w(\bar{C}_{AB},\xi,\eta)}{ \partial \xi} + 2 k\xi^{r-1}  w(\bar{C}_{AB},\xi,\eta) \right],
\end{split}
\end{equation}
\begin{equation}
\label{eq:MM2DB}
\begin{split}
 {\gamma_\eta} \eta 
 + m\eta^{r-1}  [ \gamma_\xi \xi^2 + \gamma_\eta \eta^2 ]
= - \frac{A_0}{2} \left[  \frac{\partial w(\bar{C}_{AB},\xi,\eta)}{ \partial \eta} + 2 m\eta^{r-1}  w(\bar{C}_{AB},\xi,\eta) \right].
\end{split}
\end{equation}
Consistent with \eqref{eq:ab} for $N_0 = 0$ \cite{clayton2017g,clayton2016,clayton2017z}, $\beta  =  \alpha-2$ is assumed in \eqref{eq:MM2DA} and \eqref{eq:MM2DB}, reducing the number
of requisite parameters to fifteen; $\alpha$ and $\beta$ enter the governing equations only through their difference.
Boundary conditions on internal state, are, for homogeneous conditions,
\begin{equation}
\label{eq:bcuni}
 \xi (X^1 = \pm L_0, \, X^2 = \pm  W_0)  = \xi_\text{H},
\qquad \eta (X^1 = \pm L_0, \, X^2 = \pm  W_0)  = \eta_\text{H}.
\end{equation}
Alternative conditions to \eqref{eq:LM2D}--\eqref{eq:bcuni} are considered for heterogeneous stress-free states in \S5.5.3.
 
\subsubsection{Uniaxial extension}
First consider homogeneous uniaxial-stress extension in either the $X^1$- or $X^2$-direction.
From symmetry of the loading mode and material model, 
shear stresses vanish identically: $P^1_2 = 0$, $P^2_1 = 0$.
Similarly, $F^1_2 = 0$,  $F^2_1 = 0$, and $\bar{C}_{12} = \bar{C}_{21} = 0$.
 The homogeneous deformation fields are
 \begin{equation}
 \label{eq:uniFJC}
 \varphi^1 = \lambda_1 X^1, \quad
 \varphi^2 = \lambda_2 X^2; \quad F^1_1 = \lambda_1, \quad F^2_2 = \lambda_2; \quad
 \bar{C}_{11} = (\lambda_1)^2, \quad \bar{C}_{22} = (\lambda_2)^2; \quad
 J = \lambda_1 \lambda_2.
 \end{equation}
At any single given load increment, stretch ratios are the constants $\lambda_1 > 0$ and $\lambda_2 > 0$. 

Mechanical boundary conditions are, for extension along $X^1$ with $\lambda_1 \geq 1$,
\begin{equation}
\label{eq:bcuni1}
\varphi^1(X^1 = \pm L_0) = \pm \lambda_1 L_0, \qquad p_2 (X^2 = \pm  W_0) = P^2_2  (X^2 = \pm  W_0) = 0.
\end{equation}
In this case, $P^2_2 = 0 \, \forall \, X \in \mathcal{M}$, and the sole non-vanishing stress component in \eqref{eq:LM2D} is $P^1_1$.
Note that $\lambda_2$ is unknown a priori. Given $\lambda_1$ from the first of \eqref{eq:bcuni1}, values consistent with \eqref{eq:bcuni} are obtained by solving \eqref{eq:LM2D} for $a = A = 2$ with $P^2_2 = 0$, \eqref{eq:MM2DA}, and \eqref{eq:MM2DB} simultaneously for $\lambda_2$, $\xi$, and $\eta$ as functions of $\lambda_1$. Axial stress $P^1_1$ is then found afterwards using  \eqref{eq:LM2D} with $a = A = 1$.

For axial loading along $X^2$ with $\lambda_2 \geq 1$,
\begin{equation}
\label{eq:bcuni2}
\varphi^2 (X^2 = \pm W_0) = \pm \lambda_2 W_0, \qquad p_1 (X^1 = \pm  L_0) = P^1_1  (X^1 = \pm  L_0) = 0.
\end{equation}
Now $P^1_1 = 0 \, \forall \, X \in \mathcal{M}$, and the sole non-vanishing stress component in \eqref{eq:LM2D} is $P^2_2$. Given $\lambda_2$ from the first of \eqref{eq:bcuni2}, values consistent with \eqref{eq:bcuni} are obtained by solving \eqref{eq:LM2D} for $a = A = 1$ with $P^1_1 = 0$, \eqref{eq:MM2DA}, and \eqref{eq:MM2DB} simultaneously for $\lambda_1$, $\xi$, and $\eta$ as functions of $\lambda_2$.
Axial stress $P^2_2$ is found afterwards using \eqref{eq:LM2D} with $a = A = 2$.

Values of all baseline parameters are listed in Table~\ref{table1}. Identical values of those constants shared among 1-D and 2-D theories are found to aptly describe the experimental data for stretching along $\vec{n}_1$, in conjunction with natural choice $\gamma_\xi = 1$.  The 2-D theory features additional parameters to account for orthotropic anisotropy
(e.g., stiffer response along $\vec{n}_2$, with peak stress occurring at lower stretch) as well as an areal bulk modulus $\kappa_0$ absent in the 1-D theory. 

\begin{rmk}
Adherence to physical observations dictates $a_2 > a_1$, $b_2 > b_1$, and $\kappa_0  > \mu_0$.
Since degradation is more severe, and toughness lower for stretching along $\vec{n}_2$, $m > k$ and $\gamma_\eta < \gamma_\xi$.
The standard choice \cite{clayton2014,gultekin2019} $\varsigma = \vartheta = 2$ in \eqref{eq:yaniso} was found sufficient to describe test data.
\end{rmk}

\begin{figure}
	\centering
	\subfigure[axial stress]{\includegraphics[width=0.44\textwidth]{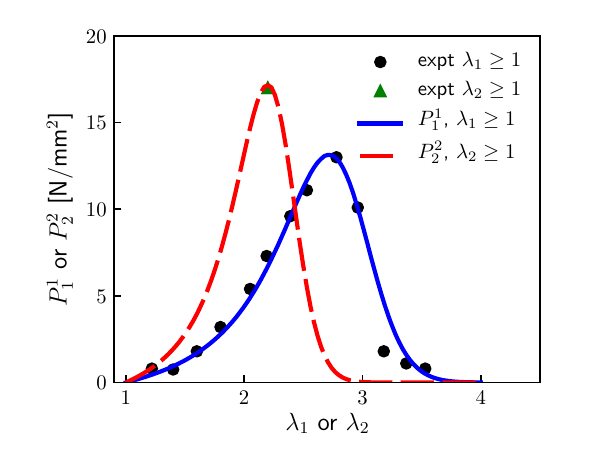}\label{fig4a}} \quad
	\subfigure[structure variables]{\includegraphics[width=0.44\textwidth]{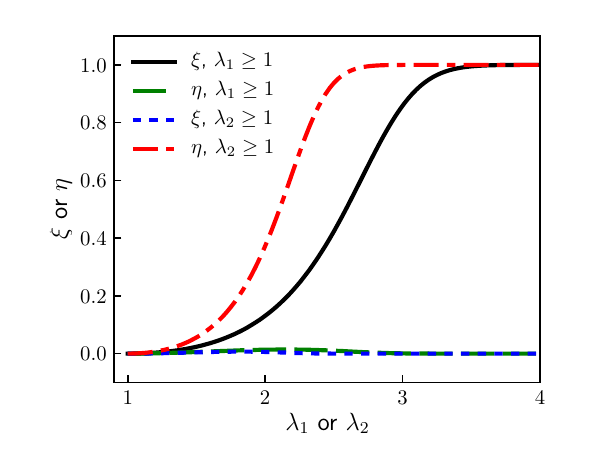}\label{fig4b}} \\
	\vspace{-3mm}
	\subfigure[axial stress]{\includegraphics[width=0.44\textwidth]{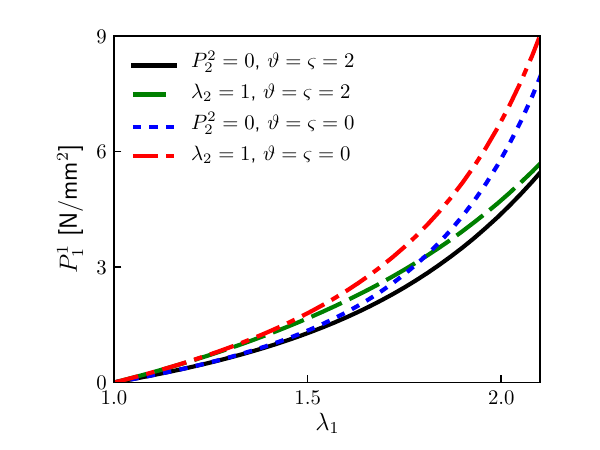}\label{fig4c}} \quad
	\subfigure[axial stress]{\includegraphics[width=0.44\textwidth]{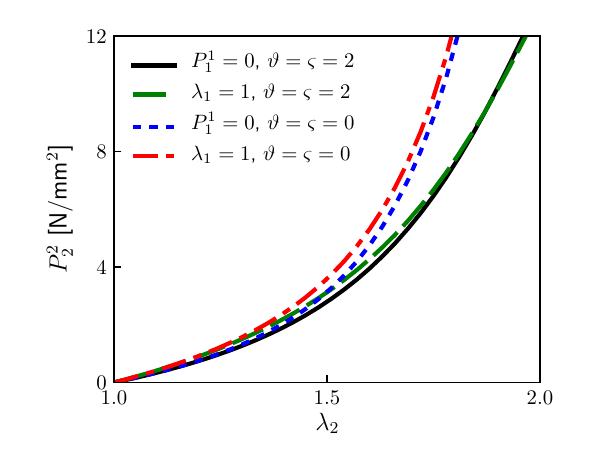}\label{fig4d}} 
	\vspace{-2mm}
\caption{Uniaxial extension and tearing of skin for imposed axial stretch $\lambda_1 \geq 1$ or $\lambda_2 \geq 1$, 2-D model: 
(a) stress $P^1_1$ or $P^2_2$ (baseline parameters, Table~\ref{table1}) with representative experimental data \cite{yang2015}  (see text \S4.5.1 for consistent definition of experimental stretch accounting for pre-stress) for straining perpendicular or parallel to Langer's lines 
(b) normalized internal structure components $\xi$ and $\eta$ (baseline parameters)
(c) stress $P^1_1$ for moderate extension $\lambda_1 \leq 2.1 $ under uniaxial stress ($P^2_2 = 0$) or uniaxial strain ($\lambda_2 = 1$) conditions
for Finsler model (baseline parameters) and ideal elastic model ($\vartheta = \varsigma = 0$)
(d) stress $P^2_2$ for moderate extension $\lambda_2 \leq 2.0$ under uniaxial stress ($P^1_1 = 0$) or uniaxial strain ($\lambda_1 = 1$) conditions
for Finsler model (baseline) and ideal elastic model ($\vartheta = \varsigma = 0$)
\label{fig4}}
\end{figure}

Model outcomes for non-vanishing stress components and internal state vector components are presented in respective Fig.~\ref{fig4a} and Fig.~\ref{fig4b}.  Experimental $P^1_1$ versus $\lambda_1$ data for loading along $\vec{n}_1$, with 
$\lambda_1 \geq 0$ prescribed in the corresponding model calculations,
are identical to $P$ versus $\sqrt{C}$ data depicted using the 1-D theory in \S4.5.1.
These data \cite{yang2015} are for relatively high-rate extension of rabbit skin along a longitudinal direction, parallel to the backbone of the torso and perpendicular to Langer's lines. Nonlinear elastic parameters should be viewed as instantaneous dynamic moduli in a pseudo elastic representation \cite{fung1993,lim2011,claytonMOSM20}, since loading times are brief relative to stress relaxation times \cite{yang2015}.
Single-experiment data of similar fidelity for transverse extension, parallel to Langer's lines, to complete load drop were not reported, but a range of maximum stress and strain were given for extension along $\vec{n}_2$ \cite{yang2015}.
A representative peak stress $P^2_2$ and corresponding stretch $\lambda_2$ based on such data \cite{yang2015} are included in Fig.~\ref{fig4a}. According to such data \cite{yang2015}, the material is stiffer, and ruptures at a higher stress ($\approx \frac{4}{3} \times$) but lower strain ($\approx \frac{2}{3} \times$), in the transverse $\vec{n}_2$-direction.

\begin{rmk}
For loading along $\vec{n}_1$, $\xi \rightarrow 1$ and $\eta \rightarrow 0$ for $\lambda_1 \gtrsim 3.5$, meaning most
internal structure evolution correlates with degradation in this direction, with small transverse effects of $\eta$.
Analogously, loading along $\vec{n}_2$ gives $\eta \rightarrow 1$ and $\xi \rightarrow 0$ for $\lambda_2 \gtrsim 3$.
The rate of increase of $\eta$ with $\lambda_2 > 1$ is more rapid than the rate of increase of $\xi$ with $\lambda_1 > 1$, since
the skin degrades sooner and fails at a lower strain for stretching parallel to Langer's lines.
The present diffuse model is an idealization characteristic of experiments when there is no sharp pre-crack \cite{fung1993,munoz2008,yang2015,mitsuhashi2018,ani2012}.
\end{rmk}

Shown in Fig.~\ref{fig4c} and Fig.~\ref{fig4d} are predictions at modest stretch along $\vec{n}_1$ or $\vec{n}_2$
under uniaxial stress conditions identical to those of Fig.~\ref{fig4a} as well as uniaxial strain, whereby $\lambda_2 =1$ or
$\lambda_1 = 1$ is enforced using the scheme of \S5.5.2 rather than respective $P^2_2 = 0$ or $P^1_1 = 0$.
Predictions for the ideal elastic case ($\vartheta = \varsigma = 0 \Rightarrow \xi = \eta = 0$) are shown for comparison.
Results are stiffer for the ideal elastic case since degradation commensurate with structure change is omitted.
In agreement with other data \cite{lanir1974}, skin is elastically stiffer in uniaxial strain relative to uniaxial stress.
Choosing a higher value of $k_0 = \kappa_0/\mu_0 > 1$ in \eqref{eq:w2d} would further increase this difference if merited

\subsubsection{Biaxial extension}
Now consider homogeneous biaxial-stress extension in the $X^1$- and $X^2$-directions.
From symmetry, $P^1_2 = 0$, $P^2_1 = 0$, $F^1_2 = 0$,  $F^2_1 = 0$, and $\bar{C}_{12} = \bar{C}_{21} = 0$.
 The homogeneous deformation fields are
 \begin{equation}
 \label{eq:uniFJCbi}
 \varphi^1 = \lambda_1 X^1, \,  \,
 \varphi^2 = \lambda_2 X^2; \quad F^1_1 = \lambda_1,  \, \, F^2_2 = \lambda_2; \quad
 \bar{C}_{11} = (\lambda_1)^2, \,  \, \bar{C}_{22} = (\lambda_2)^2; \quad
 J = \lambda_1 \lambda_2.
 \end{equation}
Stretch ratios are $\lambda_1 > 0$ and $\lambda_2 > 0$, constants over $\mathcal{M}$. 
Mechanical boundary conditions are
\begin{equation}
\label{eq:bcunibi}
\varphi^1(X^1 = \pm L_0) = \pm \lambda_1 L_0,  \qquad
\varphi^2 (X^2 = \pm W_0) = \pm \lambda_2 W_0.
\end{equation}
With $\lambda_1$ and $\lambda_2$ prescribed by \eqref{eq:bcunibi}, equilibrium equations 
 \eqref{eq:MM2DA} and \eqref{eq:MM2DB} are solved simultaneously for $\xi$ and $\eta$ as functions of 
 $\lambda_1,\lambda_2$, giving homogeneous values of fields consistent with \eqref{eq:bcuni}.
 Then $P^1_1$ and $P^2_2$ are obtained afterwards with \eqref{eq:LM2D}
 for $a = A = 1$ and $a = A = 2$.
 
\begin{figure}
	\centering
	\subfigure[normal stresses]{\includegraphics[width=0.44\textwidth]{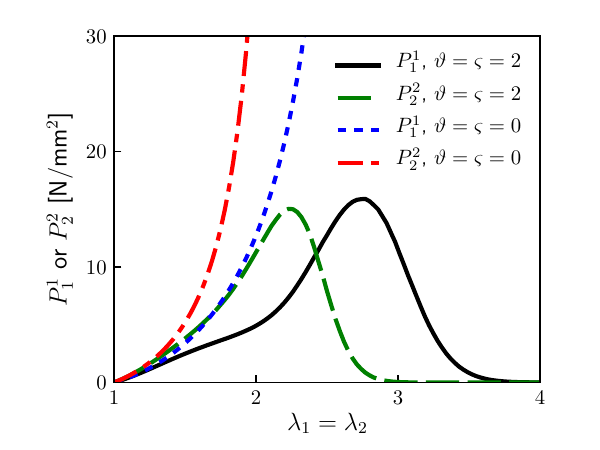}\label{fig5a}} \quad
	\subfigure[structure variables]{\includegraphics[width=0.44\textwidth]{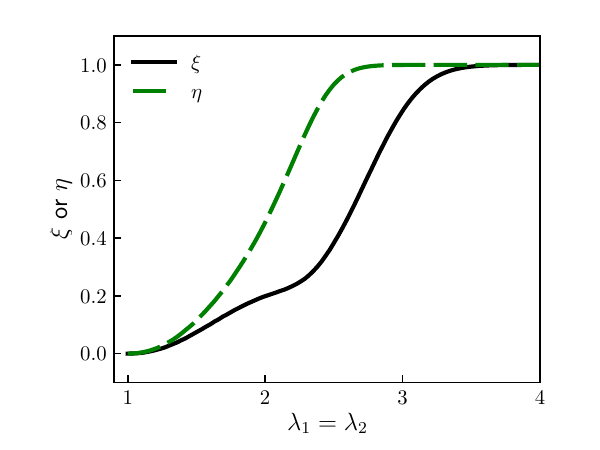}\label{fig5b}} 
	\vspace{-2mm}
\caption{Equi-biaxial extension and tearing of skin, 2-D model: 
(a) stress components from Finsler model (baseline parameters, $\vartheta = \varsigma = 2$) and
ideal elastic model ($\vartheta = \varsigma = 0$)
(b) normalized internal structure components $\xi$ and $\eta$
\label{fig5}}
\end{figure}

 Model predictions for equi-biaxial stretching, $\lambda_1 =\lambda_2$,
 are produced using the baseline material parameters of Table~\ref{table1}, obtained for the 2-D theory in \S5.5.1. 
 In Fig.~\ref{fig5a}, stresses also include those for the ideal elastic case ($\vartheta = \varsigma = 0 \Rightarrow \xi = \eta = 0$)
that are noticeably higher for $\lambda_1 >  1.5$ and increase monotonically with stretch.
 For the Finsler theory, under this loading protocol ($\lambda_1 = \lambda_2$), $P^2_2$ increases more rapidly than $P^1_1$ with increasing $\lambda_1$,
 reaching a slightly lower peak value at significantly lower stretch. Elastic stiffness during the lower-stretch loading phase is higher in the $\vec{n}_2$-direction due to the preponderance of aligned collagen fibers, 
 but degradation associated with internal structure evolution is more rapid due to the lower toughness of skin when torn in this direction.
 The latter phenomenon is evident in Fig.~\ref{fig5b}, wherein $\eta(\lambda_1) > \xi(\lambda_1)$ for $\lambda_1 \in [1.1,3.9]$.
\begin{rmk}
Experimental data on failure of skin focus
on uniaxial extension \cite{yang2015,jood2018}. Known biaxial data (e.g, \cite{lanir1974,fung1993}) do not report stretch magnitudes sufficient to cause tearing, so direct validation does not appear possible. Should skin prove to be more stiff and damage tolerant in equi-biaxial stretch experiments, $w$ of \eqref{eq:w2d} can be modified so the tangent bulk modulus proportional to $k_0$ increases more strongly with $J$ and does not degrade so severely with structure evolution.
\end{rmk}

\subsubsection{Stress-free states}
Protocols of \S5.3.2 now apply. Two boundary value problems are addressed that parallel the 1-D analysis of \S4.5.2. 
External boundary conditions are
$\xi = 0$ and $\eta = 0$ everywhere along $\partial \! \mathcal{M}$. Stress $P^A_a = 0$ everywhere in $\mathcal{M}$, so mechanical traction $p_a = P^A_a N_A = 0$ over $\partial \! \mathcal{M}$. For the generalized Finsler metric in \eqref{eq:G2D}, restrict $r > 1 \Rightarrow
\chi_1 (\xi = 0) = \chi_2 (\eta = 0) = 0$ in \eqref{eq:cartan2d}.

In the first problem, assume the specimen is stretched uniaxially along the $\vec{n}_1$-direction (i.e., along $X^1$, perpendicular to Langer's lines) until localized failure occurs. The skin ruptures completely across the midspan at $X^1 = 0$, such that $\xi(0,X^2) = 1$. 
In this ruptured state, $\bar{C}_{AB} = \delta_{AB}$ everywhere on $\mathcal{M}$ for all components except $\bar{C}_{11}$, which can differ from $\delta_{AB}$ only along the
 line $X^1 = 0$. The solution for $\eta(X^1,X^2)$ is $\eta(X^1,X^2) = 0$, for which
 \eqref{eq:micro2db} is trivially satisfied.  From symmetry, the remaining unknown field $\xi$ depends only on $X = X^1$, and $\xi(-X) = \xi(X)$. With this partial solution, the remaining governing equation \eqref{eq:micro2da} has vanishing right side and reduces to the generally nonlinear but autonomous second-order ODE
 \begin{equation}
\label{eq:free2da}
\begin{split}
\gamma_\xi \, \frac{\d^2 \xi}{\d X^2} 
= \frac{\gamma_\xi \xi} {l^2} \left( 1 + k \xi^r \right). 
\end{split}
\end{equation}
Dividing by $\gamma_\xi > 0$, \eqref{eq:free2da} is identical to \eqref{eq:1dA} with $N_0 = 0$,
$\lambda = \xi^2$, and $\chi = \chi_1 = k \xi^{r-1}$.
Solutions \eqref{eq:auto2} and \eqref{eq:autosol} hold verbatim. 
Normalized energy per unit area normal to the $X^1$-direction is
 \begin{equation}
\label{eq:surfsol2d}
\bar{\gamma}_1 = 
 \frac{1 }{ 2  \Upsilon_0}  \int_{-L_0}^{L_0} \psi  \sqrt{G} \, \d X 
= \frac{\gamma_\xi }{ 2 l }  \int_{-L_0}^{L_0}
\{  \xi^2 +  l^2  (\d \xi / \d X) ^2  \} 
\, \text{exp} [ ({k}/{r}) \xi^r]  \, \d X,
\end{equation}
identical to \eqref{eq:surfsol} when $\gamma_\xi = 1$ and $N_0 = 0$.
Given $\gamma_\xi = 1$, $k= 0.2$, $r = 2$, and $\Upsilon_0 = 0.47$ kJ/m$^2$ (Table~\ref{table1}), outcomes of the 2-D theory here match those of
the 1-D theory in Fig.~\ref{fig3a} and Fig.~\ref{fig3b} with $N_0 = 0$ and $\bar{\gamma}_1 = \bar{\gamma}$. Toughness $ 2 \bar{\gamma}_1 \Upsilon_0 = 1.0$ kJ/m$^2$ is consistent with experiment \cite{per1997,tubon2022,sree2022}.

In the second problem, assume the specimen is stretched along $\vec{n}_2$ (i.e., along $X^2$, parallel to Langer's lines). The skin ruptures completely across the midspan at $X^2 = 0$, with $\eta(X^1,0) = 1$. 
Now, $\bar{C}_{AB} = \delta_{AB}$ everywhere for all components except $\bar{C}_{22}$, which can differ from $\delta_{AB}$ only along $X^2 = 0$. The solution for $\xi(X^1,X^2)$ is $\xi = 0$, for which
 \eqref{eq:micro2da} is trivially obeyed.  From symmetry, $\eta$ depends only on $X = X^2$, and $\eta(-X) = \eta(X)$. Balance law \eqref{eq:micro2db} reduces to 
 \begin{equation}
\label{eq:free2db}
\begin{split}
\gamma_\eta \, \frac{\d^2 \eta}{\d X^2} 
= \frac{\gamma_\eta \eta} {l^2} \left( 1 + m \eta^r \right). 
\end{split}
\end{equation}
Dividing by $\gamma_\eta > 0$, \eqref{eq:free2db} matches \eqref{eq:1dA} with $N_0 = 0$,
$\nu = \eta^2$, $\chi = \chi_2 = m \eta^{r-1}$, and the obvious change of variables.
Solutions \eqref{eq:auto2} and \eqref{eq:autosol} hold. 
Normalized energy per unit area is
 \begin{equation}
\label{eq:surfsol2db}
\bar{\gamma}_2 = 
 \frac{1 }{ 2  \Upsilon_0}  \int_{-L_0}^{L_0} \psi  \sqrt{G} \, \d X 
= \frac{\gamma_\eta }{ 2 l }  \int_{-L_0}^{L_0}
\{  \eta^2 +  l^2  (\d \eta / \d X) ^2  \} 
\, \text{exp} [ ({m}/{r}) \eta^r]  \, \d X
\end{equation}
 for free surfaces normal to the $X^2$-direction, matching \eqref{eq:surfsol} if $\gamma_\eta = 1$
 and $N_0 = 0$. Given $\gamma_\eta = 0.84$, $m= 0.3$, $r = 2$, and $\Upsilon_0 = 0.47$ kJ/m$^2$ (Table~\ref{table1}), profiles of $\eta(X)$ for this problem are very similar to those of
$\xi(X)$ from the 1-D theory in Fig.~\ref{fig3a}.
Energy for $N_0 = 0$ in Fig.~\ref{fig3b} transforms as $\bar{\gamma}_2 \approx \gamma_\eta \bar{\gamma}$, and $ 2 \bar{\gamma}_2 \Upsilon_0 = 0.85 $ kJ/m$^2$ is within experimental ranges of 0.5 to 2.5 kJ/m$^2$ \cite{per1997,tubon2022,sree2022}.
\begin{rmk}
Since $ 2 \bar{\gamma}_2 \Upsilon_0 < 2 \bar{\gamma}_1 \Upsilon_0$, the model predicts that skin is more brittle in directions parallel to Langer's lines than perpendicular to Langer's lines, in concurrence with experiment \cite{ani2012,yang2015}. Collagen fibers are less coiled initially in directions parallel to Langer's lines \cite{jood2018}, giving the skin lower compliance and less potential strain accommodation at rupture in those directions.
\end{rmk}
\begin{rmk}
All parameters in Table~\ref{table1} have clear physical or geometric origins; none are ad hoc.
Constant $l$ is the critical fiber sliding distance or crack opening displacement for rupture.
Ratios $\frac{k}{r}$ and $\frac{m}{r}$ are associated with remnant strain contributions
in orthogonal $\vec{n}_1$- and $\vec{n}_2$-directions along primary initial fiber directions (e.g., perpendicular and parallel to Langer's lines). The isotropic shear modulus and bulk modulus for the matrix, consisting of ground substance and elastin, are $\mu_0$ and $\kappa_0$.
Nonlinear elastic constants $a_1$ and $b_1$ control stiffening due to collagen fiber elongation in the $\vec{n}_1$-direction,
while $a_2$ and $b_2$ control stiffening due to fiber elongation in the $\vec{n}_2$-direction.
Loss of elastic stiffness due to fiber rearrangements and damage processes in matrix, fibers, and their interfaces,
in respective $\vec{n}_1$- and $\vec{n}_2$-directions, is modulated by $\vartheta$ and $\varsigma$.
Isotropic surface energy is $\Upsilon_0$, with factors $\gamma_\xi$ and $\gamma_\eta$
scaling the fracture toughness in respective $\vec{n}_1$- and $\vec{n}_2$-directions.
\end{rmk}

\section{Conclusion}
A theory of finite-deformation continuum mechanics with a basis in generalized Finsler
geometry has been developed and refined. 
Elements of an internal state vector represent evolving microstructure features and can
be interpreted as order parameters.
Dependence of the material metric on internal state affects how distances are measured
in the material manifold and how gradients (i.e., covariant derivatives) are resolved.
A new application of the theory to anisotropic soft-tissue mechanics has been
presented, whereby the internal state is primarily associated with collagen fiber rearrangements and
breakages. The material metric contains explicit contributions from sliding or opening modes
in different material directions. Solutions to boundary value problems
for tensile extension with tearing in different directions agree with experimental data and microscopic observations on skin
tissue, providing physical and geometric insight into effects of microstructure. \vspace{5mm}
\\
\textbf{Funding:} This research received no external funding.
\vspace{1mm}
\newline 
\textbf{Conflicts of interest:} The author declares no conflicts of interest.
\vspace{1mm}
\newline
\textbf{Data availability statement:} Not applicable; this research produced no data.


\section{Appendix A: Variational derivatives}
\setcounter{figure}{0} \renewcommand{\thefigure}{A.\arabic{figure}}
\setcounter{table}{0} \renewcommand{\thetable}{A.\arabic{table}}
\setcounter{equation}{0} \renewcommand{\theequation}{A.\arabic{equation}}
\setcounter{subsection}{0} \renewcommand{\thesubsection}{A.\arabic{subsection}}

The variational derivative $\delta(\cdot)$
of \S3.3.1 invokes $(\varphi^a,D^A)$ with $a,A = 1, \ldots, n$ as the total set of $2n$ varied independent parameters
or degrees-of-freedom. 

\subsection{Deformation gradient and director gradient}
The first of \eqref{eq:variations} follows
from \eqref{eq:varphiC}, \eqref{eq:defgradC}, and commutation of $\delta (\cdot)$
and $\partial_A (\cdot)$ operators since the variation is performed at fixed $X^A$:
\begin{equation}
\label{eq:varF}
\delta F^a_A (\varphi(X),X) = \delta (\partial_A \varphi^a(X)) = \partial_A (\delta \varphi^a(X))
= \delta_A (\delta \varphi^a(X)) = (\delta \varphi^a(X))_{|A},
\end{equation}
with $\vec{F}$ treated as a two-point tensor. 
\begin{rmk}
The third equality in \eqref{eq:varF} follows from $N^B_A(X,D) \bar{\partial}_B \varphi^a(X) = 0$. The leftmost and rightmost equalities interpret $\varphi^a (X)$ and $\delta \varphi^a(X)$, respectively, as point functions rather than vector fields \cite{marsden1983,truesdell1960}.
\end{rmk}
Denote by $f(X,D)$ a generic differentiable function of arguments $\{X^A,D^A\}$
in a coordinate chart on $\mathcal{Z}$. The 
variation of $f(X,D)$ is defined by
the first of the following:
\begin{equation}
\label{eq:varderiv}
\delta f(X,D) = f(X,D)|_A \delta (D^A) = \bar{\partial}_A f (X,D) \delta (D^A),
\end{equation}
where $(\cdot)|_{A}$ is the vertical covariant derivative (e.g., as in \eqref{eq:excov}).
For the choices $V^A_{BC} =0$ and $Y^A_{BC} = 0$ of \eqref{eq:connrec}, 
$f(X,D)|_A = \bar{\partial}_A f (X,D)$ and the rightmost form is obtained,
consistent with prior definitions \cite{clayton2017g,claytonMMS22}. This is used
with \eqref{eq:gradDi} 
to obtain the second of \eqref{eq:variations}:
\begin{equation}
\label{eq:vargradD}
\begin{split}
\delta D^A_{|B} & = \delta (\partial_B D^A) - \delta N^A_B + \delta (K^A_{BC}) D^C + K^A_{BC} \delta (D^C) \\ & =
[\partial_B  \delta (D^A) - N^C_B \bar{\partial}_C \delta (D^A) + K^A_{BC} \delta (D^C)]
- \bar{\partial}_C N^A_B \delta (D^C) + \bar{\partial}_D K^A_{BC} D^C \delta(D^D) \\
& = [\delta(D^A)]_{|B} - (\bar{\partial}_C N^A_B - \bar{\partial}_C K^A_{BD} D^D) \delta(D^C),
\end{split}
\end{equation}
where it is assumed per \eqref{eq:funcC} that $\bar{\partial}_C \delta[D^A(X)] = \bar{\partial}_C [\delta(D^A) (X)] = 0$
on $\mathcal{M}$ and $\mathcal{Z}$.

\subsection{Volume form}
Two definitions have been set forth in prior work for the variation of the volume form $d \Omega (X,D)$.
The first quoted here sets \cite{claytonMMS22}
\begin{equation}
\label{eq:omegavar1}
\begin{split}
\delta (d \Omega) &  = [{\delta{\sqrt G}}/{\sqrt{G}}] d \Omega = (\ln \sqrt{G} )|_A \delta (D^A) d \Omega =
{\textstyle{\frac{1}{2}}} G^{BC} G_{CB}|_A  \delta (D^A) d \Omega \\ & = 
(C^B_{AB} - Y^B_{AB})  \delta (D^A) d \Omega = C^B_{AB} \delta (D^A) d \Omega,
\end{split}
\end{equation}
where the first equality is a definition and \eqref{eq:Gids2} and \eqref{eq:varderiv} have been used subsequently.
\begin{rmk}
According to \eqref{eq:omegavar1}, the magnitude of the volume form is varied locally over 
$n$-dimensional base space in \eqref{eq:varG} with $\alpha = 1$ prior to application of the divergence theorem \eqref{eq:stokes} used to procure
\eqref{eq:linmom} and \eqref{eq:dirmom} from \eqref{eq:intpart} of \S3.3.3. The choice \eqref{eq:omegavar1} was used in the most recent theory \cite{claytonMMS22} and implied in a prior numerical implementation \cite{clayton2018c}.
\end{rmk}

\noindent The second definition quoted here was used in the original theoretical derivations \cite{clayton2017g,clayton2016}:
\begin{equation}
\label{eq:omegavar2}
\begin{split}
\delta (d \Omega) &  = [{\delta{\sqrt \mathcal{G}}}/{\sqrt{\mathcal{G}}}] d \Omega = (\ln \sqrt{ \mathcal{G}} )|_A \delta (D^A) d \Omega = (\ln \sqrt{ G^2} )|_A \delta (D^A) d \Omega \\ & = 
G^{BC} G_{CB}|_A  \delta (D^A) d \Omega =
2 (C^B_{AB} - Y^B_{AB})  \delta (D^A) d \Omega = 2 C^B_{AB} \delta (D^A) d \Omega.
\end{split}
\end{equation}
In derivation of \eqref{eq:omegavar2}, the determinant of the Sasaki metric as defined in \eqref{eq:detGS} has been used along with \eqref{eq:Gids2} and \eqref{eq:varderiv}.
\begin{rmk}
The definition given by the first equality in \eqref{eq:omegavar2} is notionally consistent with
other earlier theory \cite{saczuk1996,saczuk1997,stumpf2000}. In the present viewpoint with \eqref{eq:omegavar2}, the magnitude of the volume form is varied locally in $2n$-dimensional total space $\mathcal{Z}$ via \eqref{eq:varG} with $\alpha = 2$ before integrating over 
 base $n$-dimensional space $\mathcal{M}$ in \eqref{eq:intpart} of \S3.3.3.
 \end{rmk}
 \begin{rmk}
Definition \eqref{eq:omegavar1} corresponds to $\alpha = 1$ and definition \eqref{eq:omegavar2} to $\alpha = 2$ in
\eqref{eq:varG}.
The only ramification in the governing Euler-Lagrange equations is
 scaling of local free energy density by a factor of one or two through $\alpha \psi C^A_{CA}$
in the micro-momentum balance, in either form \eqref{eq:dirmom} or \eqref{eq:altmic}.
Macroscopic momentum is unaffected by the definition of $ \delta ( d \Omega)$.
\end{rmk}


\section{Appendix B: Toward residual stress and growth}
\setcounter{figure}{0} \renewcommand{\thefigure}{B.\arabic{figure}}
\setcounter{table}{0} \renewcommand{\thetable}{B.\arabic{table}}
\setcounter{equation}{0} \renewcommand{\theequation}{B.\arabic{equation}}
\setcounter{subsection}{0} \renewcommand{\thesubsection}{B.\arabic{subsection}}

\subsection{Macroscopic momentum}

Consideration of residual stress begins with examination of the balance of 
linear momentum in the form \eqref{eq:altmac}, repeated and reorganized for convenience:
\begin{equation}
\label{eq:altmacB}
\partial_A P^A_a  +  P^B_a \gamma^A_{AB} - P^A_c \gamma^c_{ba} F^b_A =
 -  \{ [ \bar{\partial}_B P^A_a  + P^A_a \bar{\partial}_B (\ln \sqrt{G}) ] \partial_A D^B
 + P^A_c (\gamma^c_{ba} - \Gamma^c_{ba}) F^b_A \}.
\end{equation}
\begin{rmk}
Terms on the left side of \eqref{eq:altmacB} are
standard for nonlinear elasticity theory \cite{marsden1983}.
If the free energy $\psi$ does not depend on $D^A$ or $D^A_{|B}$,
then the stress $P^A_a = \partial \psi / \partial F^a_A$ is also conventional,
presuming $\psi$ is such that in the undeformed state, $C_{AB} = G_{AB} \Rightarrow P^A_a = 0$.
In that case, when the right side of \eqref{eq:altmacB} vanishes,
the body manifold $\mathcal{M}$ should not
contain residual stresses when $F^a_A = \partial_A \varphi^a$ for
regular motions $\varphi^a(X)$ (e.g., in the absence of topological changes).
\end{rmk}
\begin{rmk}
Departures from classical nonlinear elasticity
arise when (i) $\psi$ has dependencies on $D^A$ or $D^A_{|B}$, (ii)
when $P^A_a$ or $G$ depends on $D^A$ along with heterogeneous 
state field $\partial_A D^B \neq 0$, or (iii)
when a different connection than the Levi-Civita connection
is used for $\Gamma^c_{ba}$ (i.e., $\Gamma^c_{ba} \neq \gamma^c_{ba}$
due to $d$-dependence of spatial metric $g_{ab}$).
Each of these departures could potentially induce stresses $P^A_a \neq 0 $ in 
a simply connected body externally unloaded via $p_a = P^A_a N_A =  0$ everywhere on
its oriented boundary $\partial \! \mathcal{M}$ (i.e., residual stresses).
\end{rmk}

Analysis of a particular version of the general theory offers more insight.
First assume in \eqref{eq:mdecc} that
$\hat{g}^a_b \rightarrow \delta^a_b$ such that 
$g_{ab} (x,d) \rightarrow g_{ab}(x) = \bar{g}_{ab}(x)$: the spatial metric
tensor $\vec{g}$ is Riemannian rather than Finslerian.  Then
$\gamma^a_{bc} = \Gamma^a_{bc}$.
Now use the osculating Riemannian interpretation 
of the Finslerian material metric $\vec{G}$ offered by Corollary 2.1.1 manifesting
from \eqref{eq:funcC}:
\begin{equation}
\label{eq:osc}
\tilde{G}_{AB} (X) = G_{AB}(X,D(X)), \qquad \tilde{G}(X) = \det (\tilde{G}_{AB}(X)),
\end{equation}
\begin{equation}
\label{eq:osc2}
\tilde{\gamma}^A_{BA} = \partial_B (\ln \sqrt{ \tilde{G} } ) =  
\partial_B (\ln \sqrt{ {G} } ) + \bar{\partial}_A  (\ln \sqrt{ {G} } )  \partial_B D^A =
\gamma^A_{BA} + \bar{\partial}_A  (\ln \sqrt{ {G} } )  \partial_B D^A,
\end{equation}
\begin{equation}
\label{eq:osc3}
\qquad 
\tilde{P}^A_a (X) = P^A_a (X,D(X)), \qquad
\partial_B \tilde{P}^A_a = \partial_B P^A_a + \bar{\partial}_C P^A_a  \partial_B D^C.
\end{equation}
Substituting \eqref{eq:osc2} and \eqref{eq:osc3} into \eqref{eq:altmacB} gives, with $\gamma^c_{ba} = \Gamma^c_{ba}$,
\begin{equation}
\label{eq:alt2}
\partial_A \tilde{P}^A_a  +  \tilde{P}^B_a \tilde{\gamma}^A_{AB} - \tilde{P}^A_c {\gamma}^c_{ba} F^b_A = 0.
\end{equation}
\begin{rmk}
Expression \eqref{eq:alt2} has the standard appearance for static equilibrium in classical continuum mechanics,
but stress $\tilde{P}^A_a$ and connection $\tilde{\gamma}^A_{BC}$
both implicitly depend on internal state $D^A$. The former, $\tilde{P}^A_a$, possibly depends on $D^A_{|B}$ in addition to $D^A$ if state gradient $D^A_{|B}$ appears in the arguments of $\psi$. Coefficients $\tilde{\gamma}^A_{BC}$ are those of the Levi-Civita connection of $\tilde{G}_{AB}$ via \eqref{eq:LC1o}.
\end{rmk}
\noindent Now neglect dependence on internal state gradient in the energy density,
require $D$-dependence to arise only through $G_{AB}$, and assume the body is homogeneous (with mild abuse of notation):
\begin{equation}
\label{eq:psiosc}
\psi = \psi(F^a_A,D^A) = {\psi}(F^a_A, G_{AB}(X,D)) = \tilde{\psi}(F^a_A, \tilde{G}_{AB}(X))
= \tilde{\psi}(C_{AB} (F^a_A, g_{ab}),  \tilde{G}_{AB}(X)).
\end{equation}
Recall from \eqref{eq:lengthC} that $C_{AB} = F^a_A g_{ab} F^b_B$. As a simple example, take, where $n = \dim \mathcal{M}$,
\begin{equation}
\label{eq:mooney}
\tilde{\psi} = {\frac{\mu_0}{2}} ( C_{AB} \tilde{G}^{AB} - n)
\Rightarrow \tilde{P}^A_a = \frac{\partial \tilde{\psi}}{ \partial F^a_A} =  \mu_0 g_{ab} \tilde{G}^{AB} F^b_B,
\quad
\frac{\partial \tilde{\psi}}{ \partial \tilde{G}_{AB} } = - \frac{\mu_0}{2}  \tilde{G}^{AC}  \tilde{G}^{BD} C_{CD},
\end{equation}
and where $\mu_0 > 0$ is a constant (e.g., an elastic shear modulus).
Now assume that spatial manifold $\mathfrak{m}$ is Euclidean \cite{yavari2010,ozakin2010} such that the Riemann-Christoffel curvature tensor from $\gamma^a_{bc}$ (and thus derived from $g_{ab}$) vanishes identically.
\begin{rmk}
In this case, \eqref{eq:alt2}, the last of \eqref{eq:psiosc}, and the example \eqref{eq:mooney}
are consistent with the geometric theory of growth mechanics of Yavari \cite{yavari2010}
in the setting of quasi-statics. Incompressibility can be addressed by appending linear momentum to include 
contribution from an indeterminant pressure to be determined by boundary conditions under isochoric constraint $J=1$ \cite{marsden1983}.
Otherwise, $\tilde{\psi}$ can be augmented with term(s) to ensure $C^A_B \rightarrow \delta^A_B \Rightarrow \tilde{P}^A_a = 0$ (e.g., \eqref{eq:w1d} for $n = 1$).
\end{rmk}
The Riemann-Christoffel curvature tensor from $\tilde{\gamma}^A_{BC}$ (and thus $\tilde{G}_{AB}$) need not vanish in general:
\begin{equation}
\label{eq:RCurv}
\tilde{\mathcal{R}}^A_{BCD} = 
\partial_B \tilde{\gamma}^A_{CD} - \partial_C \tilde{\gamma}^A_{BD}
+ \tilde{\gamma}^A_{BE}  \tilde{\gamma}^E_{CD}
- \tilde{\gamma}^A_{CE}  \tilde{\gamma}^E_{BD}.
\end{equation}
\begin{rmk}
In Riemannian geometry, $\tilde{\gamma}^A_{BC}$ are symmetric, differentiable, and obey \eqref{eq:LC1o};
\eqref{eq:RCurv} has $\frac{1}{12} n^2(n^2-1)$ independent components \cite{schouten1954}. For $n = 3$, $\tilde{\mathcal{R}}^A_{BCD}$ contains six independent components, determined
completely by the metric and Ricci curvature $\tilde{\mathcal{R}}^A_{ABC}$ \cite{yavari2010,clayton2014}.
For $n=2$, $\tilde{\mathcal{R}}^A_{BCD}$ contains only one independent component, determined completely
by the scalar curvature $ \tilde{\kappa} = \frac{1}{2} \mathcal{R}_{AB} \tilde{G}^{AB}$.
For $n=1$, $\tilde{\mathcal{R}}^A_{BCD}$ always vanishes (i.e., a 1-D manifold is always flat in this sense).
\end{rmk}
\noindent When $\tilde{\mathcal{R}}^A_{BCD}$ is nonzero over a region of $\mathcal{M}$, then
no compatible deformation $\tilde{F}^A_a(X)$ exists that can push-forward $\tilde{G}_{AB}$
to match the Euclidean metric $g_{ab}(\phi(X))$ that would render corresponding
regions of $\mathcal{M}$ and $\mathfrak{m}$ isometric. In other words,
the push-forward $g_{ab} = \tilde{F}^A_a \tilde{G}_{AB} \tilde{F}^B_b$
where $\tilde{F}^A_a = \partial_a \zeta^A$ does not exist, $\zeta^A$
being (nonexistent) Euclidean coordinates on $\mathcal{M}$.
In such cases, $\mathcal{M}$ would always have to be deformed (e.g., strained)
to achieve its spatial representation $\mathfrak{m}$, since no isometry exists
between the two configurations.
\begin{rmk}
If an intrinsically curved body manifold in the reference state $\mathcal{M}$ is stress-free
per the constitutive prescription (e.g., \eqref{eq:mooney} augmented per Remark B.1.4 or another standard elasticity
model), then
the intrinsically flat body in the current state $\mathfrak{m}$ would be necessarily strained and stressed, even if external traction $p_a$ vanishes along its boundary. Thus, this particular rendition of the generalized Finsler theory supplies residual stress from a non-Euclidean material metric tensor $\tilde{G}_{AB}$ in a manner matching other works that use Riemannian geometry \cite{yavari2010,ozakin2010}.
\end{rmk}

In the full version of the generalized Finsler theory \cite{clayton2017g,claytonMMS22}, as discussed following \eqref{eq:altmacB}, residual stresses could emerge from additional sources to those discussed
under the foregoing assumptions of a Euclidean spatial metric, a conventional
hyperelastic energy potential, and an osculating Riemannian material metric with non-vanishing curvature.
A number of different curvature forms can be constructed from the various
connections and derivatives of Finsler geometry and its generalizations \cite{bejancu1990,bao2000}.
Further analysis, beyond the present scope, is needed to relate these
geometric objects to physics in the continuum mechanical setting, including residual stresses.
\begin{rmk}
Deformation gradient $F^a_A$ could be decomposed into a product of two
mappings \cite{clayton2017z}: $F^a_A (X) = \partial_A \varphi^a(X) = (F^E)^a_\alpha (X) (F^D)^\alpha_A (D(X))$.
In this case, the strain energy potential is written to emphasize the elastic
deformation $\vec{F}^E$, with the state-dependent deformation $\vec{F}^D$
accounting explicitly for inelastic deformation mechanisms, including growth \cite{rodriguez1994,lubarda2002}.
In this setting, residual stresses can arise if $(\vec{F}^E)^{-1}$ and thus $\vec{F}^D$ do not fulfill certain integrability conditions: neither two-point tensor  $(\vec{F}^E)^{-1}$ nor $\vec{F}^D$ is always integrable to a vector field \cite{clayton2014}.
\end{rmk}

\subsection{Micro-momentum and growth}

Now consider the internal state-space equilibrium equation, \eqref{eq:altmic}, first under the foregoing assumptions
used to derive \eqref{eq:alt2}. Furthermore, take $N^A_B = N^A_B(X)$, $K^A_{BC} = \gamma^A_{BC} (X)$,
and $\alpha = 1$.
Then, with these assumptions, in the osculating Riemannian interpretation of Corollary 2.1.1, \eqref{eq:altmic} is
\begin{equation}
\label{eq:altmicB}
\begin{split}
  \partial_A \tilde{Z}^A_C +   \tilde{Z}^B_C \tilde{\gamma}^A_{AB}  - \tilde{Z}^A_B \gamma^B_{AC} - 
  Q_C  = 
   \psi  \bar{\partial}_C (\ln \sqrt{G}) - R_C,
\end{split} 
\end{equation}
\begin{equation}
\label{eq:tildeZ}
\tilde{Z}^A_B(X) = Z^A_B (X,D(X)) = \frac {\partial \psi} {\partial D^B_{|A}} (X,D(X)),
\qquad 
Q_A(X,D(X)) = \frac { \partial \psi}{\partial D^A} (X,D(X)),
\end{equation}
where \eqref{eq:tildeZ} follows from \eqref{eq:deltapsi}.
Use energy density $\psi$ of \eqref{eq:psiosc}, so $\tilde{Z}^A_B =  0$ identically.
Choose the volumetric source term $R_C =  \psi  \bar{\partial}_C (\ln \sqrt{G})$, which here
represents the local change in energy density per unit reference volume due to
effects of growth on the local volume form $d \Omega (X,D)$, since now, per \eqref{eq:omegavar1} of Appendix A, $ \psi \delta (d \Omega) = \psi [ \bar{\partial}_C (\ln \sqrt{G}) \delta(D^C)] d \Omega = R_C \delta (D^C) d \Omega$. 
\begin{rmk}
Physical justification exists in the context of growth mechanics for biological systems: 
$R_C$ can account for the affect on energy density from changes in mass due to tissue growth \cite{yavari2010,lubarda2002}. Thus \eqref{eq:altmicB} further reduces to, with \eqref{eq:psiosc},
to a form very similar to the equilibrium case of Yavari \cite{yavari2010} (e.g., matching equation (2.73) of ref.~\cite{yavari2010} with vanishing time derivative, if  here $\bar{\partial}_A G_{BC}$ is arbitrary):
\begin{equation}
\label{eq:altmicB2}
Q_A = \frac{ \partial \psi}{\partial D^A } = \frac{\partial \psi} {\partial G_{BC}} \frac{\partial G_{BC}} {\partial D^A} = 
\frac{\partial \tilde{\psi}} {\partial \tilde{G}_{BC}} \frac{\partial G_{BC}} {\partial D^A} = 0.
\end{equation}
\end{rmk}

To see how internal state components $\{D^A\}$ can represent growth, consider the case $n=2$ (i.e., 2-D $\mathcal{M}$ such as a biological membrane), by which $\{ D^A \} \rightarrow (D^1,D^2)  = ( l_1 \xi^1, l_2 \xi^2)$, where $l_{1,2} > 0$ are normalization constants
that render the $\xi^A$ dimensionless. Choose a polar (i.e., cylindrical $\{ X^A \} \rightarrow (R,\Theta$)) coordinate system on
a region of $\mathcal{M}$ with \eqref{eq:mdec} applying, such that $\bar{\vec{G}} = \text{diag}(1,R^2)$.
Assume a generalized Finslerian contribution $\hat{\vec{G}} = \text{diag}( \text{exp}(h_1 ( \xi^1)), \text{exp}(h_2(\xi^2))$, where $h_1(D(X)) = h_1(D^1(R,\Theta)/l_1)$ and
$h_2(D(X)) = h_2(D^2(R,\Theta)/l_2)$ are differentiable functions of their arguments.
 In matrix form, the second of \eqref{eq:mdec} becomes, in this example of anisotropic growth,
\begin{equation}
\label{eq:Gg}
[G_{AB}] =
\begin{bmatrix}
G_{RR}& 0 \\
0 & G_{\Theta \Theta} 
\end{bmatrix}
=
 [\hat{G}^C_A][\bar{G}_{CB}] 
= 
\begin{bmatrix}
\text{exp}(h_1 (\xi^1)) & 0 \\
0 & R^2 \text{exp}(h_2(\xi^2))
\end{bmatrix}.
\end{equation}
A more specific case is now studied in detail. Denote by $\chi(R)$ a radial growth function.  Then set
\begin{equation}
\label{eq:Yav0}
\xi = \xi^1 = \frac{D^1}{l_1} = \frac{D^2}{l_2} = \xi^2, \quad \xi = \xi(R); \qquad h = h_1 = -h_2 = 2 \chi, \quad h = h (\xi(R)) = 2 \chi(R).
\end{equation}
This produces metric $\tilde{G}_{AB}(X)$ of Yavari (ref.~\cite{yavari2010}, eq.~(2.101)) for anisotropic growth of an annulus:
\begin{equation}
\label{eq:Yav1}
\begin{split}
[{G}_{AB}(R,\xi)] & = 
\begin{bmatrix}
\text{exp}(h (\xi(R))) & 0 \\
0 & R^2 \text{exp}(-h(\xi(R)))
\end{bmatrix} 
\\
\quad \Rightarrow \quad [\tilde{G}_{AB}(R)] & = 
\begin{bmatrix}
\text{exp}(2 \chi(R)) & 0 \\
0 & R^2 \text{exp}(-2\chi(R))
\end{bmatrix}.
\end{split}
\end{equation}
\begin{rmk}
In this special case given by \eqref{eq:Yav1}, internal state changes preserve volume via $\det (G_{AB}(X,D)) = R^2$ being independent of $\chi$, $\xi$, and $D$, so $C^B_{AB} = 0$. 
\end{rmk}
Now apply the energy potential \eqref{eq:mooney}, such that for internal state equilibrium, 
\eqref{eq:altmicB2} becomes, defining $\dot{h}(\xi) = \d h(\xi) / \d \xi$,
\begin{equation}
\label{eq:Yav2}
Q_A  = - \frac{\mu_0}{2}  \tilde{G}^{BD} \tilde{G}^{CE} C_{DE}
\bar{\partial}_A G_{BC}
= 0 \, \Rightarrow \,
\begin{cases}
2  l_1 Q_1(\xi,R) = - \mu_0  \text{exp}(-h(\xi)) C_{RR} \dot{h}(\xi) = 0, \\
2  l_2 Q_2(\xi,R) =  \mu_0  R^{-2} \text{exp}(h(\xi)) C_{\Theta \Theta} \dot{h}(\xi) = 0. \\
\end{cases}
\end{equation}
Thus, equilibrium of internal state is only ensured for this particular strain energy function and material metric
when $\dot{h} = 0$.
A sample function with three equilibrium states at $\xi = 0, \frac{1}{2}, 1$ is the double well:
\begin{equation}
\label{eq:dwell}
h = \xi^2 (1 - \xi)^2, \qquad \dot{h} = 2 \xi ( 1 - \xi) ( 1 - 2 \xi).
\end{equation}
Now revisit the Levi-Civita connection and curvature for the metric $\tilde{\vec{G}}$ in \eqref{eq:Yav1}.
Denote $h'(\xi(R)) = [\d h(\xi(R)) / \d \xi] [\d \xi(R) / \d R] = \dot{h}(\xi) \xi'(R)$.
From \eqref{eq:LC1o}, the $\tilde{\gamma}^A_{BC}$, have the non-vanishing components
\begin{equation}
\label{eq:gammaY}
\tilde{\gamma}^R_{RR} = \frac{h'}{2}, \quad
\tilde{\gamma}^R_{\Theta \Theta} = \text{exp}(-2h) \left( \frac{R^2 h'}{2} - R\right), \quad 
\tilde{\gamma}^\Theta_{R \Theta} = \tilde{\gamma}^\Theta_{ \Theta R} 
= \frac{1}{R} - \frac{h'}{2}.
\end{equation}
Recalling $\tilde{\kappa}$ is the scalar curvature, the non-vanishing covariant components of 
$\tilde{\mathcal{R}}_{BCDE} = \tilde{\mathcal{R}}^A_{BCD} \tilde{G}_{AE} $
are, from \eqref{eq:RCurv},
\begin{equation}
\label{eq:RCpolar}
\begin{split}
\tilde{\mathcal{R}}_{R \Theta R \Theta} & =  \tilde{\mathcal{R}}_{ \Theta R \Theta R }
= - \tilde{\mathcal{R}}_{R \Theta  \Theta R} = - \tilde{\mathcal{R}}_{ \Theta R R \Theta} = - R^2 \tilde{\kappa} 
= - [\partial_R \tilde{\gamma}^R_{\Theta \Theta} + \tilde{\gamma}^R_{\Theta \Theta} (\tilde{\gamma}^R_{RR}  - \tilde{\gamma}^\Theta_{R \Theta})] \tilde{G}_{RR}
\\ & = - \frac{\d}{\d R} \left[
\text{exp}(-2h)  \left( \frac{R^2 h'}{2} - R \right)
\right] \text{exp}(h)
+ 
 \left( \frac{R^2 h'}{2} - R \right)
\left(  \frac{1}{R} - h' \right) \text{exp}(-h) 
\\ & =
-\frac{R}{2} \text{exp}(-h) \left[ R \{ h'' - (h')^2 \} + 4h' \right]  
\\ & = -\frac{R}{2} \text{exp}(-h) \left[ R \left( \frac{ \d^2 \xi } { \d R^2} + \frac{\d \xi}{\d R} \frac{\d} {\d R} \right) \frac{\d h }{ \d \xi} 
- R \left( \frac{\d h }{ \d \xi}   \frac{\d \xi}{\d R} \right)^2 + 4 \frac{\d h }{ \d \xi} \frac{\d \xi}{\d R} \right] .
\end{split}
\end{equation}
Take the annular material manifold $\{ \mathcal{M}: R \in [R_0,R_1], \Theta \in [0,\Theta_1] \}$, $R_1 > R_0 > 0$, $\Theta_1 < 2 \pi$.
Since $R>0$ and for bounded $h$, the local flatness condition from \eqref{eq:RCpolar} is
\begin{equation}
\label{eq:flatR}
R \{ h'' - (h')^2 \} + 4h' = 0 \quad \leftrightarrow \quad R \left( \frac{ \d^2 \xi } { \d R^2} + \frac{\d \xi}{\d R} \frac{\d} {\d R} \right) \frac{\d h }{ \d \xi} 
- R \left( \frac{\d h }{ \d \xi}   \frac{\d \xi}{\d R} \right)^2 + 4 \frac{\d h }{ \d \xi} \frac{\d \xi}{\d R} = 0.
\end{equation}
\begin{rmk}
The first of \eqref{eq:flatR} is a second-order nonlinear ODE for radial distribution of the generic function $h = h(R)$.
The second is a second-order nonlinear ODE for $\xi= \xi(R)$ that could be solved if intermediate functional form $h(\xi)$ is known a priori (e.g., \eqref{eq:dwell}).
Trivial solutions are $ h(R) = \text{constant}$ and $\xi(R) = \text{constant}$.
General non-trivial analytical solutions are not obvious. Given appropriate boundary conditions, determination of particular non-trivial solutions for flatness, if they exist, would appear to require numerical methods.
\end{rmk}



\bibliography{refs}

\end{document}